\documentclass[12pt, preprint]{aastex}

\usepackage{graphicx}
\usepackage{epsfig}
\usepackage{natbib}
\usepackage{rotating}

\slugcomment{To appear in PASP}

\shorttitle{Giant Planet Imaging}

\shortauthors{Beichman et al.}

\graphicspath{{converted_graphics/}}

\begin{document}

\title{Imaging  Young Giant Planets From Ground and Space} 
\author{Charles A. Beichman} 
\affil{NASA Exoplanet Science Institute, Jet Propulsion Laboratory, California Institute of Technology}
\author{John Krist, John T. Trauger}
\affil{Jet Propulsion Laboratory}
\author{Tom Greene}
\affil{NASA Ames Research Center}
\author{Ben Oppenheimer, Anand Sivaramakrishnan}
\affil{American Museum of Natural History}
\author{Ren\'e Doyon}
\affil{Universit\'e de Montr\'eal}
\author{Anthony Boccaletti}
\affil{Observatoire de Meudon}
\author{Travis S. Barman}
\affil{Lowell Observatory}
\author{Marcia Rieke}
\affil{University of Arizona}

\date{\today}  

\begin{abstract}

High contrast imaging can find and characterize gas giant planets around nearby young stars and the closest M stars, complementing radial velocity and astrometric searches by exploring orbital separations inaccessible to indirect methods. Ground-based coronagraphs are already probing within 25 AU of nearby young stars to find objects as small as $\sim$ 3 Jupiter masses. This paper compares near-term and future ground-based capabilities with high contrast imaging modes of the James Webb Space Telescope (JWST). Monte Carlo modeling reveals that JWST can detect planets with masses as small as 0.2 M$_{Jup}$ across a broad range of orbital separations. We present new calculations for planet brightness as a function of mass and age for specific JWST filters and extending to 0.1 M$_{Jup}$.


\end{abstract}

\maketitle

\section{Introduction} 
 
As coronagraphic and adaptive optics technologies improve, the number of directly imaged planets is increasing, most recently with 4 companions being detected in orbit around two nearby A stars. Because the 3 planets around HR8799 \citep{marois} and the single planet around Fomalhaut \citep{kalas} are young, their internal reservoirs of gravitational energy generate enough luminosity to make these objects visible \citep{saumon}. In addition, there is an as-yet-unconfirmed planet seen once around $\beta$ Pic \citep{lagrange}. These young planets plus earlier discoveries, e.g. 2M1207-3932b \citep{chauvin} and GQ Lup b \citep{neuhauser}, are confirmed to be companions via their common proper motion with their host star and in a number of cases by orbital motion as well. While it is possible to use estimated ages and evolutionary tracks to distinguish in a gross sense between planets ($<$13 M$_{Jup}$, the deuterium burning limit), brown dwarfs ($13<M<70$ M$_{Jup}$, the hydrogen burning limit) and low mass stars ($>$70 M$_{Jup}$), it is difficult to assign a reliable mass to directly imaged companions. In some cases dynamical estimates based on the configuration of the debris disk constrain the planet mass, e.g. $<$3 M$_{Jup}$ for Fomalhaut-b \citep{kalas,chiang}.

The relationships between near-IR brightness, age, and mass are uncertain and dynamical mass determinations are difficult for planets on long period orbits, particularly in the absence of a dust disk. What is needed to anchor the models of young planets are objects of known age with a combination of imaging (giving luminosity and effective temperature) plus dynamical information (giving mass). This combined information will come from direct imaging and dynamical mass measurements from ground-based RV or astrometry from either the ground \citep{vanbelle,pott} or space using SIM-Lite \citep{beichman01,tanner} or GAIA \citep{sozetti}. Detections of transiting young planets would be extremely valuable, but the variability of young host stars may make these planets hard to detect and the extreme environment of ''hot Jupiters'' may make it difficult to draw general conclusions. 

Direct imaging has opened a new region of the Mass--Semi Major Axis (SMA) parameter space for planets (Figure~1) and has given rise to new theoretical challenges. The existence of giant planets at separations larger than $\sim 10$ AU is difficult to account for in standard core accretion models \citep{pollack, ida05,sdr09} and a different formation mechanism, gravitational fragmentation in the disk \citep{boss} may be operating. Alternatively, a combination of the two mechanisms may be responsible for these distant planets with outward migration or planet-planet scattering moving planets formed in dense inner regions onto orbits as distant as 100 AU \citep{veras}.

There have been a number of investigations of coronagraphic imaging of planets, particularly in the context of designs for the Terrestrial Planet Finder (TPF-C), including Agol (2007), Beckwith (2009) and Brown (2009). These authors have investigated the challenging task of finding planets, both gas or ice giants and terrestrial planets, through their {\it reflected light}. The reflected light signal depends on the inverse square of the star-planet separation, the planetary albedo, and orbital phase function with resulting planet-star contrast ratios as small as $10^{-8}$ to $10^{-11}$ (Jupiters and Earths at 1 AU, respectively). The goal of these authors has been to either optimize the design of TPF in terms of aperture size \citep{beckwith} or to optimize   search strategies for various  TPF designs \citep{agol}. This paper addresses a more near term and far less challenging problem, namely the detection of {\it self-luminous} gas giants using telescopes and instruments that either are or will become operational in the next 5-10 years. The contrast ratios for self-luminous giant planets are far more favorable, $10^{-6}$ to $10^{-8}$  and the details of the calculations are very different from the reflected light case.

In particular, we explore the prospects for imaging self-luminous giant planets from the ground and from space using the James Webb Space Telescope (JWST; Gardner et al 2006). This application of JWST for exoplanet research complements recent studies \citep{greene, deming} discussing the role of JWST for transit spectroscopy. We investigate how imaging surveys might yield statistical information on the distribution of planets as functions of mass and orbital location. In what follows we describe two samples of stars suitable for direct searches, nearby young stars and nearby M stars ($\S2$), introduce a number of instruments suitable for planet surveys ($\S3$), describe a plausible population of exoplanets ($\S4$), and utilize a Monte Carlo simulation to predict the yield of surveys under different scenarios ($\S5$). We explicitly examine the prospects for finding planets for which both imaging and dynamical observations might become available ($\S6$).

\section{The Stellar Sample}

The two most important factors from an observational standpoint  in searching for planets are star-planet contrast ratio and angular resolution. Young gas giant planets generate enough luminosity via gravitational contraction to be bright in the near-IR \citep{saumon, burrows2003}, making ages less than $\sim$ 1 Gyr an important characteristic of appropriate target stars. Because the Inner Working Angles (IWA) of typical observing systems are limited to a few tenths of an arcsecond ($\S3$) or a few tens of AU at the distance of typical young stars, proximity of target stars is another important criterion. These astrophysical and observational factors, youth and proximity, lead to two natural populations for study: the closest stars with ages less than  1 Gyr and the closest M stars for which a Jupiter mass planet of even a few Gyr would be detectable and for which the inner few AU become accessible.  The samples of stars discussed below are meant to be representative enough to allow the detectability of planets to be investigated as a function of distance  and age, the two most critical variables for all imaging investigations, for a wide variety of instruments. Samples developed for individual projects will have be made  more rigorously  in terms of age, mass, metallicity, distance, binarity, cluster environment, etc. as appropriate to a specific set of scientific goals. For example, for simplicity in the present study we exclude binary stars despite the obvious interest in the question of planets in such systems. For close binaries it is beyond the scope of this paper to calculate properly the coronagraphic response interior to the IWA at the 10$^{-7}$ level relevant to some of the instruments. For more widely separated binaries, the presence of an un-obscured bright companion in the camera field of view (2-15\arcsec) can wreak havoc with instrument performance. The results presented here represent lower limits to the performance on binaries.

\subsection{A Sample of Young Stars}

Extremely young objects, 1-5 Myr old, are found in well known star forming clusters associated with nearby molecular clouds such as Taurus and Chameleon (Table~\ref{clusters}). The closest of these associations are 100-140 pc away so that a classical coronagraph on a 5-8 m telescope could probe only beyond 15-25 AU at 1.6 $\mu$m (80 AU at 4.4 $\mu$m). An improved view of the inner parts of young planetary systems requires closer stars (or eventually a larger telescope like the proposed 30-40 m facilities). We have supplemented the youngest stars with ``adolescent'' stars having ages between 10 and 1 Gyr. These have been identified via X-ray emission, isochronal analysis, and common proper motion and can be as close to the Sun as 25 pc \citep{zuckerman}. Depending on the wavelength and instrument, these systems can be probed to within a 5-10 AU of their host stars. 

We have utilized a number of compilations of infant and adolescent stars to assemble a target sample. First, $\sim$200 stars chosen for an astrometric survey for gas giants with the Space Interferometer Mission (SIM-Lite) \citep{beichman01, tanner} encompass both classical and weak-lined T Tauri stars with masses from 0.2-2 M$_\odot$, ages from 1 Myr up to 100 Myr and distances from 25-140 pc. Second, the Spitzer FEPS survey \citep{meyer} includes over 300 stars of F,G,K spectral types with ages of roughly 10 Myr to 1 Gyr \citep{hillenbrand}. Third, a group of A stars selected for debris disk observations with Spitzer \citep{rieke05} includes clusters out to several hundred pc. We restricted the A-star sample to 150 pc and added additional single A0-A9 (IV/V or V) stars with credible ages to obtain a more complete sample of almost 200 stars out to 50 pc. The properties of the various samples are given in Tables 1 and 2 and illustrated in Figures~\ref{stellarprops} and ~\ref{HmagHisto}.

\subsubsection{Influence of Stellar Properties on Incidence of Planets}

Various stellar properties may affect the likelihood of a star developing and retaining a planetary system, including stellar mass, metallicity, and the presence of disks.

High stellar mass may enhance the probability of a star having one or more gas giant planets. There are theoretical grounds for expecting this effect \break  \citep{ida05,sdr09} as well as observational hints from observations of sub-giants with 1-2 M$_\odot$ precursors. Johnson (2007) shows a factor of 3 increase in the incidence of RV-detected planets between host stars with 0.5-1.5 M$_\odot$ and those with masses $>$1.5 M$_\odot$. Conversely, low mass M stars appear to have a smaller incidence of gas giant planets as determined from RV studies \citep{butler} and initial coronagraphic surveys \citep{mccarthy, oppenheimer2001,oppenheimer} and \citep{metchev}. We have taken these effects into account in our modeling ($\S4$) by a) increasing the incidence of higher mass planets around stars with mass greater than 1.5 M$_\odot$ and b) restricting the incidence of high mass planets around low mass stars. It is well known that higher metallicity enhances the probability of mature stars to have Jupiter mass giant planets \citep{valenti}, but results for planets of lower mass suggest that this effect is not important for Neptune mass planets \citep{sousa}. Almost nothing is known about whether or how these effects operate at the larger orbital distances probed by imaging surveys. Although Agol (2007) shows that biasing a survey to high metallicity can improve a survey's yield by 14-19\%, we do not include this effect in the models considered herein. If one simply wants to maximize the probability of finding (high mass) planets, then one might take these trends into account by focusing on high mass, high metallicity stars. However, understanding these dependencies (particularly if the population of distant planets differs from interior planets) must be addressed via unbiased surveys. 

Debris or protostellar disks can present a challenge for planet searches. They may serve as marker for the presence of planets, e.g. Fomalhaut and possibly $\beta$ Pic, but may also mask the presence of planets if the diffuse emission is bright enough. For most of the coronagraphic targets considered here, we show that the search for self-luminous planets is unaffected by diffuse emission. 

\begin{itemize}
 \item {\it Target selection}. We excluded target stars with large amounts of nebulosity and/or optically thick disks, i.e. high values of disk to stellar luminosity ($L_d/L_*$). This restriction excludes obscured or partially obscured objects which are typically the youngest protostars still possessing primordial, gas rich disks. For example, the detection of planets of even Jovian mass would be extremely difficult within the disk of AB Aur which has a disk with $L_d/L_*\sim 0.6$ \citep{tannirk}. The SIM sample was explicitly chosen to exclude objects with nebulosity by inspection of imaging data \citep{tanner}. The Spitzer data for the FEPS sample revealed only six objects with high optical depth disks, i.e. $L_d/L_*>0.01$ \citep{hillenbrand}. The other 26 sources in the FEPS sample with prominent Spitzer disks had an average value of $log(L_d/L_*)= -3.8$ with a dispersion of $\pm0.5$, comparable to the range bounded by $\beta$ Pic and Fomalhaut. 
 \item {\it Residual diffuse emission} We developed a very simple model of diffuse emission appropriate to the stars and spatial scales examined in these simulations ($\sim$ 5-100 AU) and then calculated the flux density of spurious point sources relative to the brightness of the star. As will be discussed below ($\S3.5$), the level of spurious objects due to clumps in the local background is well below the level of residual scattered starlight in the instruments considered here for self-luminous planets and typical disks, i.e. $log(L_d/L_*)= -3.8$.
 \end{itemize} 

Finally, it is well known that stellar ages are difficult to estimate. We investigated this effect on the simulations by allowing the age to vary with a log-normal distribution having a dispersion of a factor of two around the nominal age. The average properties of the detected planets were not appreciably affected by this variation. Younger ages made some planets more easily detectable, while older planets fell below detection limits. The derived properties, especially the mass, of a planet detected around any particular star will depend, of course, on the uncertain age of the parent star.

\subsection{A Sample of Nearby M Stars}

The very closest stars to the Sun also offer the prospect of finding self-luminous planets. Nearby M stars are advantageous since the parent star is 5-10 magnitudes fainter than higher mass stars, leading to a more favorable contrast ratio for self luminous planets, and because their proximity to the Sun can expose planets located within a few AU of the star. Unfortunately, field M stars are typically older than 1 Gyr, implying that their planets will be faint. Further, these visually faint stars are relatively poor targets for ground-based Adaptive Optics systems relying on visible stellar photons for wavefront correction. Nevertheless, a dozen objects of potentially planetary mass have been found around M stars either via RV (GJ 876b) or  imaging (2MASS 1207b), so it is useful to consider what objects might be detectable with imaging. We have assembled a list of 196 M stars (M0-M9V) within 15 pc that are either single or whose companions are at least 30$\arcsec$ distant according to the SIMBAD and NStED databases. We added to the sample AU Mic (GJ 803) and AT Mic which, although they are in multiple systems, are young and therefore potential hosts of bright planets. The AU Mic debris disk \citep{plavchan} has a disk to star luminosity ratio of $log(L_d/L_*)=-3.4$ which is not enough to impede detection of most planets ($\S3.3$). We derived very approximate ages for the stars using their X-Ray luminosity (when available) and a X-ray-age relationship derived from Preibisch and Feigelson (2005; their Figure 4). For the 100 nearby M stars in our sample without this information, we adopted a representative age of 5 Gyr \citep{zapatero}.

\section{The Instruments}

The key parameters of an instrument used for direct imaging of planets are its Inner and Outer Working Angles (IWA and OWA), its starlight rejection as a function of angular separation from a target star,  its optical efficiency and sensitivity. We discuss these parameters for a number of ground-based and space-based instruments. For ground-based instruments we include the currently operational Near Infrared Infrared Coronagraph (NICI) on the Gemini Telescope \citep{biller09} which is comparable in performance to the Subaru/HiCIAO instrument \citep{suzuki} as well as next generation coronagraphs in development for the Palomar telescope (P1640), the Gemini Planet Imager and SPHERE on the VLT, and an idealized coronagraph on a 30 m telescope (TMT). We include a ground-based 5 $\mu$m capability based on the Multiple Mirror Telescope (MMT). These ground-based capabilities are contrasted with three JWST instruments: a Lyot coronagraph on NIRCam (various wavelengths from 2.1-4.6 $\mu$m), a Non Redundant Mask on the Tunable Filter Instrument (TFI/NRM), and a Four Quadrant Phase Mask on the Mid Infrared Instrument (MIRI/FQPM). This information is summarized in Tables~\ref{resolution} and \ref{CoronaProps} and Figure~\ref{ContrastRatio}a.

The IWA of a telescope of diameter, $D$, equipped with a Lyot coronagraph has a typical radial extent of $\sim2-5\lambda/D$, or typically a few tenths of an arcsec in the near-IR. For a classical Lyot coronagraph the IWA is defined unambiguously by a hard-edged mask while for a band-limited coronagraph the IWA is not a single number but can be defined as the angle at which the off-axis transmission drops below 50\%. The Outer Working Angle (OWA) of an instrument depends on such parameters as the size of the detector array in a simple imaging coronagraph, the number of actuators in a deformable mirror, or the size of a sub-aperture in an interferometric system. Table~\ref{resolution} summarizes this information and projects this angular scale out to different distances. It is clear from the table that probing the region interior to 100 AU requires target systems within 150 pc and preferably much closer. 

\subsection{Ground-based Coronagraphs}

Coronagraphs on large ground-based telescopes are evolving rapidly with advances in coronagraph design, extreme Adaptive Optics and post-coronagraph wavefront control. The current generation of coronagraphs are finding young gas giants, e.g. HR 8799, and new instruments such as Subaru/HiCIAO and Gemini/NICI will push these searches to lower masses and smaller orbits with contrast ratios of $\Delta$Mag=13-15 mag \citep{biller09,wahhaj}. The next generation instruments, including P1640 at Palomar Mountain \citep{hinkley,anand2008}, the Gemini Planet Imager (GPI) \citep{mac} and the SPHERE instrument \citep{beuzit,boccaletti} on the Very Large Telescope (VLT), will achieve contrast ratios of $\Delta$Mag$\sim$18 mag at 1$\arcsec$ and thus probe masses $\sim$ few M$_{Jup}$. P1640 will be limited to Northern Hemisphere targets whereas many of the nearest, young stars are visible only from Southern observatories. GPI and SPHERE with observe Southern targets with contrast limits comparable to P1640's but with improved angular resolution and magnitude limits due to their larger apertures. We project the performance of future coronagraphs on the next generation of 30+ m telescopes such as the TMT, GMT and E-ELT, adopting a contrast ratio floor of $10^{-8}$ in the mid-range of what has been discussed for these highly segmented telescopes \citep{mac}. It is important to note that ground-based coronagraphs operating with Extreme Adaptive Optics systems require bright target stars for the extreme wavefront control needed for contrast ratios of $<<10^{-6}$. Stars fainter than R$\sim$ 8 mag (H$\sim$5-7 mag for typical FGKM stellar colors) will have poorer coronagraphic performance (Figure~\ref{HmagHisto}). 

Finally, we note that ground-based imaging searches at 3 and 5 $\mu$m are already underway, trading increased planet brightness against higher thermal backgrounds \citep{heinze, kenworthy}. With L$^\prime$ and M band sensitivities of $\sim$ 16 mag and 14 mag (5 $\sigma$), respectively, on the MMT \citep{heinze}, surveys with instruments like Clio should be able to probe the 5-10 M$_{Jup}$ range within 10-100 AU. In the longer term, interferometry with the LBTI offers the prospect of examining nearby young stars with $<50$ mas resolution. However, the 8-10 magnitudes of difference in sensitivity between JWST (Table~\ref{CoronaProps}) and ground-based telescopes will gave JWST a substantial advantage for the imaging surveys considered here. We approximate the performance of a ground-based 5 $\mu$m coronagraph on a large telescope by adopting the characteristics of NIRCam's coronagraph but with a magnitude floor of M= 14 mag.

\subsection{High Contrast Imaging with the James Webb Space Telescope(JWST)}

Three of the instruments on JWST have capabilities for high contrast imaging. We present performance information on the coronagraphs planned for the Near-IR camera (NIRCam), the Tunable Filter Instrument (TFI), and the Mid-Infrared Instrument (MIRI). The calculations of contrast performance combine diffraction-based estimates of telescope performance including 131 nm of total wavefront error \citep{stahl} with instrument performance models provided by the instrument team members (co-authors on this paper) responsible for those modes. The reader is referred to the references quoted in each section for details. Since the JWST mirror is still being fabricated, estimates of telescope performance are subject to change. While the wavefront error is relatively large compared to standards of advanced AO systems (50 nm) or future coronagraphs designed to search for earths (1 nm for TPF-C; Trauger and Traub 2007), JWST will operate under extremely stable conditions (perhaps 10-20 nm variations over a few hours) and with extremely low backgrounds in the near- and mid-infrared where young exoplanets are bright.  JWST  achieves its (modest) imaging quality without reference  to a bright target star, making JWST  well-suited to  searching for planets around  faint stars inaccessible to ground-based telescopes. 

We do not examine the performance of JWST at wavelengths $\leq$ 2 $\mu$m for two reasons. First, JWST's coronagraphic performance at short wavelengths will depend critically on the (as yet) poorly known wavefront errors of the mirror making such predictions premature. Second, at short wavelengths, 8-30 m ground-based telescopes with extreme AO will have significant advantages over JWST for bright stars in imaging situations where scattered starlight dominates the noise and where large collecting areas can overcome modest sky backgrounds.

\subsubsection{NIRCam}

The NIRCam instrument \citep{marcia05} includes a coronagraph with 5 focal plane masks (Figure~\ref{nircamlayout}). Three round spots or ``Sombrero''-shaped masks and two wedge-shaped masks are optimized for design wavelengths of 2.1, 3.35 and 4.3 and 4.6 $\mu$m (Table~\ref{CoronaProps}; Green et al. 2005; Krist et al. 2007). The occulting spots are apodized, but only quasi-band limited \citep{kuchner} since the wavefront error in the JWST telescope is sufficiently large that the coronagraphic performance is dominated by the telescope scattering not by diffraction. The performance of the 5 masks is predicted on the basis of a full diffraction calculation assuming nominal performance of the segmented JWST primary and using appropriate Lyot stops and occulting spots \citep{krist}. Figure~\ref{nircamperformance} shows contrast ratios after speckle suppression has been carried out using roll subtraction ($\pm5\deg$ is allowed during JWST operations) and assuming a random position offset error of 10 mas between rolls and random wavefront variations of 10 nm. At 1\arcsec\  from the central star, JWST should achieve almost 12 magnitudes of suppression while at 4\arcsec\ the suppression will approach 18 mag. For the survey simulations described here we have used the predicted performance of the 4.3 $\mu$m (design wavelength) spot with an inner working angle of $6\lambda/D\sim 850$ mas. Comparable results are obtained for the 4.6 $\mu$m wedge occulter. Examination of Figure~\ref{nircamperformance} suggests that the spot occulter performs better at larger separations while the wedge works better at smaller angles. As discussed below, this expectation is confirmed in the simulations, although the differences are small. We examine NIRCam performance in two filters, F444W and F356W using sensitivity limits appropriate to the difference of two 1 hr exposures (Rieke et al 2005; http://ircamera.as.arizona.edu/nircam/features.html), degraded by a factor of two for the lower throughput of NIRCam (20\% vs 80+\%) with the coronagraphic pupil mask.

\subsubsection{The JWST Non Redundant Mask (NRM)}

The Fine Guidance System on JWST incorporates a near-IR science camera equipped with a Tunable Filter Instrument (TFI; Doyon et al. 2008). In addition to standard coronagraphic imaging modes, the TFI provides an important complement to the NIRCam coronagraph. The Non-Redundant Mask (NRM) is a true interferometer which will take advantage of JWST's extreme stability to make high contrast images at high angular resolution \citep{anand}. By masking out all but 7 sub-apertures each with projected size of $D_s\sim$1 m across JWST's 6.5 m pupil it is possible to create 21 independent baselines (Figure~\ref{NRMLayout}; Sivaramakrishnan et al. 2009) to observe with resolution $\sim0.5\lambda/D\sim0.07\arcsec$ over a field of view (radius) of $\sim0.6 \lambda/D_s\sim0.55\arcsec$ (Table~\ref{resolution}). Careful calibration of fringe visibilities with respect to reference stars should result in contrast ratios of $\Delta$Mag $\sim$ 12.5 mag \citep{anand}, a major improvement over typical ground-based values of 4-5 mag \citep{lloyd}. If visibility calibration proves impractical, the contrast performance will be a factor of $\sim$10 worse, i.e. $\Delta$Mag $\sim$ 10 mag. An important problem still to be addressed is the effect of detector stability on NRM performance in the presence of the unattenuated photon fluxes from bright central stars. Shot-noise and possible flat-field noise due to pixel-to-pixel variations of $>10^{-5}$ will limit contrast ratios for stars with 4.4 $\mu$m magnitudes of $\sim5$ mag or brighter. 

\subsubsection{The JWST Mid-Infrared Imager (MIRI)}

The Mid-Infrared instrument (MIRI) on JWST is equipped with three Four Quadrant Phase Masks (FQPM) operating in narrow bands (R$\sim$20) at 10.65, 11.4, and 15.5 $\mu$m as well as with a conventional coronagraph operating at 23 $\mu$m \citep{rouan,bocca}. The latter will predominantly be used for the study of disks since its IWA will be relatively coarse ($>2.2\arcsec$). With contrast ratios in the range of $\Delta$Mag= 8-12.5 mag at an IWA (radius) of $1\lambda/D\sim 0.36\arcsec$ at 11.4 $\mu$m,  the MIRI FQPM will be able to probe within a 10-20 AU of closest young host stars. The IWA for this instrument does not have a sharp edge so that companions interior to the nominal IWA would be visible but highly attenuated at $<1\ \lambda/D$. The MIRI/FQPM offers angular resolution between that of NIRCam and TFI/NRM, but with the advantage of a much wider field of view than TFI/NRM, up to 13$\arcsec$. The contrast curve shown in Figure~\ref{ContrastRatio} assumes subtraction of a Point Spread Function (PSF) reference star for speckle suppression, pointing jitter of 7 mas and a 20 nm variation in wavefront error between observations. A version of the FQPM has been in operation on the NACO instrument on the VLT since 2003 \citep{boccaletti2004} and the MIRI prototype has been tested in the laboratory \citep{baudoz} given confidence that the contrast goals described here can be achieved.

\subsection{Noise from Diffuse Emission}

As noted above, bright nebulosity and/or a disk around a young star is a potential source of noise for planet searches. Thermal emission from dust will be negligible at the angular separations ($>>1$ AU at 10s of pc), stellar luminosities, and short wavelengths ($<$ 5$ \mu$m) considered here; only MIRI observations for the closest, most luminous A stars might be affected by thermal emission. We thus focus on the effects of scattered light using observations of Fomalhaut \citep{kalas} and $\beta$ Pic \citep{golimowski} to develop a simple model for the brightness of possible sources produced in clumps in diffuse scattered light at large radii. Let the radial dependence of surface brightness be modeled as $I(r)=I_0(r_0)\left( \frac{r}{r_0}\right) ^{-2}\left( \frac{r}{r_0}\right) ^{-\beta}$ where the $r^{-2}$ term comes from the increased stellar illumination and the r$^{-\beta}$ term is due to the increasing surface density of dust as one moves closer to the star. We derived similar values of V$\sim$R band surface brightness of $I_0(100 AU)$ =20.6 mag arcsec$^{-2}$ and $\beta=2$ for Fomalhaut (face-on) to $I_0(100 AU)$ =19.2 mag arcsec$^{-2}$ and $\beta=1$ for $\beta$ Pic (edge-on), both normalized to $L_d/L_*=10^{-3.8}$ which is the average disk luminosity for sources in the FEPS sample \citep{hillenbrand}. The brightness of a spurious point source ($5\sigma$) from a clump in the disk emission, relative to the brightness of the star, $F_*$, can be written $F_{disk}/F_*=5 \eta I(r) \Omega/F_*$ where $\Omega\sim(\lambda/D)^2$ is the solid angle of a diffraction limited beam and $\eta=10\%$ is the fraction of the diffuse emission in a clump (as opposed to smooth emission which could be subtracted out). Figure~\ref{ContrastRatio}b includes curves based on the Fomalhaut profile for a star at 50 pc with $L_d/L_*=10^{-3.8}\, {\rm and} \ 10^{-3}$ for two instrumental cases: 1.65 $\mu$m and D=5 m; and 4.4 $\mu$m and D=6.5 m. The figure suggests that {\it for appropriately selected stars} spurious sources from scattered light will not be a significant problem beyond 0.5\arcsec \ and only a marginal problem at smaller separations for JWST/NRM or ground-based telescopes. Note that we have made the conservative assumption that the scattering efficiency of the dust grains is flat rather than falling off at wavelengths $>$ 1 $\mu$m.

\section{A Population of Planets}

The combination of RV studies and transit observations has given us a good understanding of the incidence of gas and icy giant planets with masses of a few Jupiter masses down to a few tens of Earth masses located from a few stellar radii out to 5 AU \citep{cumming}. Within this orbital range approximately 10-15\% of solar type stars have gas giants (M$>0.3$ M$_{Jup}$) and perhaps  double that fraction if one extends the mass range to 0.01 M$_{Jup}$ \citep{lovis}. The exact fraction of stars with (hot) Neptune-sized planets remains in dispute but transit data from the CoRoT and Kepler satellites will soon resolve this issue. Very little is known about the incidence of planets in the outer reaches of planetary systems because of small RV amplitudes, vanishing transit probabilities, and long orbital timescales. Imaging and microlensing \citep{gould, bennett} provide probes of these systems with imaging offering the prospect of detailed follow-up observations. Previous imaging surveys on 8m class telescopes provide statistical constraints of $<25\%$ for the incidence of relatively massive planets ($>2$ M$_{Jup}$) on relatively large separations, 40-200 AU \citep{lafren, biller}.

For our simulation we have adopted a simplified model for the distribution of planets in the Mass-SMA plane. We assume that every star has (only) one planet drawn from a distribution with an incidence $dN/dM\propto M^{-1}$ between 0.1 and 10 M$_{Jup}$. The maximum mass for M stars is capped at 2 M$_{Jup}$ to reflect the under-abundance of massive planets for these stars \citep{johnson07}. Reflecting the growing evidence for an increased incidence of planets orbiting massive stars \citep{johnson}, we enhance the number of massive planets around stars with M$>$1.5 M$_\odot$ over the simple $M^{-1}$ power law. For these stars we added a log-normal distribution of planets with a mean of 2 M$_{Jup}$ and a factor of two dispersion in mass. The exact nature of this enhancement did not make a large difference in the simulation results. Based on the current census of exoplanets, we allowed $\eta=$20\% of the trials to place a planet between 0.1 and 5 AU. For the remaining $1-\eta$=80\% of stars, we drew from a distribution in SMA (denoted by $a$) with $dN/da\propto a^{-1}$ between 5 and 200 AU which favors closer-in planets and thus represents a more difficult case for direct imaging. Figure~\ref{YoungStarsInputHisto} shows the distribution of planets in the Mass-SMA plane and is similar to those adopted by Lafreni\`ere et al (2007). We also investigated $dN/da\propto a^0$ to examine what range in orbital distributions might be detectable for comparison with alternative formation and/or migration mechanisms. Orbital eccentricities were drawn from a probability distribution function between $0<\epsilon<0.8$ derived from the observed distribution of eccentricities for 269 radial velocity planets with periods greater than 4 days (Cumming et al 2004; Schneider 2009 and refs. therein). 

Our calculations require predictions of the brightness of planets at various wavelengths as functions of mass and age. One widely used group of models are the CONDO3/DUSTY models \citep{baraffe} which follow the evolution of a contracting planet. These models combine the evolution of T$_{eff}$ and Radius with a detailed atmospheric model to predict the appearance of planets across a wide range of wavelengths. Baraffe (private communication) extended these models to include planets with masses as low as 0.1 M$_{Jup}$ for this paper. We used filter profiles for JWST/NIRCam and JWST/MIRI to produce magnitudes for planets in these passbands to augment what was already available for ground-based filters (Appendix I). As the color-magnitudes diagrams indicate (Figure~\ref{colorcolor}), the predominant effect governing the appearance of a planet is its effective temperature with considerable overlap in colors as objects of different mass pass through a particular temperature. The [4.4]-[11.4] color-magnitude diagram spreads out the effects between mass and age on T$_{eff}$ and luminosity and may be useful in breaking these otherwise degenerate parameters.

There are, however, a number of caveats that should be considered when using these models. 

\begin{itemize}
 \item First, the physics underlying these models becomes unreliable at effective temperatures below 100 K. While this is not an issue for the {\it young} planets considered in $\S$6.1, the lack of good models for $\sim$ 1 M$_{Jup}$ planets older than a few Gyr is a problem for the analysis of planets orbiting older M stars ($\S$6.2). As will be discussed below, the JWST instruments have the sensitivity needed to observe 1 M$_{Jup}$ planets orbiting the nearest M stars at separations  of a few AU. The lack of good models at the low temperatures of these objects makes these results qualitative. 
 \item Second, the Baraffe calculations are based on a so-called ``hot start'' evolution which ignores the effects of core accretion. These effects have recently been identified as important for the earliest evolutionary phases of these planets \citep{marley}. There can be significant differences between the luminosity and effective temperature between a planet forming through core accretion with an associated accretion shock vs. simply following the gravitational contraction of a pre-existing ball of gas of the same mass (the ``hot start'' model). At very young ages, the core accretion systems can be 5-100 times fainter than simple hot start model prediction. This effect is illustrated in Figure~\ref{Fortneyfig} for planets of 2 and 10 M$_{Jup}$ in the 5 $\mu$m M band for the CONDO3 models used in this paper \citep{baraffe} and for the core accretion models \citep{fortney}. The differences can be significant for young, massive planets: up to 3-5 magnitudes in M ([4.4 $\mu$m]) brightness at an age of 1 Myr for a 10 M$_{Jup}$ planet. The differences are more modest, 1-2 magnitudes, for older, lower mass planets, e.g. 1-2 M$_{Jup}$ at ages of 10 Myr. We discuss a limited comparison between hot-start and core accretion models below.

 \end{itemize}

Irradiation by a central star can greatly modify a planet's appearance \citep{burrows2003, baraffe}, but is of limited importance for the systems considered here because of the large planet-star separations detectable with direct imaging. Furthermore, for young stars, the effect of irradiation at separations larger than a few AU is small in comparison to the planet's internal energy. In the case of NRM or FQPM imaging or observations with a 30-40 m telescope, the planets are close enough to their host stars ($\leq$5 AU) that stellar irradiation can become modestly important. In this case we combined planet's intrinsic effective temperature, $T_{Eff, int}$, with the additional energy from the star of luminosity $L$ at a separation, $a$, assuming an albedo=0.1 and complete redistribution of the absorbed radiation to arrive at a new, higher $T_{Eff,new}$ for the planet. We then selected as our model for the planet's emission the model with the same mass but for a younger object having the newly calculated, elevated $T_{Eff,new}$:

\begin{equation}
T_{Eff, ext}=270 (1-Albedo)^{0.25}\, L_{*,\odot}^{0.25}\,a_{AU}^{-0.5}\, K
\end{equation}

\begin{equation}
T_{Eff,new}=\left ( T_{Eff, int}^4 + T_{Eff, ext}^4\right ) ^{0.25} 
\end{equation}

\section{The Monte Carlo Simulation}

For a given instrument configuration (telescope, instrument, wavelength, contrast ratio as a function of off-axis angle) we selected a particular stellar sample: young stars (comprising the SIM YSO, FEPS, and A star lists, $\S2.1$) or the nearby M stars ($\S$2.2). We drew planets of random mass, semi-major axis and eccentricity according to the distributions described above. The planet's separation from its host star was weighted according to the time spent in the appropriate Keplerian orbit and randomized over all possible starting points and orientations. It should be noted that highly eccentric planets spend much of their time near apoastron and may thus peek outside the Inner Working Angle of an instrument and be detectable for a fraction of their orbital period \citep{agol,brown}. For the large orbital separations of  relevance here, the orbital periods are typically so long so that this effect is a static one over the duration of an individual survey, in contrast with the planets considered by Agol (2007) and Brown (2009) which investigate the changing effects of orbital motion on the changing visibility of habitable zone planets ($\sim$ 1-3 AU) orbiting stars within 10-15 pc. 

Each star served as a seed for 1,000 Monte Carlo runs. A planet was scored as a detection if it: 1) lay between the inner and outer working angles; 2) was above the $5\sigma$ sensitivity limit for an observation consisting of the difference between two 1 hour long integrations to account for a differential technique for speckle suppression, e.g. roll or PSF subtraction; and 3) was brighter than the $5\sigma$ floor set by the contrast ratio appropriate to the apparent star-planet separation and stellar magnitude. Scores were kept for each star and for each Mass-SMA bin. The simulations were run for different instrument configurations and for planet distributions with $dN/da\propto a^0$ and $a^{-1}$ (Table~\ref{detectionsummary} et seq.).

Detection of a companion to a bright star in a single sighting is not, of course, adequate to claim that a faint adjacent source is a planet. Verification of the planetary nature of the object requires observations at different epochs to detect common proper motion, orbital motion, or differential parallactic motion relative to nearby reference stars \citep{zimmerman}. This step will be a critical part of any realistic survey.

\section{Discussion}

\subsection{Results For Young Stars}

Table~\ref{detectionsummary} summarizes how the different instruments probe the planet Mass-SMA parameter space with results given for two different assumptions about the distribution of planets, $dN/da\propto a^{-1}$ and $a^0$. In what follows we concentrate on the $\alpha=-1$ case. The effect of uncertain ages ($\times2$ dispersion around the nominal age) was  investigated for one ground-based and one space-based  instrument, but  did not  make a significant difference to the outcome so long as the  average value is preserved.

Table~\ref{detectionsummary} presents Monte Carlo results averaged over the entire sample of 641 ``Young Stars'' as well as for the 25 stars achieving the highest detectability scores, i.e. the fraction of Monte Carlo draws resulting in a detected planet. Logarithmic averages of the mass and true SMA (not the apparent orbital separation) of the detected planets, as well as the average of the minimum detectable mass and SMA for each star, were calculated for each simulation. Average values of mass and semi-major axis for all detected planets are summarized  by instrument in Figure~\ref{DetectSummary} where the symbol size is proportional to the fractional detection rate and the ``error bars'' give the 1 $\sigma$ dispersion in mass and semi-major axis of the detected planets. Symbols are shown for the average over all stars in the young star sample and  for the best 25 stars  detected by each instrument. 

Detailed information for each instrument and planet population is given in Tables~\ref{detectionsummary} et seq. Columns (1)-(2) identify the sample and instrument; columns (3) and (4) give the number of stars with any detection of a planet and the number of stars with planets detected more than $>25$\% of the time; column (5) gives the fraction of times a planet was detected, averaged over all stars having at least 1 detection; columns (6) and (7) give average and minimum values of detected planet mass; columns (8) and (9) give average and minimum values of detected planet semi-major axes; column (10) gives average age of detected planets; columns (11) and (12) present detectability scores for imaged planets that were also detectable using either RV or SIM-Lite astrometry ($\S6.2$). The average values of mass and SMA include estimates of the dispersion in these quantities. Table~\ref{Bestdetectionsummary} repeats this information but averages over only the $\leq$25 stars with the highest fraction of detections. Listings of the stars with the highest scores are presented in Appendix II for reference. It should be noted that two recently imaged A stars with planets, Fomalhaut and HR 8799, both finished high in the rankings, e.g. with scores $\sim$ 30\% for NIRCam and $\sim 5\%$ for GPI, but were not among the top 25 targets in the simulations.

Figure~\ref{DetectSummary} and Tables~\ref{detectionsummary}-\ref{Bestdetectionsummary} suggest that ground-based coronagraphy with the next generation of instruments (P1640, GPI, SPHERE) should routinely detect planets larger than about 3-5 M$_{Jup}$ within 20-50 AU with favorable cases yielding planets as small as 1 M$_{Jup}$ or close as 15 AU. This information is shown graphically in Figure~\ref{groundHP1640} et seq for a number of instruments. In these and subsequent plots, the contours represent probability of detection of a planet in a specific Mass-SMA bin, i.e. the number of planets detected in that bin divided by the total number of planets generated in that bin (Figure~\ref{YoungStarsInputHisto}). The initial discoveries of the planets orbiting HR 8799 and Fomalhaut (50-100 AU) are encouraging and suggest that with instrumental improvements, detections of planets much closer to the stars should become possible in this mass range. These results are consistent with predictions for GPI \citep{mac}.  M-band observations from the ground suffer from high thermal backgrounds making such surveys somewhat unfavorable despite the brightness of young planets at this wavelength. Performance improvements possible with the LBT(I) will enhance the number of planets relative to the MMT values.

Given our assumptions about the planetary systems and the optimization of the survey, success rates for the next generation of ground-based surveys (GPI \& P1640) could be as high as 25-35\% for an optimized H-band survey  and 10\% for an optimized M-band survey. Eventually, an advanced coronagraph on TMT (Figure~\ref{groundHTMT}) could push this detection threshold up to $\sim70$\% at lower masses ($\sim$1-2 M$_{Jup}$) and with minimum separations as small as a few AU for the most favorable stars. 

JWST will detect lower mass planets than is possible from the ground, with success rates of up to 40\% for the best stars (Table~\ref{detectionsummary}). Operating at 3.6 or 4.4 $\mu$m (Figure~\ref{nircam}) NIRCam will have a broad plateau of $>$30\% detection probability outside of 50 AU and $>$50\% outside of 100 AU for masses down to 0.2 M$_{Jup}$. Interior to 50 AU the probability of detection drops rapidly except for the most massive planets. The NIRCam performance is similar at 3.6 and 4.4 $\mu$m with the decrease in brightness of the planets offsetting the improved resolution at the shorter wavelength. The table confirms that the performance differences between NIRCam's Spot and Wedge-shaped masks are small with a slight advantage for the Wedge to  find closer-in planets. For the most favorable 25 stars, i.e. the closest and/or youngest, NIRCam can detect planets as small as 0.1 M$_{Jup}$ or as close in as 15 AU. However, as a Lyot coronagraph operating on a telescope of modest size, NIRCam is not sensitive to the inner reaches of planetary systems. 

The NRM imager (Figure~\ref{NRMdetect}) operates with a small Inner Working Angle and may find planets with an average orbital separation for the 25 best stars of 30 AU for masses as low as 0.8 M$_{Jup}$. Planets as small as 0.1 M$_{Jup}$ and orbital separations as small as 5-10 AU could be detected in the most favorable cases. This performance is limited to a small Outer Working Angle and relies critically on achieving a stable visibility calibration. Without this calibration the predicted contrast ratio is $\sim$10$\times$ worse and the TFI/NRM success ratio drops by a factor of 2$\sim$3. 

MIRI coronagraphy, as illustrated by the performance of the 11.4 $\mu$m FQPM (Figure~\ref{MIRIdetect}), will complement NIRCam and NRM imaging with its small inner working angle (1$\lambda/D$) coupled with a large field of view (13$\arcsec$). For the 25 best stars, MIRI will have a 70\% success rate in finding planets with average masses of 1-2 M$_{Jup}$ at average separations of 60 AU; planets as small as 0.10 M$_{Jup}$ and separations as small as $<5$ AU are possible. 

It must be emphasized that the ``success rates'' described above  depend  on each telescope and instrument combination achieving its nominal performance (contrast ratio and sensitivity)  and on the assumptions implicit in the population of planets, e.g.  at least 1 planet per system with a particular distribution of masses and orbits. Until each instrument is brought into operation, these results must be considered highly preliminary. This is particularly the case for  the innovative modes on JWST, e.g. TFI/NRM and MIRI FQPM where large extrapolations in performance are being made compared with the current state of the art.

These results, summarized by stellar host properties (Figure~\ref{StarProps} and ~\ref{SpHisto}) reveal interesting differences between the instruments. The top portion of Figure~\ref{StarProps} compares the performance of NIRCam and TFI/NRM at 4.4 $\mu$m. The NIRCam coronagraph does the best job on the closest stars whose more mature planets require the highest contrast ratio. Some of the best targets for TFI/NRM are relatively distant young stars where NRM's high angular resolution brings luminous 10 Myr old planets into view that are hidden from other instruments. The bottom panel of this figure adds MIRI into the comparison which largely displaces TFI/NRM by doing a good job of finding the youngest planets at all stellar distances. Figure~\ref{StarProps} shows only the highest scoring instrument for each star. In fact there is good overlap in instrument scores in most cases, suggesting that it will be possible to characterize these planets at many wavelengths leading, possibly, to determinations of  $T_{eff}$ and radius (Fig~\ref{colorcolor}). 

The distributions of spectral types with high detection fractions is shown for representative ground-based (P1640 at 1.65 $\mu$m) and space-based instruments (NIRCam at 4.4 $\mu$m). The spectral types of the entire young star sample is shown in blue while the top ranked 100 stars in the Monte Carlo simulations are shown in red (NIRCam, scores $>$29\%) and yellow (P1640, scores $>$16\%). Highly ranked NIRCam targets span the full range of input spectral types with an average age of 10$^8$ yr whereas young (10$^{6.8}$ yr) K and M stars at the low mass end and  high mass A stars dominate the P1640 rankings.

The top panel of Figure~\ref{PlanetProp} compares the planetary detections for the three JWST instruments and shows that NIRCam does best for planets more distant than 40 AU with average masses as low as $<$1 M$_{Jup}$. MIRI operates over a comparable range of orbital distances and mass limit. TFI/NRM operates uniquely in the 10-20 AU range for the closest stars and overlaps with MIRI in the 40-50 AU range for younger, more distant stars. The two vertical bands of NRM detections highlight the two sub-samples of young stars, i.e. 25-50 pc and 100-140 pc. The bottom panel compares present and future capabilities from the ground (P1640/GPI/SPHERE vs TMT). While P1640/GPI will find planets $>$ 3-5 M$_{Jup}$ and SMA$>$25-50 AU, an eventual TMT coronagraph will be able to probe to within 20 AU for considerably lower masses. JWST's TFI/NRM and MIRI/FQPM perform well compared with TMT because they operate at 0.5-1.0$\lambda/D$ in comparison with $2.5-4\lambda/D$ for a classical coronagraph. The change in inner working angle cancels much of the advantage of shorter wavelength and larger telescope diameter. The increased brightness of planets at 4 $\mu$m compared to 1.65 $\mu$m also contributes to JWST's performance despite its smaller size.

The high success fractions for the JWST instruments suggests that at the completion of modest sized surveys, 25 $\sim 50$ stars, it should be possible to test some of  the assumptions made in this simulation: overall fraction of young stars with planets exterior to 5 AU, ``hot start'' vs. ``core accretion'' evolutionary tracks, and orbital distribution with particular emphasis on the existence of planets on distant orbits. For example, the data obtained with JWST should suffice to distinguish between $dN/da\propto$ $a^0$ and $a^{-1}$ with a significant difference in the predicted average SMA between the two cases (Table~\ref{Bestdetectionsummary}). Figure~\ref{Cumulative} shows the cumulative yield of planets from different instruments in surveys of the most highly ranked stars. A survey of 50 stars with P1640, JWST NIRCam, TFI/NRM and MIRI, and TMT would yield 12, 18, 21, 31 and 31 planets, respectively, for $\alpha$=-1 model and the assumption that there is one planet per star. In the case of JWST NIRCam the difference in the average value of the semi-major axis between the  $\alpha$=-1 and 0 cases is highly significant, 81$\pm3$ AU vs. 120$\pm$1 AU. For JWST and TMT, this result would remain significant with only half the stars having planets instead of the assumed 100\%. While this is not a definitive examination of parameter extraction from the simulations, this result suggests that  surveys  that can be accomplished in reasonable amounts of telescope time will  address some of the key questions about the population of planets in the outer reaches of these planetary systems.

The addition of spectroscopic follow-up observations will provide insights into the physical properties of individual objects while the addition of  dynamical measurements will make these tests of theory much more stringent ($\S$ 6.3). 

\subsubsection{``Core Accretion'' vs. ``Hot Start'' Models}

An important consideration for young planets is that the assumed ``hot start'' evolutionary models \citep{baraffe} may not be correct. As mentioned above, planets formed by ``core accretion'' may be considerably fainter \citep{marley,fortney} than the ``hot start'' models used here. We investigated these differences for planets between 1-10 M$_{Jup}$ and ages from 1-100 Myr (Table 2 in Fortney et al. 2008) for which comparable magnitudes are available for both sets of evolutionary tracks. We calculated Monte Carlo simulations for three cases (Table~\ref{Fortneydetectionsummary}): two ground based systems, P1640 and TMT at 1.65 $\mu$m, and JWST/NIRCAM at 4.4 $\mu$m. The drop off in the number of stars with planets is most marked with P1640 at 1.65 $\mu$m. The lower temperature at each age-mass point in the Fortney models affects the planet magnitudes dramatically, lowering the number of stars with at least one planet detection from 324 to 7 out of a total sample of 364 stars. For the more sensitive TMT observations, the drop-off is not so severe, from 362 to 103 stars with planets. NIRCAM at 4.4 $\mu$m is not affected by the change in planet brightness for two reasons. First, as mentioned above the differences are muted at longer wavelengths. Second, NIRCam's sensitivity is such that it can detect planets at 1 M$_{Jup}$ for either evolutionary model. NIRCam would lose the ability to detect core accretion planets of still lower mass or orbiting older stars, but the lack of Fortney models for these cases prevents us from discussing this quantitatively.

Another effect may mitigate the ``hot start'' vs. ``core accretion'' problem. Planets located beyond 10 AU (the majority of those detectable via direct imaging) may not form via core accretion, but rather (at least in part) by gravitational fragmentation in the disk \citep{boss}. These planets may in fact be well represented by the ``hot start'' models. Slight evidence for this hypothesis comes from fact that Fomalhaut b cannot have a mass much greater than 3 M$_{Jup}$ lest it perturb the Fomalhaut ring \citep{kalas, chiang}. A core accretion planet would have to be considerably more massive than this to have the observed brightness at the assumed age of Fomalhaut. The ``hot start'' model may not prove to a bad representation for the planets on the distant orbits that direct imaging will detect.

\subsection{Results For Nearby M Stars}

Our Monte Carlo simulations (Table~\ref{Mstardetectionsummary}) suggest JWST (and to a lesser extent TMT) will be sensitive to self-luminous planets orbiting the nearest M stars at orbital distances beyond 10 -20 AU. These results are unfortunately tentative due to the inadequacy of the coldest, lowest mass models with effective temperatures below 100 K. Although we bounded the upper end of the mass range for M star planets at 2 M$_{Jup}$ due to the apparent dearth of high mass planets around low mass stars, high mass planets (5-10 M$_{Jup})$ would be easy to detect.

NIRCam is sensitive to planets orbiting nearby M stars having masses as low as 0.5 $M_{Jup}$ located at an average separation of 30$\sim$40 AU (Figure~\ref{NIRCAMdetectmstar} and Table~\ref{Mstardetectionsummary}-\ref{BestMstardetectionsummary}) with considerably closer and small mass planets being detectable for the most favorable stars ($\leq 0.5$ M$_{Jup}$ and $\sim$ 4 AU). TFI/NRM should be able to detect planets of comparable mass but on orbits as proximate as 1-5 AU for younger, more distant M stars with X-ray derived ages $<10^8$ yr. MIRI/FQPM should detect planets as small as 0.5 M$_{Jup}$ and $\sim$40 AU with detections of higher mass objects possible at distances as close in as 3-5 AU (Figure~\ref{MIRIdetectmstar}). With such small orbits these planets might also be detectable with radial velocity or astrometric measurements. A 0.5 M$_{Jup}$ planet in a 5 AU orbit (22 yr period) around a 0.25 M$_\odot$ star at a distance of 10 pc would have radial velocity and astrometric amplitudes of 13 m s$^{-1}$ and 100 $\mu$as, respectively. The combined imaging and dynamical observations would anchor planetary evolutionary models for ages of $\sim$ 1 Gyr or more ($\S6.3$). 

From the ground at 1.65 $\mu$m, the sensitivity of the current (NICI) or even next generation of ground based coronagraphs (P1640, GPI, SPHERE) is probably inadequate to find mature planets around nearby M stars. The average score for these observing systems ranges from 3-10\% for the few of the youngest M stars with any detections at all. The greater collecting area and angular resolution of the TMT improves the prospects for success (up 20\% in the most favorable cases), but the intrinsic faintness of older planets  at short wavelengths will be difficult to overcome. 

As indicated in the tables, we investigated both $\alpha=-1$ and  $\alpha=0$ power-law distributions of orbits. In contrast with the case of young stars, there is relatively little effect of the changing orbital separation, suggesting that angular resolution is not dominant factor preventing detection of these planets, rather that sensitivity is the bigger problem. In fact, the broader distribution ($\alpha=0$) led to a modest {\it decrease} in the number of planets detected for the simple reason of running out of field of view in the instruments considered in the simulations, e.g. a 10\arcsec\ field at 10 pc corresponds to 100 AU which is only half of the 200 AU outer limit considered herein.

Finally, we note that the nearest M stars offer an alternative prospect for imaging planets, i.e. detection using  {\it reflected} starlight. Depending on the ultimate performance of the JWST telescope at short wavelengths ($\leq$ 2 $\mu$m), the NIRCam coronagraph might be able to find such planets around  the few  M stars within 5 pc, e.g. GJ 411 (Lalande 21185) at 2.5 pc where JWST/NIRCam's Inner Working Angle of 0.28\arcsec\ at 2.1 $\mu$m corresponds to 0.7 AU (Table~\ref{resolution}). The  brightness of a Jupiter at separation $a$ in reflected light corresponds to a contrast ratio of $10^{-7}\sim 10^{-8}  {\rm a}_{AU}^{-2}$ (depending on albedo and phase function) which will be difficult to achieve at 1\arcsec\ (Figure~\ref{nircamperformance}). More probably, it will take an extremely capable  AO coronagraph, e.g. a $10^{-8}$ system on a 30 m telescope on the ground or a TPF-C telescope in space  will be able to push into the  domain of reflected light systems. A discussion  of these prospects is beyond the scope of this paper.  The reader is referred to Agol (2007) or Brown (2009) for a detailed examination of detection of planets via reflected light.

\subsection{Obtaining Masses via Dynamical Measurements}

A dynamical technique is needed to determine the masses of planets detected via imaging. Estimates based on interactions with dust disks provide an indication of planet mass, e.g. Fomalhaut \citep{chiang}, but radial velocity or astrometry can provide more definitive information, particularly if a near-complete orbit can be monitored.  But both techniques are challenging and no young planets on even the closest $<0.1$ AU orbits, i.e ``Hot Jupiters'', have yet been found definitively via RV. Setiawan et al. (2008) have claimed an RV detection of a planet orbiting TW Hya, but this claim has been called into question as being due to large scale photospheric variations \citep{huelamo}. The result remains controversial. Similarly, Prato et al. (2008) identified potential ``hot Jupiters'' orbiting DN Tau and V836 Tau based on visible spectroscopy, but used follow-up IR spectroscopy to demonstrate that the variations were due to photospheric variability not planets.  Astrometry can find young gas giant planets, e.g. a Saturn-mass planet in a 5 AU orbit at 140 pc would have an astrometric amplitude of 12 $\mu$as and would be readily detectable with SIM-Lite \citep{beichman01, tanner, unwin} and larger planets orbiting more nearby stars would be detectable with GAIA \citep{sozetti} or ground-based interferometry at the 100 $\mu$as level \citep{vanbelle,pott}. 

We investigated the prospects for indirect detection by positing {\it single measurement} capabilities of 1 m s$^{-1}$ and 4 $\mu$as as appropriate to ground-based RV studies and for the Space Interferometer Mission (SIM). In both cases we assumed 250 observations spread over 10 years and required final SNR=5.8 relative to amplitude of the reflex motion \citep{traub}. For planet periods greater than the observational duration, we degraded the noise performance according to (Period/10 yr)$^3$. To account for photospheric variability, rotation and other deleterious effects, we parameterized the stellar RV and astrometric jitter between 1 Myr and 5 Gyr as power-laws between 100 m s$^{-1}$ and 1 s$^{-1}$ and 4$\times$(140 pc/Dist) $\mu$as to 1$\times$(140 pc/Dist) $\mu$as \citep{valeri}. Radial velocity measurements were not considered for stellar types earlier than F0 due to the difficulty of finding suitable spectral lines.

Columns (11) and (12) of Table~\ref{detectionsummary} and Table~\ref{Bestdetectionsummary} give the fraction of planets detected by imaging that might also be detected via RV or astrometric measurements. The most favorable systems are those found at the smallest separations, e.g. those imaged using TFI/NRM, MIRI/FQPM or TMT. Up to 30\% of TMT imaging detections could, in the most favorable cases, also be detected via SIM astrometry. The majority of the mass measurements will have to be obtained with precision astrometry, because of the competing selection effects of improved imaging and astrometric detectability with orbital radius on the one hand and of decreasing RV amplitude with increasing radius on the other. 

Precise mass measurements will not be possible for the planets located at 10s of AU from their host stars. While the absolute astrometric shifts are large (hundreds of $\mu$as), the orbital time scales (10s to 100s of years) are  too long to cover a significant portion of an orbit during SIM's mission lifetime. However, measurements of orbital accelerations over a 5-10 year baseline can give a useful indicator of planet mass. While the linear terms of the stellar reflex motion are absorbed into the host's proper motion, the first (quadratic) term in the deviation from stellar proper motion due a long period orbit is given approximately by:

\begin{equation}
\rm{Non-linear \, Deviation}\sim \frac{2\, \pi^2\, \rm{A}}{Dist} \, \frac{\rm{a}_{Jup}^2} {\rm{a}^2} \, \frac{\rm{M_p}}{\rm{M_{Jup}}} \, \, \, t_{obs}^2 = 7.6 \, \frac{25 \, pc}{Dist}\, \frac{{M_p}}{M_{Jup}} \, (\frac{50 AU}{a})^2 \, (\frac{t_{obs}}{5\, yr})^2 \, \mu \rm{as}
\end{equation}

\noindent where A=500 $\mu$as, $a$ is the semi major axis and $t_{obs}<< T_{Period}$ is the measurement duration, or roughly 38 $\mu$as for a 5 M$_{Jup}$ planet in a 50 AU orbit at 25 pc. This deviation from stellar proper motion and parallax would be detectable by SIM-Lite. The interpretation of the measurements would be complicated by the unknown eccentricity and orientation, but the results would help to constrain the masses of distant planets detected with imaging, particularly in the absence of a visible dust disk. A full discussion of recovery of planet parameters from incomplete orbits is beyond the scope of this paper. The reader is referred to Traub et al (2009) for more information.

\section{Conclusions}

This article has combined estimates of the performance of a number of instruments, evolutionary tracks of planets, and possible populations of planets to assess the potential for the direct detection of planets using high contrast imaging. The results are necessarily speculative given the considerable uncertainties in each area: projections of instrument and telescope performance tend to be optimistic, some of the fundamental physics underlying the evolutionary tracks remains in question, and the number of planets on distant orbits and the mechanisms for getting them there are largely unknown. Nevertheless we can draw some robust conclusions from this analysis:

\begin{itemize}
 \item The early successes of imaging of planets around HR 8799 and Fomalhaut are harbingers of many results to come. Our  Monte Carlo results confirm other predictions  that  coronagraphs with extreme Adaptive Optics and post-coronagraph wavefront control, e.g. P1640, GPI and SPHERE, will be able to detect young planets with average masses of 5 M$_{Jup}$ within $<50$ AU of  young stars and, in favorable cases, find planets as close in as 5-10 AU and with masses as small 1-2 M$_{Jup}$. The requirement for Extreme AO systems to have bright host stars will limit these searches to closer stars and earlier spectral types. The success rates for these instruments will improve with the technology, from 15\% for the current generation (NICI et al) to 30\% for the next generation (P1640, GPI, SPHERE) and perhaps 70\% for an optimistic TMT instrument. These results depend, of course, on our assumptions of instrument and planet properties.  At longer wavelengths, 3-5 $\mu$m, ground-based telescopes like the MMT and LBT will similarly be able to detect $\sim$5 M$_{Jup}$ planets with the intrinsic planetary brightness offsetting the higher thermal background at these wavelengths. 

\item The Monte Carlo results show that JWST's NIRCam coronagraph can find planets with an average mass of 1.5  M$_{Jup}$ at separations of $\sim 80 $ AU with  masses of a few tenths of a  Jupiter mass and separations of 50 AU possible in the most favorable cases. The TFI's Non Redundant Mask imager will probe a comparable mass range in the inner portions of young stars in the regions like Taurus while MIRI's Four Quadrant Phase Mask (FQPM) coronagraph will complement NIRCam and TFI/NRM over a broad range of planet masses and separation yielding information on planet radius and temperature. The performance of these instruments depends critically on JWST meeting its performance goals (wavefront error, etc) and are of necessity speculative until the telescope is on orbit. However, these results will be of interest to those planning  various instrument campaigns.
 
\item JWST's sensitivity will allow it to search for  1-2 M$_{Jup}$ objects orbiting  nearby M stars ($<$2 Gyr) at orbits of a few to a few tens of AU should such planets exist. The intrinsic faintness of these few Gyr old objects will make this a very challenging experiment for ground-based telescopes operating at 1-2 $\mu$m, particularly given the faintness of many of their host stars and the demands of Extreme AO imaging.

\item An Extreme Adaptive Optics coronagraph operating at 1.65 $\mu$m on a 30-40 m telescope could find $\sim1$ M$_{Jup}$ planets within $<5$ AU of young parent stars as well as provide high spectral observations in the near-IR for objects initially detected by JWST at longer wavelengths. 

\item In addition to studying the physical conditions of these planets, there is a special premium for discovery and characterization of the closest-in systems using JWST/NRM and MIRI and TMT since these objects might also be detected by dynamical techniques using RV, or astrometry with ground-based interferometers, GAIA or SIM-Lite. Combined imaging and dynamical data will anchor evolutionary models of young planets and thereby help to put models for the formation and subsequent evolution of planets on a more sound theoretical footing.

\end{itemize}

\section{Acknowledgments} 

This work used archive information drawn from the NSTeD, 2MASS and SIMBAD archives. Some of the research described in this publication was carried out at the Jet Propulsion Laboratory, California Institute of Technology, under a contract with the National Aeronautics and Space Administration. The work of A. Sivaramakrishnan is supported in part by the National Science Foundation Grant No. AST-0804417. We gratefully acknowledge  I. Baraffe in extending the CONDO3 models to lower masses. CAB is extremely grateful to Ben Oppenheimer, Dave Latham and Dimitar Sasselov for their hospitality during sabbatical sojourns at the American Museum of Natural History and the Center for Astrophysics. We thank Peter Lawson for providing Figure 1 and Dr. Sally Dodson-Robinson for useful discussions. Finally, we acknowledge the extensive efforts of an anonymous referee whose careful reading and numerous suggestions greatly improved the content and presentation of this paper.

\clearpage

\begin{table}
\caption{Properties of Nearby Clusters}
\tiny
\begin{tabular}{lcc|lcc}
Cluster & Age (Myr) & Distance (pc)&Cluster & Age (Myr)& Distance (pc)\\ \hline
$\beta$ Pic$^{1, 2}$ & 10-12& 31$\pm$21&Pleiades$^{3, 4}$&120&135$\pm2$\\
Tucanae-Horologium$^{1, 2}$ & 30 & 48$\pm$7 & Chamealeon$^1$&6&108$\pm$9\\
Taurus-Aurigae$^6$ &range $\sim$ 1-10&140$\pm$10 & & &\\
TW Hya$^{1, 2}$ & 8&48$\pm$13&Upper Sco, Sco Cen$^{5,7}$ & 2-5&130$\pm$10\\ \hline
\end{tabular}
\tablecomments{Characteristic distances to clusters can be misleading since clusters may have considerable depth along the line sight. Ages are derived with respect to pre-main sequence evolutionary tracks (e.g. Siess et al. 2000), lithium abundances and kinematics and their absolute accuracy is probably no better than 50-100\%, particularly for extremely young objects. However, our knowledge of the relative ages of various clusters is  considerably better. References: $^1$Torres et al 2007; $^2$Zuckerman and Song 2004; $^3$Pan, Shao, Kulkarni 2004; $^4$Martin, Dahm, Pavlenko 2001;$^5$ Wilking et al. 2008;$^6$ Elias 1978; $^7$ Preibisch \& Zinnecker (1999). }
\label{clusters}
\end{table}

\clearpage

\begin{table}
\caption{Numbers of Stars in Stellar Samples }
\begin{tabular}{lcclcc}\hline
SIM Young Star Project & 217\\
Spitzer FEPS Project & 306\\
Spitzer plus nearby A stars & 188\\
Nearby M stars ($<15$ pc)& 196 \\ \hline
\end{tabular}
\label{samples}
\end{table}

\clearpage

\begin{table}
\caption{Inner Working Angle And Physical Resolution}
\tiny
\begin{tabular}{lrrrrrr}
Telescope (m) &5.0 & 6.5 & 6.5 & 6.5 & 8.0& 30.0\\
Wavelength ($\mu$m) &1.65 & 2.2 & 4.4 & 11.4 & 1.65 & 1.65\\ \hline
\multicolumn{7}{l}{\it Inner Working Angle (mas)}\\ \hline
NIRCAM/Wedge ($4\lambda/D$) &--- & 280 & 560 & --- & ---& ---\\
NIRCAM/Sombrero ($6\lambda/D$) &--- & 420 & 850 & --- & ---& ---\\
``MMT-like'' ($4\lambda/D$) &--- & --- & 560 & --- & ---& ---\\
TFI/Non Redundant Mask ($0.5\lambda/D$) &--- & 35 & 70 & --- & ---& ---\\
MIRI/FPQM ($ 1\lambda/D$) &--- & --- & --- & 365 & ---& ---\\
Palomar/P1640 ($ 2.5\lambda/D$) &170 & --- & --- & --- & ---& ---\\
GPI/SPHERE ($ 2.5\lambda/D$) &--- & --- & --- & --- & 105& ---\\
TMT Coronagraph ($ 2.5\lambda/D$) &--- & --- & --- & --- & ---& 30\\ \hline
\multicolumn{7}{l}{\it Physical Resolution (AU) at 10 pc.}\\
NIRCAM/Wedge ($4\lambda/D$) &--- & 2.8 & 5.6 & --- & ---& ---\\
NIRCAM/Sombrero ($6\lambda/D$) &--- & 4.2 & 8.5 & --- & ---& ---\\
``MMT-like''($4\lambda/D$) &--- & --- & 5.6 & --- & ---& ---\\
TFI/Non Redundant Mask ($0.5\lambda/D$) &--- & 0.4 & 0.7 & --- & ---& ---\\
MIRI/FPQM ($1\lambda/D$) &--- & --- & --- & 3.7 & ---& ---\\
Palomar/P1640 ($ 2.5\lambda/D$) &1.7 & --- & --- & --- & ---& ---\\
GPI/SPHERE ($ 2.5\lambda/D$) &--- & --- & --- & --- & 1.1& ---\\
TMT Coronagraph ($ 2.5\lambda/D$) &--- & --- & --- & --- & ---& 0.3\\ \hline
\multicolumn{7}{l}{\it Physical Resolution (AU) at 50 pc.}\\
NIRCAM/Wedge ($4\lambda/D$) &--- & 14 & 28 & --- & ---&---\\
NIRCAM/Sombrero ($6\lambda/D$) &--- & 21 & 42 & --- & ---&---\\
``MMT-like'' ($4\lambda/D$) &--- & ---& 28 & --- & ---&---\\
TFI/Non Redundant Mask ($0.5\lambda/D$) &--- & 1.8 & 3.7 & --- & ---&---\\
MIRI/FPQM ($1\lambda/D$) &--- & --- & --- & 18 & ---&---\\
Palomar/P1640 ($ 2.5\lambda/D$) &9 & --- & --- & --- & ---&---\\
GPI/SPHERE ($ 2.5\lambda/D$) &--- & --- & --- & --- & 5&---\\
TMT Coronagraph ($ 2.5\lambda/D$) &--- & --- & --- & --- &---& 1.5\\ \hline
\multicolumn{7}{l}{\it Physical Resolution (AU) at 140 pc.}\\
NIRCAM/Wedge ($4\lambda/D$) &--- & 40 & 80 & --- & ---&---\\
NIRCAM/Sombrero ($6\lambda/D$) &--- & 60 & 120 & --- & ---&---\\
``MMT-like'' ($4\lambda/D$) &--- & ---& 80 & --- & ---&---\\
TFI/Non Redundant Mask ($0.5\lambda/D$) &--- & 5 & 10 & --- & ---&---\\
MIRI/FPQM ($1\lambda/D$) &--- & --- & --- & 50 & ---&---\\
Palomar/P1640 ($ 2.5\lambda/D$) &24 & --- & --- & --- & ---&---\\
GPI/SPHERE ($ 2.5\lambda/D$) &--- & --- & --- & --- & 15& ---\\
TMT Coronagraph ($ 2.5\lambda/D$) &--- & --- & --- & --- &--- & 4\\ \hline
\end{tabular}
\tablecomments{$^1$The term ``Inner Working Angle'' is used to describe the off-axis angle (radius) at which the transmission of the occulting mask drops below 50\%}
\label{resolution}
\end{table}

\clearpage

\begin{table}
\caption{Illustrative Properties of Coronagraphs Used in Simulations}
\begin{tabular}{lrrrrrr}
		& Wavelength & \multicolumn{4}{c}{$\Delta$Mag$^1$ at $\theta$} &Sens. Limit$^2$\\ 
Instrument& ($\mu$m)	& $0.5\arcsec$ & $1\arcsec$ &$2\arcsec$ & $4\arcsec$&($5\sigma, 1 hr$ mag)\\ \hline
NICI &1.65 $\mu$m	&12& 14 & 14.5 & & 20\\
P1640		& 1.65 $\mu$m 	&16&18&18&18&20\\
GPI 		& 1.65 $\mu$m 	&17.5&18&18&18&21\\
"MMT-like"	& 4.3 $\mu$m 	&9.9&11.7&14.3&16.2&14\\
TMT		& 1.65 $\mu$m 	&17.5&17.5&17.5&17.5&22.5\\ \hline
NIRCam Spot & 3.35 $\mu$m	&9.9&12.4&15.1&17.8&24.8\\
NIRCam Spot & 4.4 $\mu$m	&9.9&11.7&14.3&16.2&23.6\\
TFI/NRM w. Cal.	& 4.70 $\mu$m &12.5&--&-&--&20.6\\
TFI/NRM	w/o Cal.& 4.70 $\mu$m &10.0&--&-&--&20.6\\
MIRI/4QPM 	& 11.4 $\mu$m 	&9.0&9.5&12&13&17.6\\
\end{tabular}
\tablecomments{$^1$Rejection ratios are $5\sigma$ See test for references for individual instruments. 
$^2$Sensitivity limits are 5$\sigma$ in the difference of two 3600s exposures and include a
degradation for lower coronagraphic throughput.}
\label{CoronaProps}
\end{table}

\clearpage

\begin{table}
\tiny
\caption{Young Planet Simulations (All Stars)}
\begin{tabular}{lr|rrr|rr|rr|r|rr}
(1)&(2)&(3)&(4)&(5)&(6)&(7)&(8)&(9)&(10)&(11)&(12)\\
 &$\lambda$& Num&Num &Avg. &Mass & Min & SMA&Min &Age &RV$^1$&Astr$^1$\\
Inst&($\mu$m)&Det&($>25$\%)& Score (\%) & M$_{Jup}$ &M$_{Jup}$&(AU)&(AU)& (Myr) &(\%) &(\%) \\ \hline
\multicolumn{12}{l}{641 Young Stars, Orbit $\alpha$=-1}\\ 
NICI&1.65&349&0&7.0&5.9$\pm$2.4&3.28&80$\pm$31&30&15&0.03&0.68\\
P1640&1.65&428&12&10.0&5.7$\pm$2.5&3.42&59$\pm$28&16&26&0.21&2.31\\
GPI&1.65&535&57&12.9&5.0$\pm$3.0&2.88&55$\pm$26&13&31&0.39&3.09\\
TMT&1.65&602&334&28.4&4.1$\pm$2.4&1.53&43$\pm$12&3&44&2.86&12.32\\
NIRCAM Spot&3.56&632&58&11.8&2.6$\pm$1.6&0.38&135$\pm$22&65&53&0.00&0.26\\
MMT$^3$&4.44&85&1&4.3&7.7$\pm$2.6&6.25&80$\pm$36&39&17&0.05&1.70\\
NIRCAM Spot&4.44&640&173&16.1&1.7$\pm$0.9&0.18&130$\pm$30&64&54&0.00&0.40\\
TFI/NRM&4.44&612&286&22.5&2.7$\pm$2.1&0.67&22$\pm$13&5&49&2.41&9.20\\
TFI/NRM NoCal$^2$&4.44&419&15&10.1&5.0$\pm$2.9&2.75&29$\pm$13&8&17&0.55&3.36\\
MIRI&11.40&633&240&24.4&3.1$\pm$1.9&0.57&93$\pm$26&24&53&0.73&3.73\\ \hline
\multicolumn{12}{l}{641 Young Stars, Orbit $\alpha$=-1, Randomized Ages}\\ 
P1640&1.65&412&15&11.0&5.5$\pm$2.4&3.16&61$\pm$27&16&25&0.20&2.40\\
NIRCAM&4.44&640&189&18.4&1.9$\pm$1.0&0.18&121$\pm$27&50&54&0.01&0.60\\ \hline
\multicolumn{12}{l}{641 Young Stars, Orbit $\alpha$=0}\\ 
NICI&1.65&388&27&10.7&5.2$\pm$2.9&3.03&95$\pm$32&32&14&0.01&0.20\\
P1640&1.65&421&45&12.0&5.6$\pm$2.5&3.39&82$\pm$35&20&25&0.08&0.76\\
GPI&1.65&504&101&14.1&5.4$\pm$2.6&3.09&81$\pm$35&17&35&0.15&1.04\\
TMT&1.65&600&306&27.0&4.1$\pm$2.4&1.51&81$\pm$24&3&43&1.88&5.22\\
MMT$^3$&4.44&84&0&4.6&7.7$\pm$2.4&6.16&105$\pm$36&43&15&0.01&0.68\\
NIRCAM Spot&4.44&640&354&34.0&1.6$\pm$0.8&0.14&140$\pm$25&62&54&0.00&0.17\\
TFI/NRM&4.44&609&53&12.7&2.6$\pm$2.0&0.70&31$\pm$21&6&48&1.38&3.29\\
MIRI&11.40&637&465&37.0&2.7$\pm$1.8&0.53&122$\pm$18&27&53&0.41&1.28\\
\end{tabular}
\tablecomments{Columns (1) and (2) identifies the instrument; column (3) gives number of stars with at least one planet detection; column (4) gives number of stars with at least a $>25$\% probability of detecting a planet; column (5) gives the average detectability averaged over all stars with at least one detection; columns (6) and (7) give average and minimum values of detected planet mass; columns (8) and (9) give average and minimum values of detected semi-major axes; column (10) gives average age of detected planets; columns (11) and (12) present detectability scores for imaged planets that were also detectable using either RV or SIM-Lite astrometry. $^1$Percentage of sources detected via imaging also detectable with RV or astrometry. $^2$Without visibility calibration. $^3$Approximates the performance of a coronagraph on a telescope like the MMT as similar to that of NIRCam but with a magnitude floor of M=14 mag.}
\label{detectionsummary}
\end{table}

\clearpage

\begin{table}
\tiny
\caption{Young Planet Simulations (Best 25 Stars)}
\begin{tabular}{lr|rrr|rr|rr|r|rr}
(1)&(2)&(3)&(4)&(5)&(6)&(7)&(8)&(9)&(10)&(11)&(12)\\
 &$\lambda$& Num&Num &Avg. &Mass & Min & SMA&Min &Age &RV$^1$&Astr$^1$\\
Inst&($\mu$m)&Det&($>25$\%)& Score (\%) & M$_{Jup}$ &M$_{Jup}$&(AU)&(AU)& (Myr) &(\%) &(\%) \\ \hline
\multicolumn{12}{l}{641 Young Stars, Orbit $\alpha$=-1}\\ 
NICI&1.65&349&0&15.9&4.0$\pm$2.4&0.93&89$\pm$31&24&2&0.00&0.17\\
P1640&1.65&428&12&26.1&3.3$\pm$2.5&1.41&46$\pm$28&9&8&0.22&10.18\\
GPI&1.65&535&57&35.4&2.7$\pm$3.0&0.92&47$\pm$26&8&5&0.45&12.51\\
TMT&1.65&602&334&69.7&2.4$\pm$2.4&0.45&39$\pm$12&1&13&2.10&35.62\\
NIRCAM Spot&3.56&632&58&36.5&1.4$\pm$1.6&0.11&83$\pm$22&15&39&0.03&3.28\\
MMT$^3$&4.44&85&1&11.1&5.3$\pm$2.6&3.20&51$\pm$36&12&42&0.16&5.42\\
NIRCAM Spot&4.44&640&173&37.8&1.1$\pm$0.9&0.10&75$\pm$30&18&37&0.00&2.16\\
TFI/NRM&4.44&612&286&43.1&0.8$\pm$2.1&0.12&28$\pm$13&6&3&0.61&9.30\\
TFI/NRM NoCal$^2$&4.44&419&15&25.9&2.4$\pm$2.9&0.56&35$\pm$13&7&2&0.17&4.08\\
MIRI&11.40&633&240&68.8&1.3$\pm$1.9&0.10&57$\pm$26&2&22&4.34&20.93\\ \hline
\multicolumn{12}{l}{641 Young Stars, Orbit $\alpha$=-1, Randomized Ages}\\ 
P1640&1.65&412&15&29.8&3.0$\pm$2.4&1.02&49$\pm$27&9&10&0.29&9.49\\
NIRCAM&4.44&640&189&41.8&1.1$\pm$1.0&0.10&71$\pm$27&13&36&0.26&5.04\\ \hline
\multicolumn{12}{l}{641 Young Stars, Orbit $\alpha$=0}\\ 
NICI&1.65&388&27&29.0&3.4$\pm$2.9&0.73&115$\pm$32&25&1&0.00&0.00\\
P1640&1.65&421&45&32.8&3.3$\pm$2.5&0.85&113$\pm$35&22&2&0.00&0.13\\
GPI&1.65&504&101&40.7&2.7$\pm$2.6&0.70&108$\pm$35&17&2&0.00&0.29\\
TMT&1.65&600&306&63.8&2.3$\pm$2.4&0.47&83$\pm$24&2&9&0.46&13.63\\
MMT$^3$&4.44&84&0&9.6&6.2$\pm$2.4&3.86&88$\pm$36&21&24&0.00&1.69\\
NIRCAM Spot&4.44&640&354&62.7&0.9$\pm$0.8&0.10&119$\pm$25&30&39&0.00&0.08\\
TFI/NRM&4.44&609&53&29.4&0.9$\pm$2.0&0.15&55$\pm$21&10&2&0.00&0.50\\
MIRI&11.40&637&465&79.8&1.1$\pm$1.8&0.10&101$\pm$18&2&25&3.11&7.54\\
\end{tabular}
\tablecomments{Same as Table~\ref{detectionsummary}, but for an average over only those stars with the
highest detection fraction (up to the 25 highest ranked targets).}
\label{Bestdetectionsummary}
\end{table}

\clearpage

\begin{table}
\tiny
\caption{Young Planet Simulations Hot-Start vs. Core Accretion Models}
\begin{tabular}{lr|rrr|rr|rr|r}
(1)&(2)&(3)&(4)&(5)&(6)&(7)&(8)&(9)&(10)\\
 &$\lambda$& Num&Num &Avg. &Mass & Min & SMA&Min &Age \\
Inst&($\mu$m)&Det&($>25$\%)& Score (\%) & M$_{Jup}$ &M$_{Jup}$&(AU)&(AU)& (Myr) \\ \hline
P1640-Hot Start&1.65&324&152&40.9&3.7$\pm$1.9&1.51&42$\pm$23&7&9\\
P1640-Core Accretion&1.65&7&2&16.9&3.5$\pm$1.3&2.18&15$\pm$6&4&33\\
TMT-Hot Start&1.65&362&353&93.0&2.5$\pm$1.4&1.00&44$\pm$7&1&15\\
TMT-Core Accretion&1.65&103&87&76.0&3.4$\pm$1.2&1.33&36$\pm$10&1&17\\
NIRCAM-Hot Start&4.44&364&110&47.9&2.6$\pm$1.0&1.01&69$\pm$24&14&24\\
NIRCAM-Core Accretion&4.44&364&111&48.4&2.7$\pm$0.8&1.01&70$\pm$27&13&30\\
\end{tabular}
\tablecomments{Columns are the same as in Table~\ref{detectionsummary} but for systems with ages less than $<100$ Myr and planets with masses between 1-10 M${Jup}$. }
\label{Fortneydetectionsummary}
\end{table}

\clearpage

\begin{table}
\tiny
\caption{M Star Simulations (All Stars)}
\begin{tabular}{lr|rrr|rr|rr|r|rr}
(1)&(2)&(3)&(4)&(5)&(6)&(7)&(8)&(9)&(10)&(11)&(12)\\
 &$\lambda$& Num&Num &Avg. &Mass & Min & SMA&Min &Age &RV$^1$&Astr$^1$\\
Inst&($\mu$m)&Det&($>25$\%)& Score (\%) & M$_{Jup}$ &M$_{Jup}$&(AU)&(AU)& (Myr) &(\%) &(\%) \\ \hline
\multicolumn{12}{l}{196 M Stars, Orbit $\alpha$=-1}\\ 
NICI&1.65&7&0&3.8&1.6$\pm$0.4&1.22&15$\pm$7&11&20&0.09&2.57\\
P1640&1.65&15&1&6.8&1.6$\pm$0.4&1.21&9$\pm$3&3&38&1.27&5.49\\
GPI&1.65&17&1&9.7&1.4$\pm$0.5&0.96&11$\pm$4&2&43&2.00&7.48\\
MMT$^2$&4.44&11&1&9.2&1.5$\pm$0.3&1.17&21$\pm$9&4&65&0.67&4.49\\
TMT&1.65&29&10&16.6&0.8$\pm$0.7&0.50&10$\pm$7&1&160&4.02&11.17\\
NIRCAM Spot&4.44&196&196&37.3&0.5$\pm$0.1&0.10&34$\pm$12&8&2,044&0.92&10.82\\
TFI/NRM&4.44&196&12&14.7&0.6$\pm$0.2&0.14&4$\pm$1&1&2,044&11.66&14.64\\
MIRI&11.40&196&67&25.6&0.8$\pm$0.2&0.32&48$\pm$18&4&2,044&2.91&10.15\\  \hline
\multicolumn{12}{l}{196 M Stars, Orbit $\alpha$=0}\\ 
NICI&1.65&4&0&2.1&1.4$\pm$0.3&0.93&19$\pm$9&7&13&0.10&1.50\\
P1640&1.65&12&0&4.2&1.5$\pm$0.4&1.10&10$\pm$6&2&33&1.68&3.50\\
GPI&1.65&17&0&4.7&1.4$\pm$0.5&0.97&8$\pm$5&2&43&2.21&3.85\\
TMT&1.65&29&1&9.4&1.0$\pm$0.6&0.60&12$\pm$10&0&162&3.85&5.96\\
MMT$^2$&4.44&11&0&4.6&1.5$\pm$0.3&1.18&30$\pm$16&6&65&0.65&1.45\\
NIRCAM Spot&4.44&196&97&24.0&0.5$\pm$0.1&0.10&46$\pm$16&9&2,044&0.56&3.10\\
TFI/NRM&4.44&196&0&11.5&0.6$\pm$0.2&0.13&3$\pm$1&1&2,044&10.41&11.45\\
MIRI&11.40&196&39&22.1&0.8$\pm$0.2&0.31&74$\pm$25&6&2,044&2.50&4.69\\
\end{tabular}
\tablecomments{Columns (1) and (2) identifies the instrument; column (3) gives number of stars with at least one planet detection; column (4) gives number of stars with at least a $>25$\% probability of detecting a planet; column (5) gives the average detectability averaged over all stars with at least one detection; columns (6) and (7) give average and minimum values of detected planet mass; columns (8) and (9) give average and minimum values of detected semi-major axes; column (10) gives average age of detected planets; columns (11) and (12) present detectability scores for imaged planets that were also detectable using either RV or SIM-Lite astrometry. $^1$Percentage of sources detected via imaging also detectable with RV or astrometry. $^2$Approximates the performance of a coronagraph on a telescope like the MMT as similar to that of NIRCam but with a magnitude floor of M=14 mag.}
\label{Mstardetectionsummary}
\end{table}

\clearpage

\begin{table}
\tiny
\caption{M Star Simulations (Best 25 Stars)}
\begin{tabular}{lr|rrr|rr|rr|r|rr}
(1)&(2)&(3)&(4)&(5)&(6)&(7)&(8)&(9)&(10)&(11)&(12)\\
 &$\lambda$& Num&Num &Avg. &Mass & Min & SMA&Min &Age &RV$^1$&Astr$^1$\\
Inst&($\mu$m)&Det&($>25$\%)& Score (\%) & M$_{Jup}$ &M$_{Jup}$&(AU)&(AU)& (Myr) &(\%) &(\%) \\ \hline
\multicolumn{12}{l}{196 M Stars, Orbit $\alpha$=-1}\\ 
NICI&1.65&7&0&3.8&1.6$\pm$0.4&1.22&15$\pm$7&11&20&0.09&2.57\\
P1640&1.65&15&1&6.8&1.6$\pm$0.4&1.21&9$\pm$3&3&38&1.27&5.49\\
GPI&1.65&17&1&9.7&1.4$\pm$0.5&0.96&11$\pm$4&2&43&2.00&7.48\\
MMT$^2$&4.44&11&1&9.2&1.5$\pm$0.3&1.17&21$\pm$9&4&65&0.67&4.49\\
TMT&1.65&29&10&19.2&1.0$\pm$0.7&0.59&12$\pm$7&1&106&4.63&12.92\\
NIRCAMSpot&4.44&196&196&43.3&0.5$\pm$0.1&0.10&29$\pm$12&6&478&1.00&14.24\\
TFI/NRM&4.44&196&12&24.7&0.5$\pm$0.2&0.11&5$\pm$1&1&923&15.58&24.37\\
MIRI&11.40&196&67&79.0&0.5$\pm$0.2&0.11&35$\pm$18&1&93&9.02&32.97\\  \hline
\multicolumn{12}{l}{196 M  Stars, Orbit $\alpha$=0}\\ 
NICI&1.65&4&0&2.1&1.4$\pm$0.3&0.93&19$\pm$9&7&13&0.10&1.50\\
P1640&1.65&12&0&4.2&1.5$\pm$0.4&1.10&10$\pm$6&2&33&1.68&3.50\\
GPI&1.65&17&0&4.7&1.4$\pm$0.5&0.97&8$\pm$5&2&43&2.21&3.85\\
TMT&1.65&29&1&10.8&1.1$\pm$0.6&0.58&14$\pm$10&1&114&4.43&6.88\\
MMT$^2$&4.44&11&0&4.6&1.5$\pm$0.3&1.18&30$\pm$16&6&65&0.65&1.45\\
NIRCAMSpot&4.44&196&97&34.0&0.5$\pm$0.1&0.10&63$\pm$16&13&2,630&0.00&0.91\\
TFI/NRM&4.44&196&0&18.0&0.4$\pm$0.2&0.10&3$\pm$1&1&500&14.16&17.93\\
MIRI&11.40&196&39&67.7&0.5$\pm$0.2&0.12&65$\pm$25&1&78&9.04&17.42\\
\end{tabular}
\tablecomments{Same as Table~\ref{detectionsummary}, but for an average over only those stars with the
highest detection fraction (up to the 25 highest ranked targets).}
\label{BestMstardetectionsummary}
\end{table}

\clearpage

\begin{figure}[h]
\begin{center}
\includegraphics[height=.7\textheight]{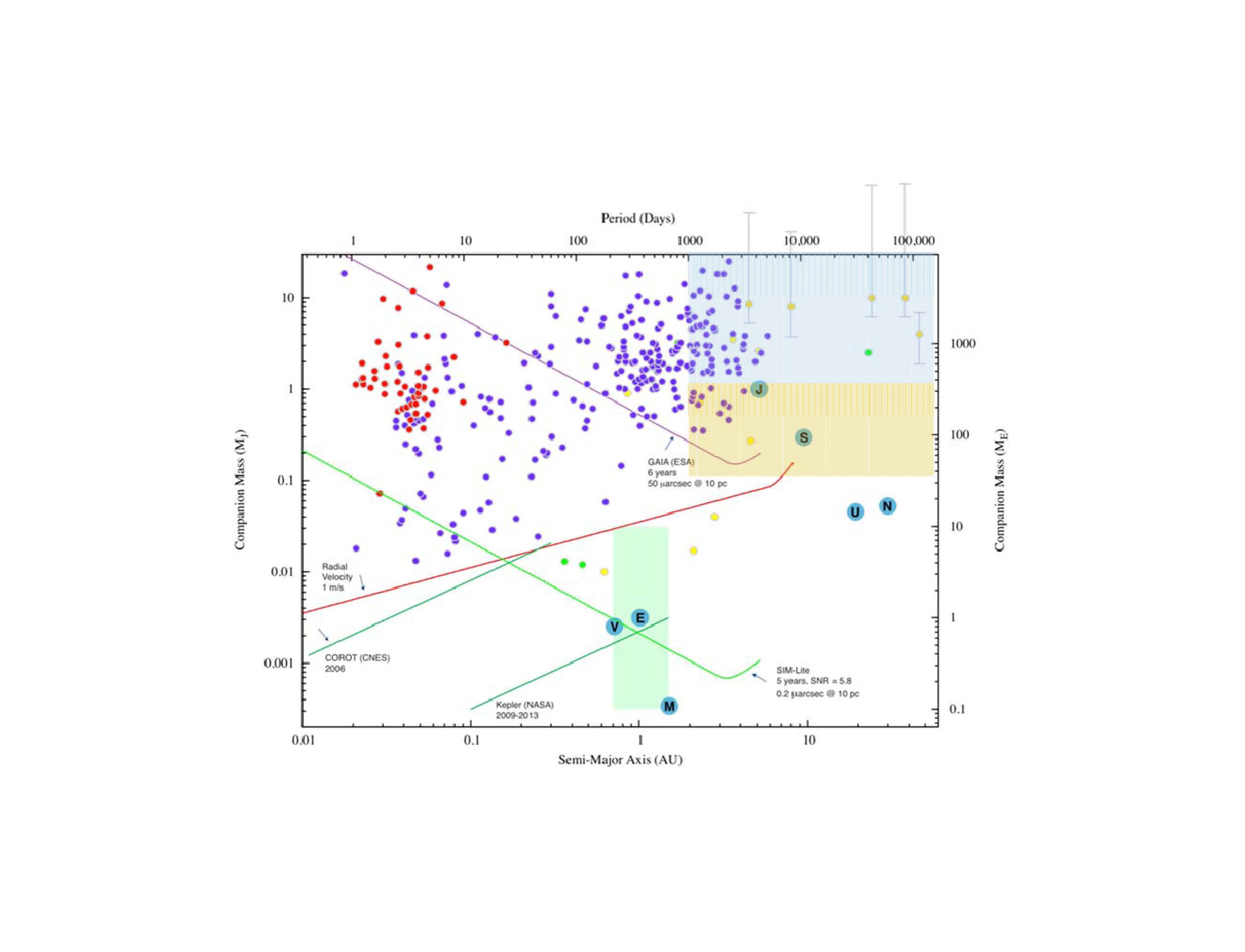} 
\caption{The distribution of detected planets (as of mid-2009 as taken from the Exoplanet Encyclopaedia, http://exoplanet.eu/) in the Mass-Semi Major Axis (SMA) plane. Different techniques dominate in different parts of this parameter space: {\it transits} in the upper left corner (red circles) ; {\it radial velocity} detections between 0.01-5 AU and above 0.01 M$_{Jup}$ (purple circles); {\it direct imaging} detections in the upper right hand corner (orange circles with error bars); a few {\it microlensing} detections (yellow circles) between 1-5 AU; and three {\it pulsar timing} planets (green points). Sensitivity limits for various techniques are shown as solid lines (RV, transit, and astrometry). The top shaded areas in the upper right shows the region that will be probed by ground-based imaging in the coming decade (upper, $>$1 M$_{Jup}$, blue) and by JWST (lower, $<$1 M$_{Jup}$, beige). Figure courtesy of Peter Lawson (JPL).}
\end{center}
 \label{PhaseSpace}
\end{figure}

\clearpage
\begin{figure}[h]
\begin{center}$
\begin{array}{c}
\includegraphics[height=.4\textheight]{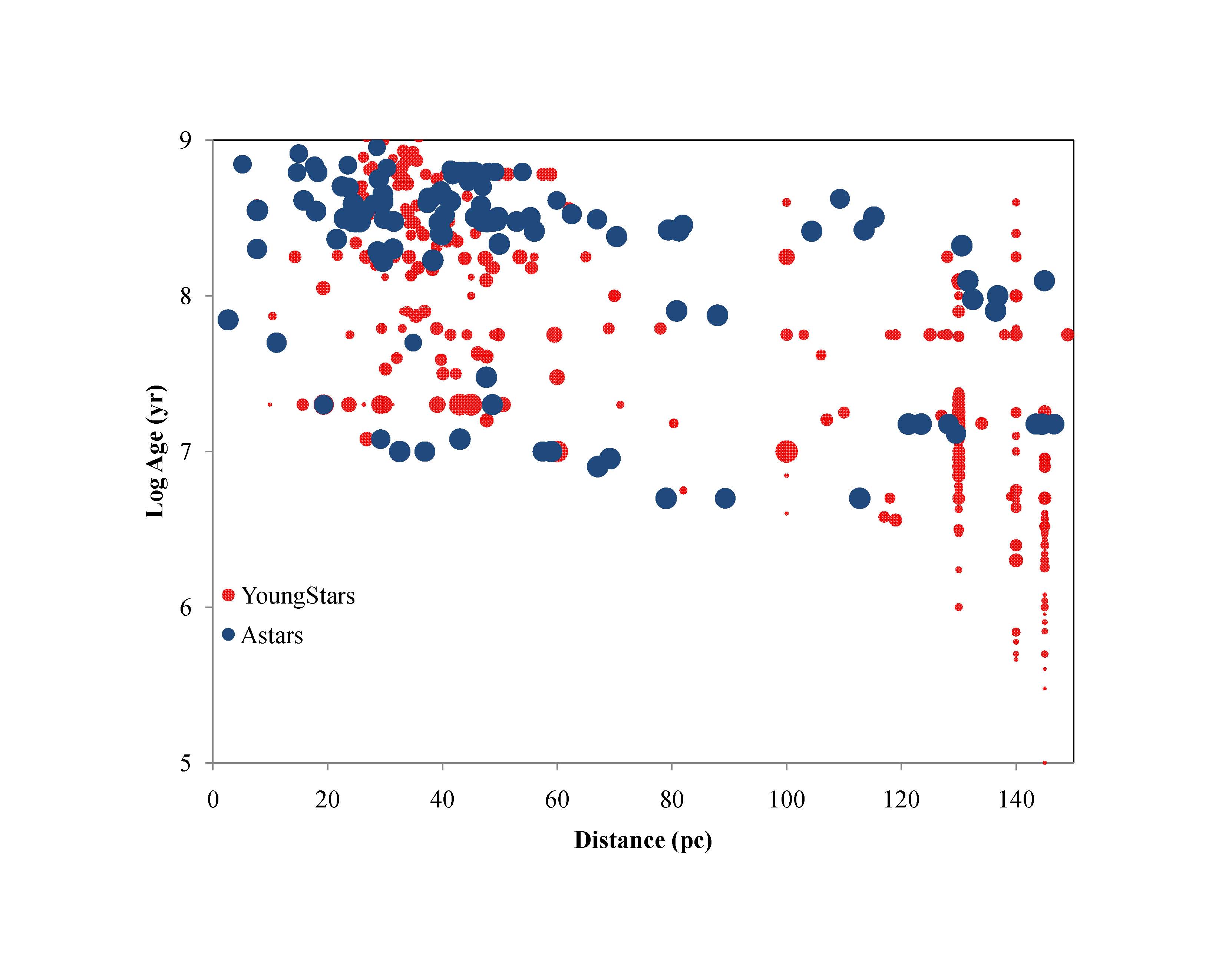} \\
\includegraphics[height=.4\textheight]{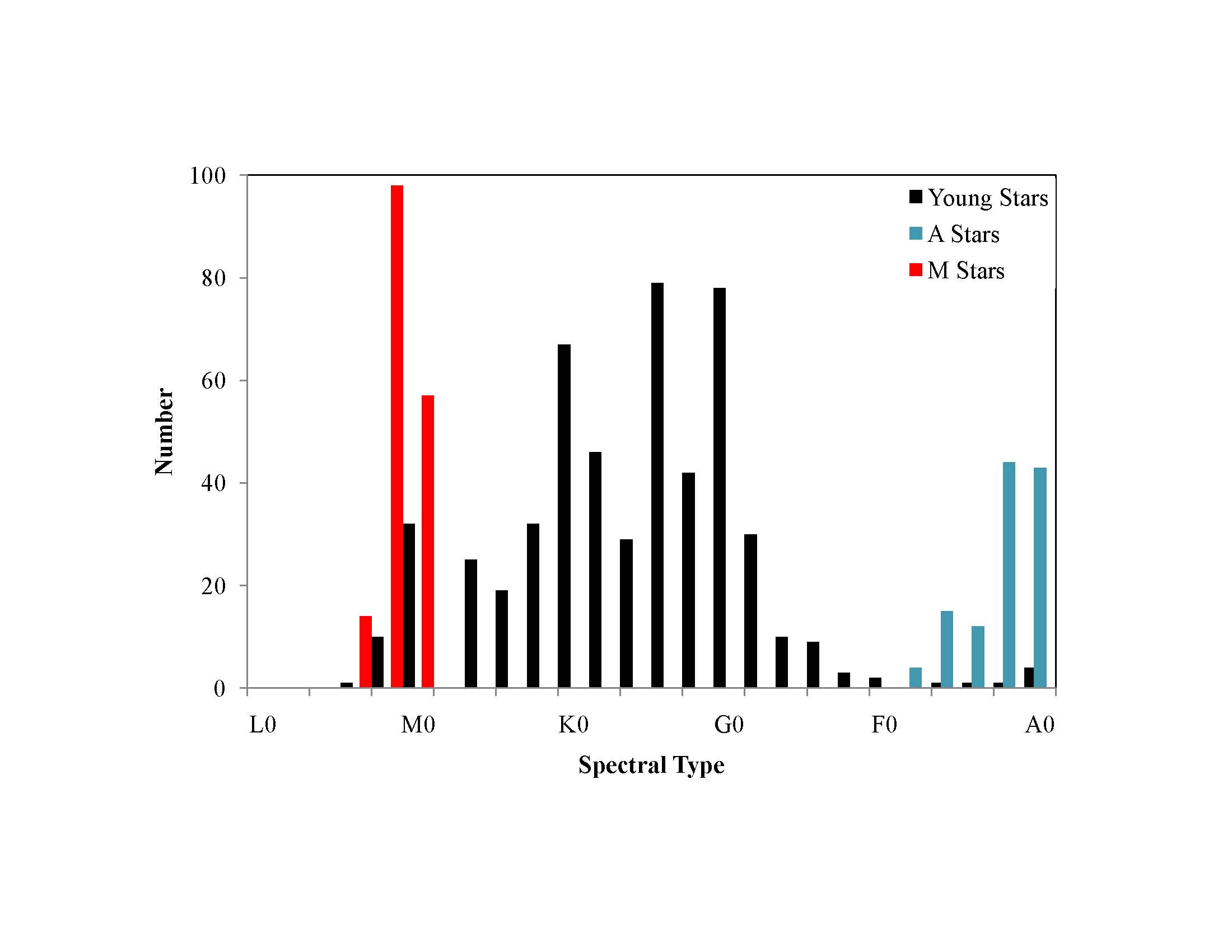}\\
\end{array}$
\end{center}
\caption{top) A sample of young A, F, G, K and M stars covers a range of ages from under 1 Myr up to 1 Gyr and distances from 5 to 150 pc. The size of the circle denotes spectral type from A stars (largest) to M stars (smallest). Bottom) The distribution of spectral types among the SIM+FEPS and  A star samples (which together comprise the ``young star'' sample) and M star sample.}\label{stellarprops}
\end{figure}

\clearpage
 \begin{figure}
 \includegraphics[height=.6\textheight]{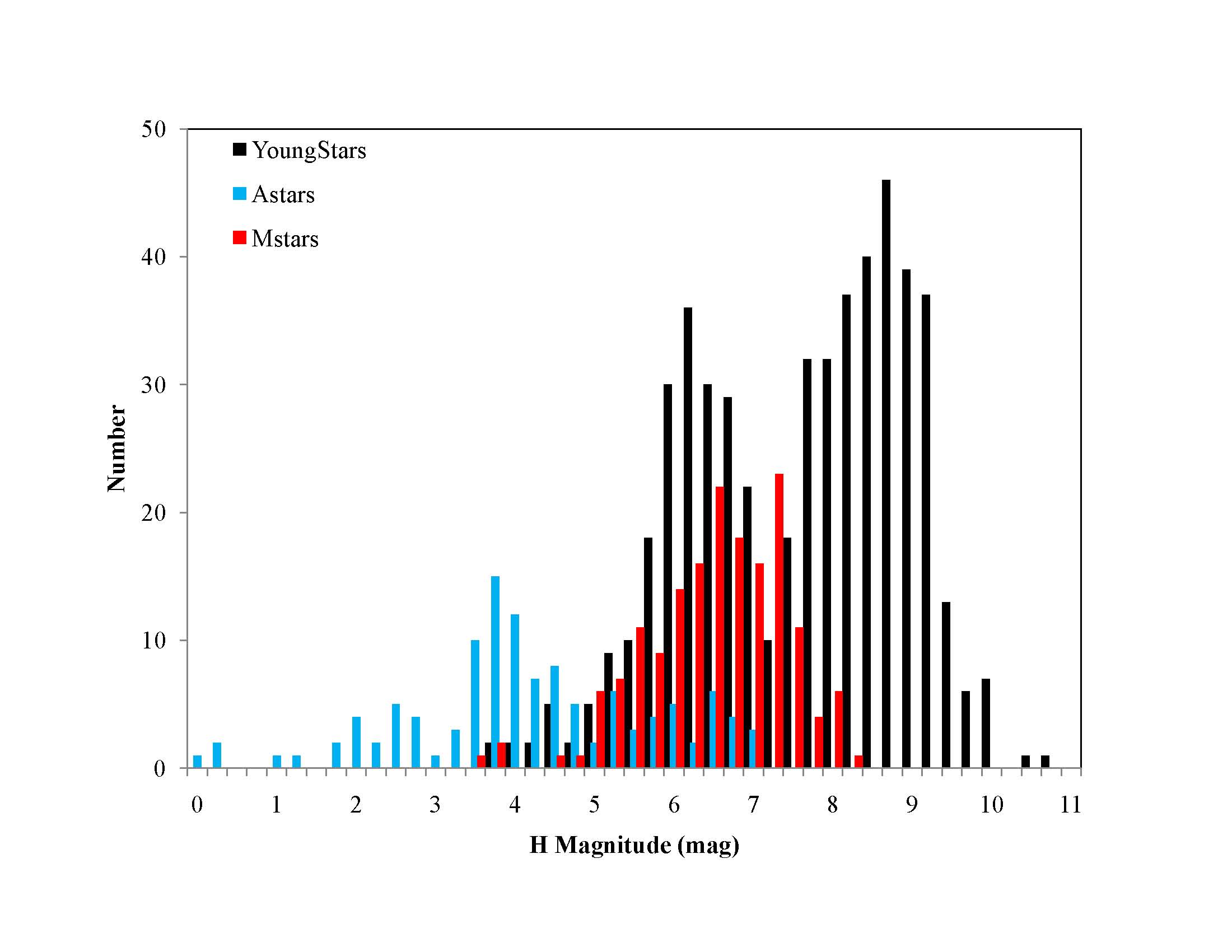}
 \caption{The stars in the various samples cover a broad range of H magnitudes which can be an important parameter when considering performance of adaptive optics systems.}\label{HmagHisto}
 \end{figure}

\clearpage
\begin{figure}
\begin{center}$
\begin{array}{c}
\includegraphics[height=.4\textheight]{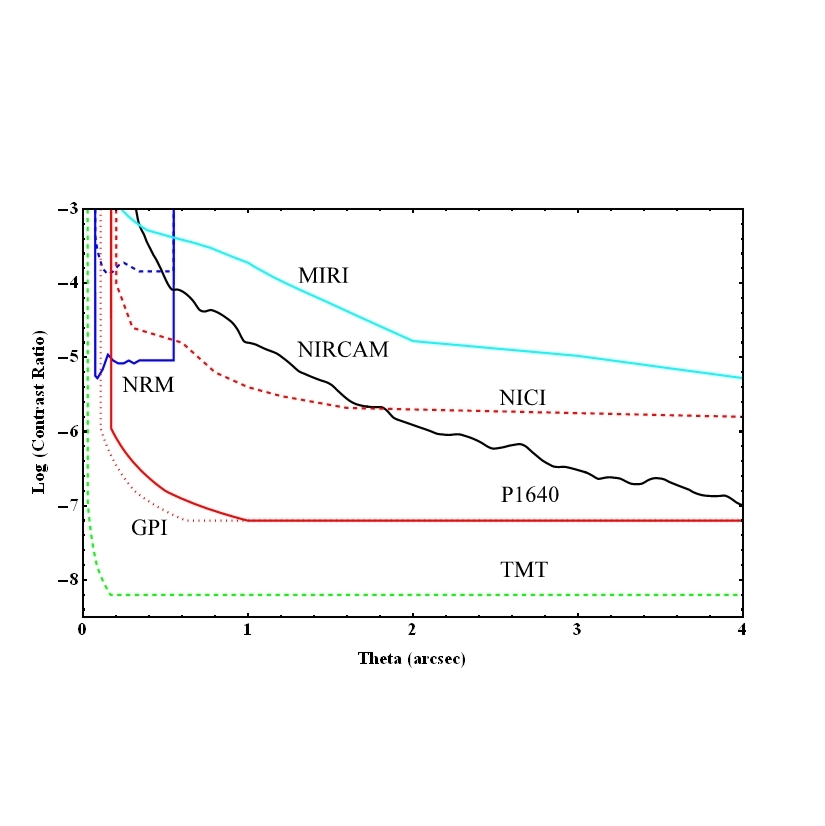} \\
\includegraphics[height=.4\textheight]{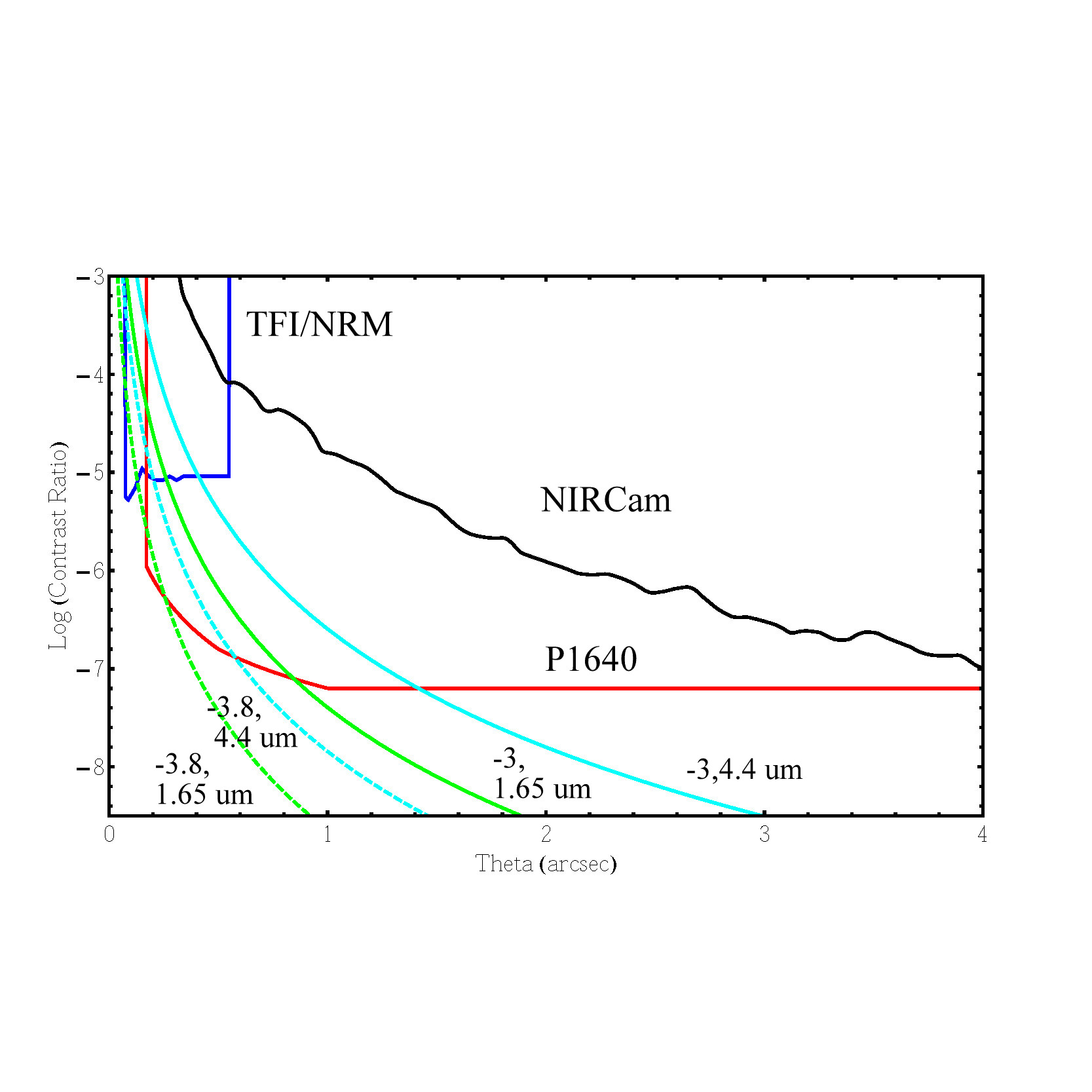}\\
\end{array}$
\end{center}
 \caption{\footnotesize top) The performance (5 $\sigma$) of 7 high contrast imaging systems is shown in terms of contrast ratio as a function of off-axis angle: MIRI with 4 Quadrant Phase Mask at 11.4 $\mu$m (top curve, cyan); NIRCam Lyot coronagraph at 4.4 $\mu$m (black); Gemini NICI instrument (red, dashed); P1640 at 1.65 $\mu$m (red) with an extension to smaller inner working angles for GPI operating n an 8 m telescope (red, dotted); an idealized coronagraph on a 30 m telescope (TMT) operating at 1.65 $\mu$m (green, dashed curve). Inside of 1\arcsec\ we show two curves for the Non Redundant Mask (NRM) at 4.4 $\mu$m with and without visibility calibration (solid and dotted blue curves). Bottom) The NIRCam, P1640, and TFI/NRM curves are repeated along with curves showing the brightness of potential spurious sources from diffuse scattered emission associated with debris disks as observed with JWST at 4.4 $\mu$m (cyan) and 1.65 $\mu$m with P1640 (green). Disks with $L_d/L_*=10^{-3}\, {\rm and} \, 10^{-3.8}$ are shown as solid and dashed, respectively. The details are described in the text. }\label{ContrastRatio}
 \end{figure}

\clearpage
\begin{figure}
 \includegraphics[height=.25\textheight]{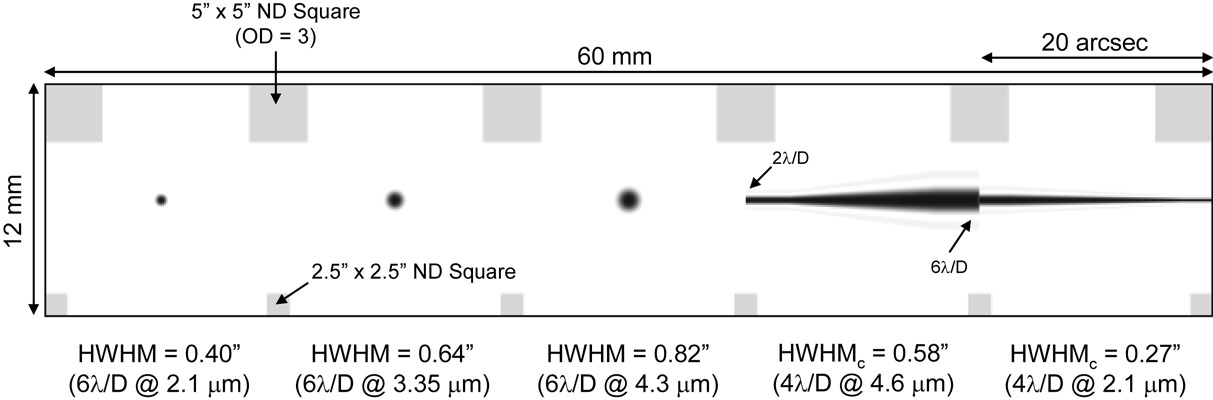} 
\caption{The layout of the coronagraphic focal plane masks in the NIRCam instrument includes 3 occulting spots plus 2 occulting wedges. Neutral density squares are placed across the top and bottom for source acquisition.}\label{nircamlayout}
 \end{figure}

\clearpage
\begin{figure}
\begin{center}
 \includegraphics[height=.6\textheight]{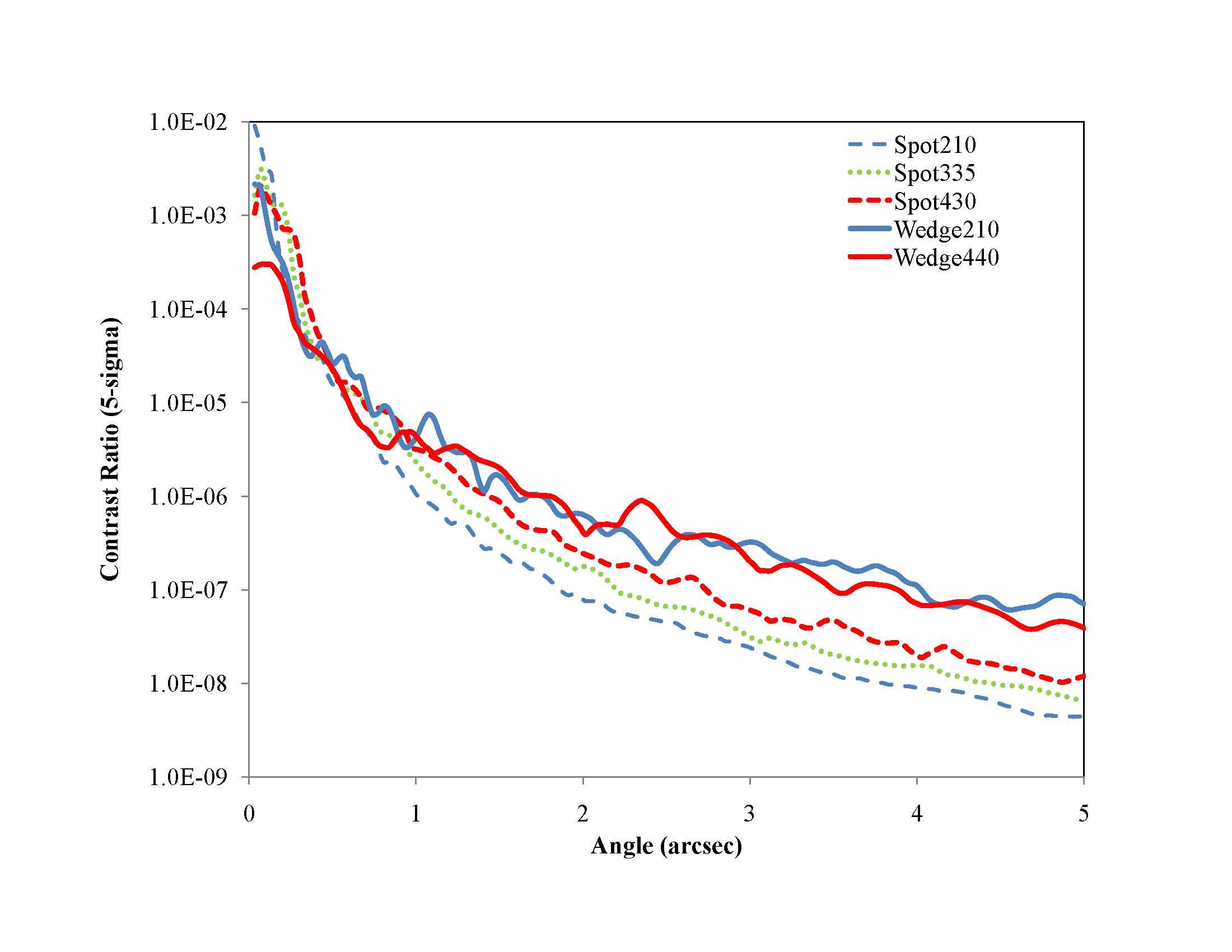} 
\end{center}
 \caption{The contrast ratio (5 $\sigma$) as a function of off-axis angle is shown for the various NIRCam coronagraph masks assuming subtraction of two rolls (+5$^{\circ}$ and -5$^{\circ}$) for speckle suppression. A position offset error of 10 mas and a wavefront error of 10 nm between rolls has been assumed. The two wedges are shown as solid lines while the (F460 in red, F210 in blue) and the three spots as dash/dotted lines (F430 in red, F335 in green and F210 in blue). }\label{nircamperformance}
 \end{figure}

\clearpage
\begin{figure}
\begin{center}
 \includegraphics[height=.4\textheight]{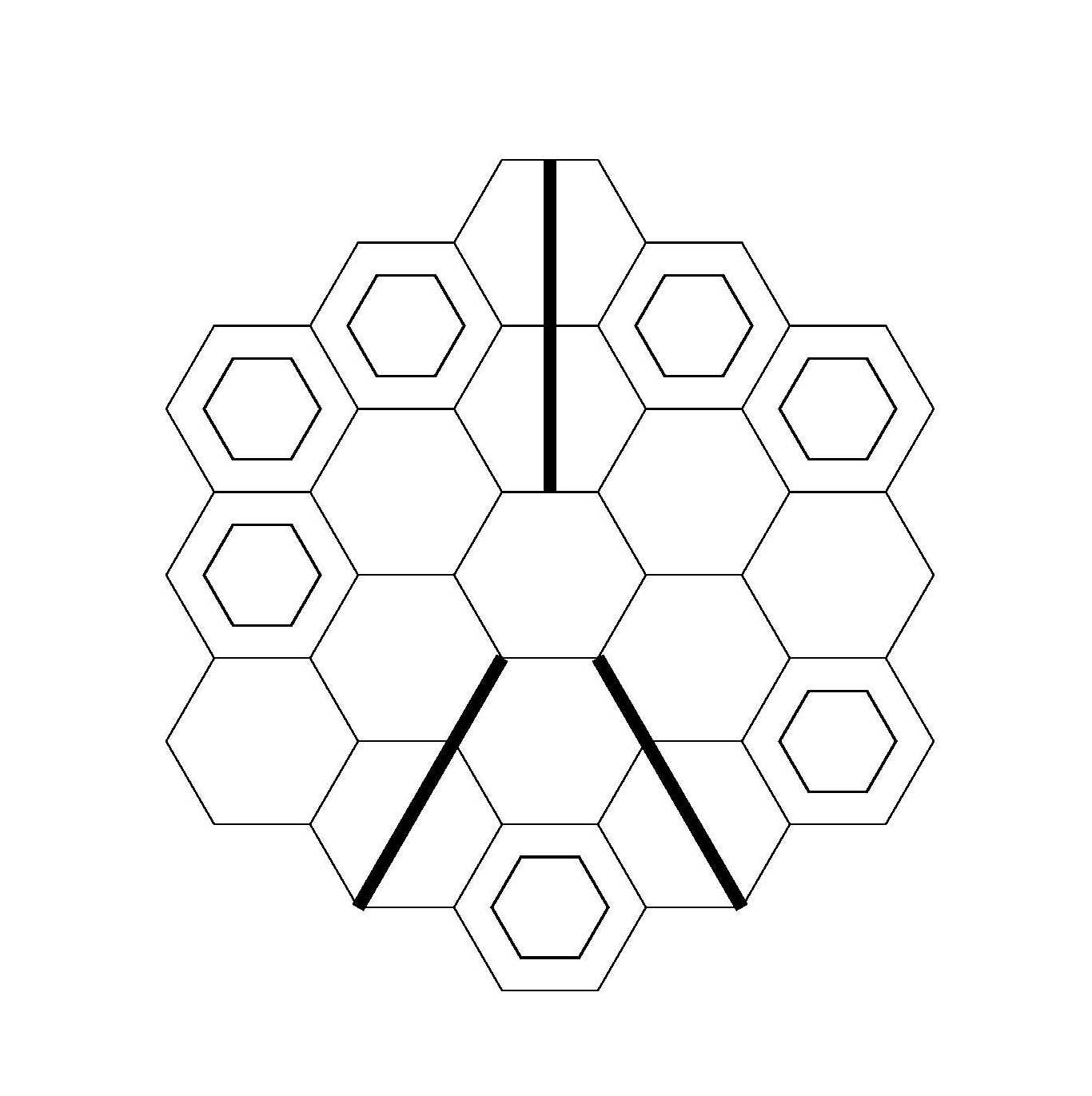}
\end{center}
 \caption{The layout of the sub-apertures projected onto the JWST primary mirror for the non-redundant mask (NRM) interferometer \citep{anand}.}\label{NRMLayout}
 \end{figure}

\clearpage
\begin{figure}
 \includegraphics[height=.6\textheight]{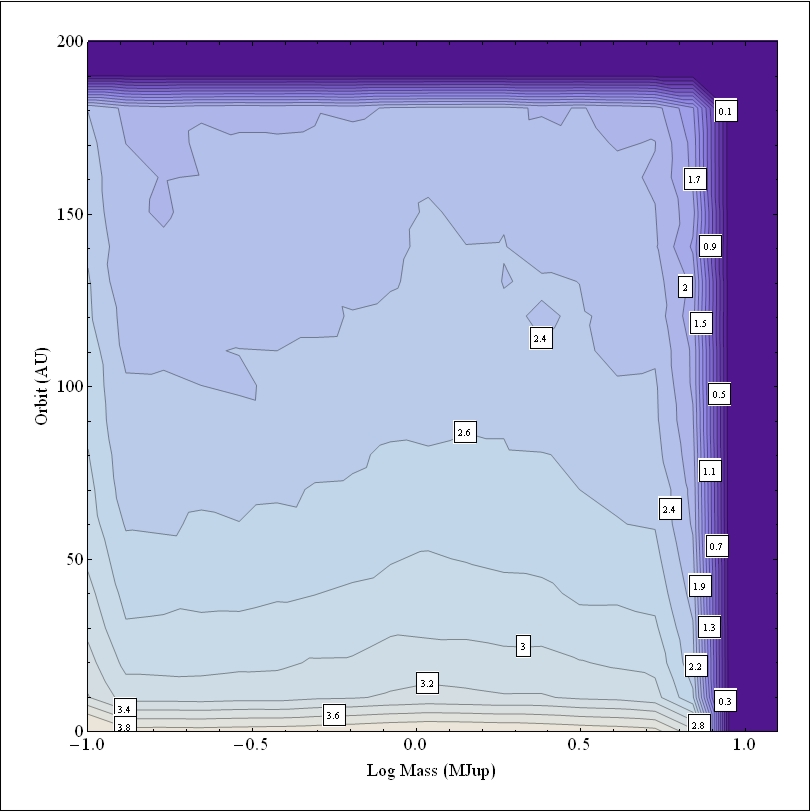} 
 \caption{ The distribution of planet masses and semi-major axes for a typical Monte Carlo run for the young stellar sample assuming $dN/da\propto\, a^{-1}$. As discussed in the text, the masses of planets orbiting M stars are capped at 2 M$_{Jup}$ compared with 10 $_{Jup}$  and an enhanced population of $\sim$ 2 M$_{Jup}$ planets has been adopted for stars more massive than 1.5 M$_\odot$. The contours represent logarithmic intervals with arbitrary normalization.}\label{YoungStarsInputHisto}
 \end{figure}

\begin{figure}
\begin{center}$
\begin{array}{c}
\includegraphics[height=.4\textheight]{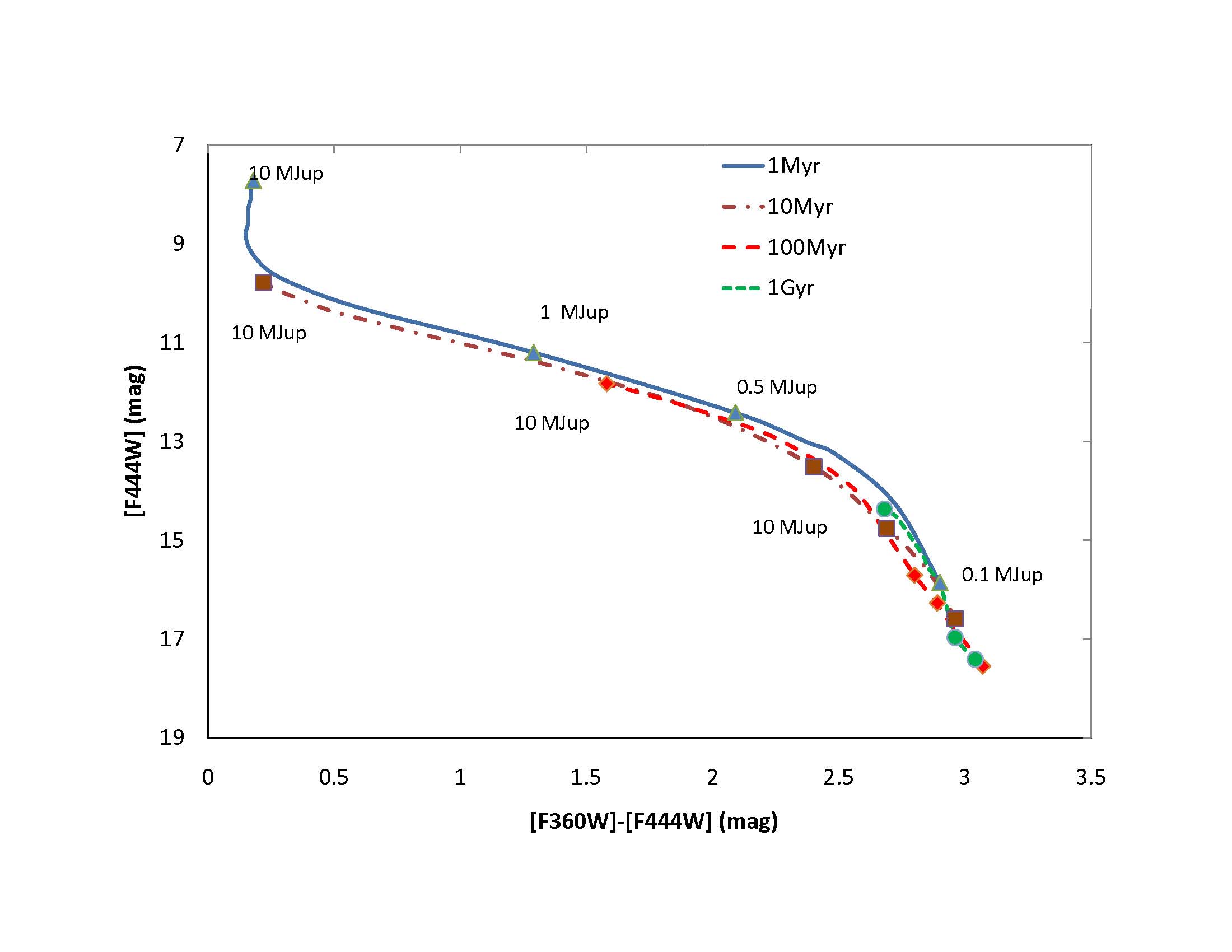} \\
\includegraphics[height=.4\textheight]{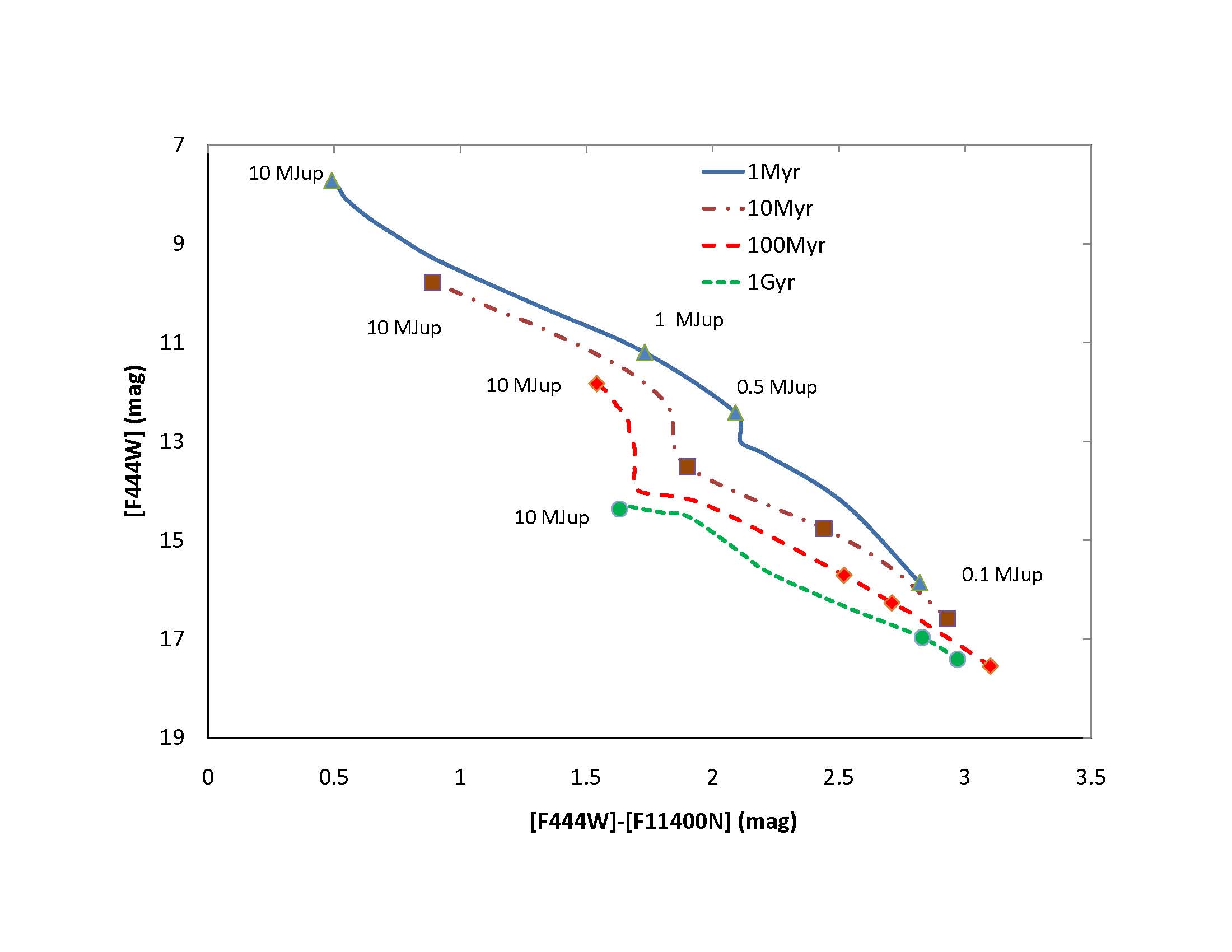}\\
\end{array}$
\end{center}
\caption {Color-magnitude diagrams for young planets using Baraffe (2003) models calculated for masses as low as 0.1 M$_{Jup}$ with effective temperatures as low as 100K. Top) models in to near-IR bands observable with NIRCam or TFI/NRM; bottom) models in bands observable with either NIRCam or TFI/NRM and MIRI. The combination of 5 and 11 $\mu$m colors appears to break the degeneracy between age and mass and may be valuable in assessing the evolutionary state of different planets.}\label{colorcolor}
\end{figure}

\clearpage
\begin{figure}
 \includegraphics[height=0.6\textheight]{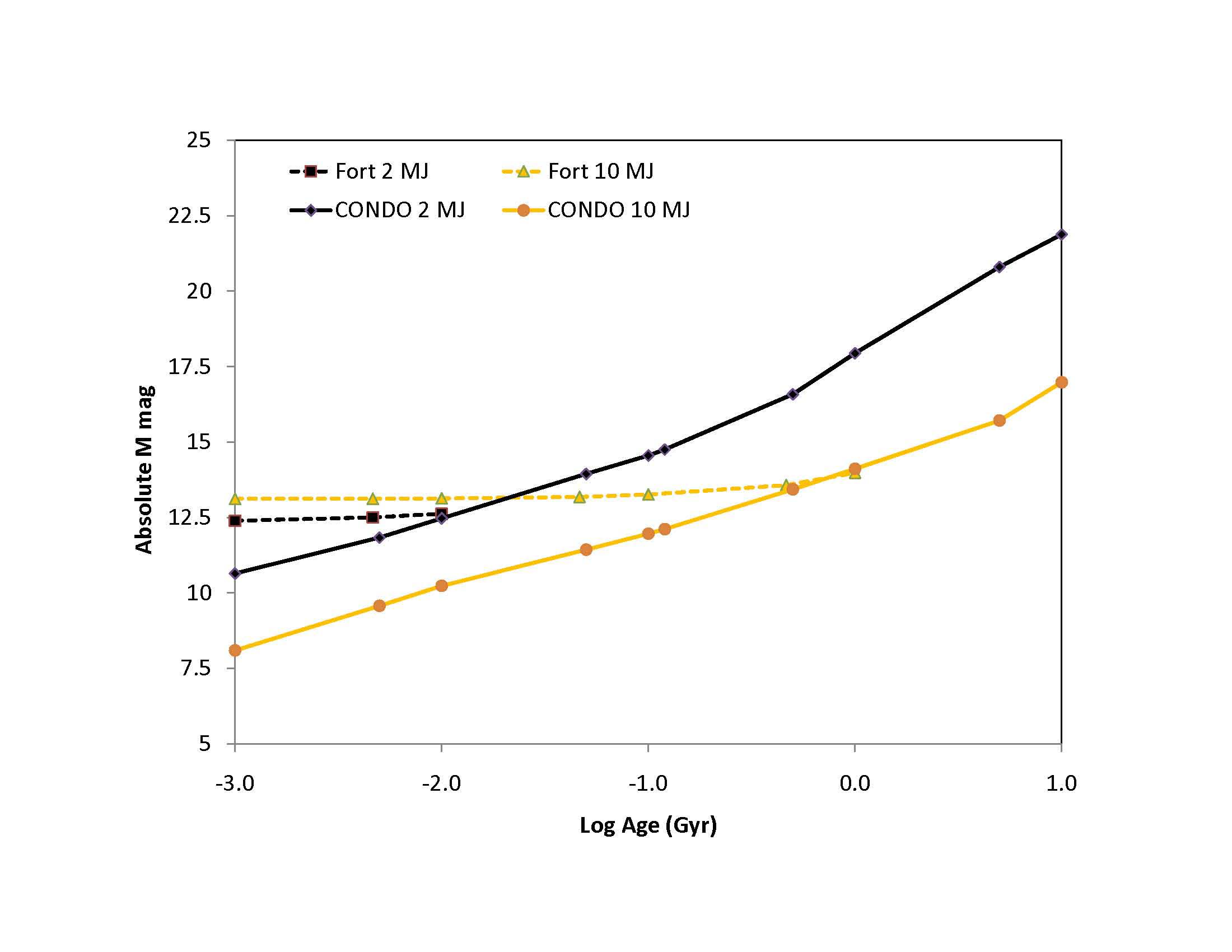}
 \caption{A comparison of two sets of evolutionary tracks for 1 and 10 M$_{Jup}$ planets. The solid curves represent the M [4.5 $\mu$m] brightness from the CONDO3 models \citep{baraffe} used in this paper. The dashed curves represent the core accretion models \citep{fortney} which are generally fainter at any given time. Thus a "core accretion" planet of a certain brightness will be a factor of two or more massive than a planet following the ``hot start''  contraction tracks.}\label{Fortneyfig}
 \end{figure}

\clearpage
\begin{figure}
\begin{center}
\includegraphics[height=.6\textheight]{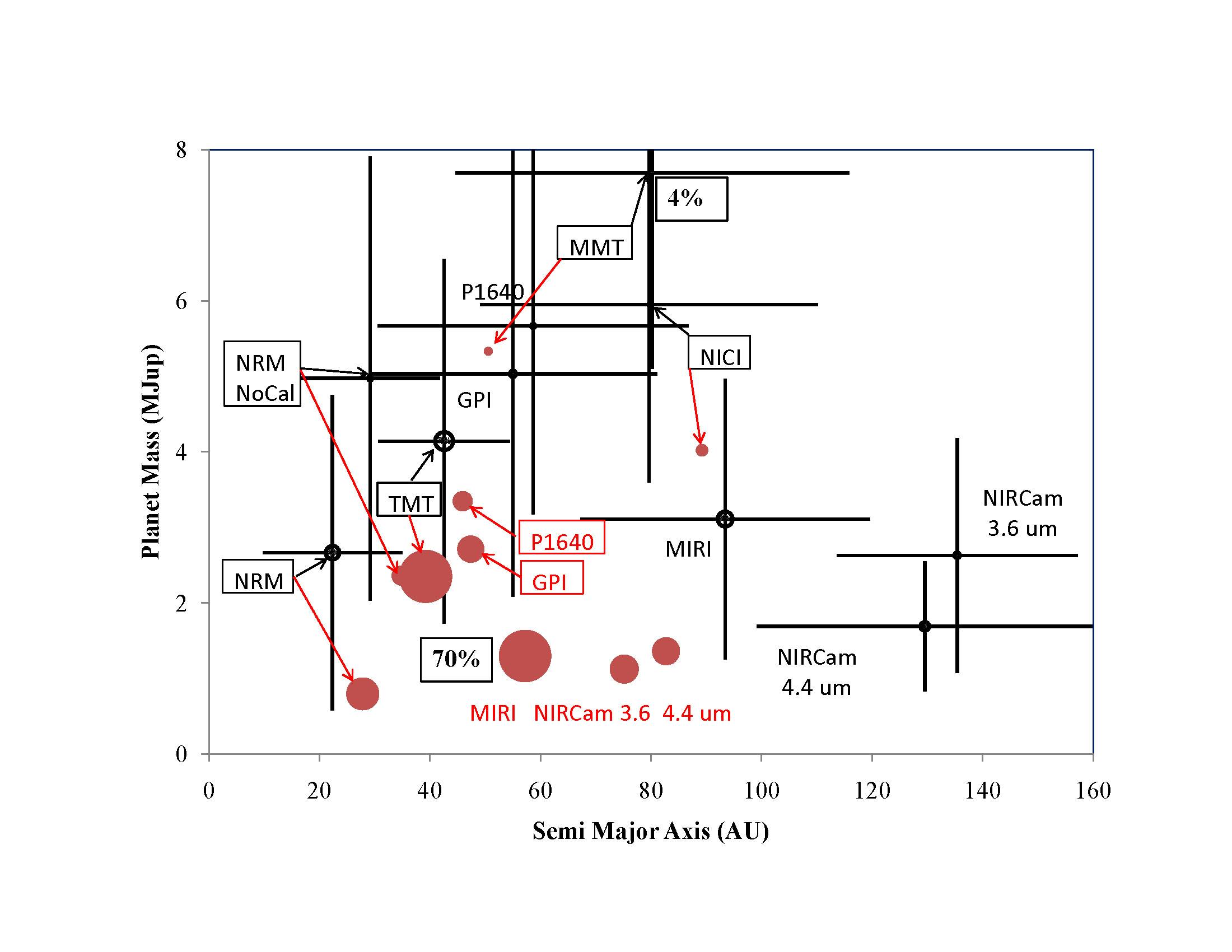} 
\end{center}
\caption{The plot shows the average values of planet mass and semi-major axis for planets detected with each instrument as presented in Table~\ref{detectionsummary}. The size of the symbol is proportional the fraction of stars around which at least 1 planet was detected; the percentages corresponding to two representative symbol sizes are indicated. The vertical and horizontal bars denote the 1 $\sigma$ dispersion in these quantities. The open circles with error bars are for the samples averaged over all stars with detections with the error bars denoting the 1$\sigma$ dispersion in the planetary properties. The red circles, with the error bars omitted for clarity, denote the planet parameters averaged over the best 25 stars in each run.}\label{DetectSummary}
\end{figure}

\clearpage
\begin{figure}
\begin{center}
\includegraphics[height=.6\textheight]{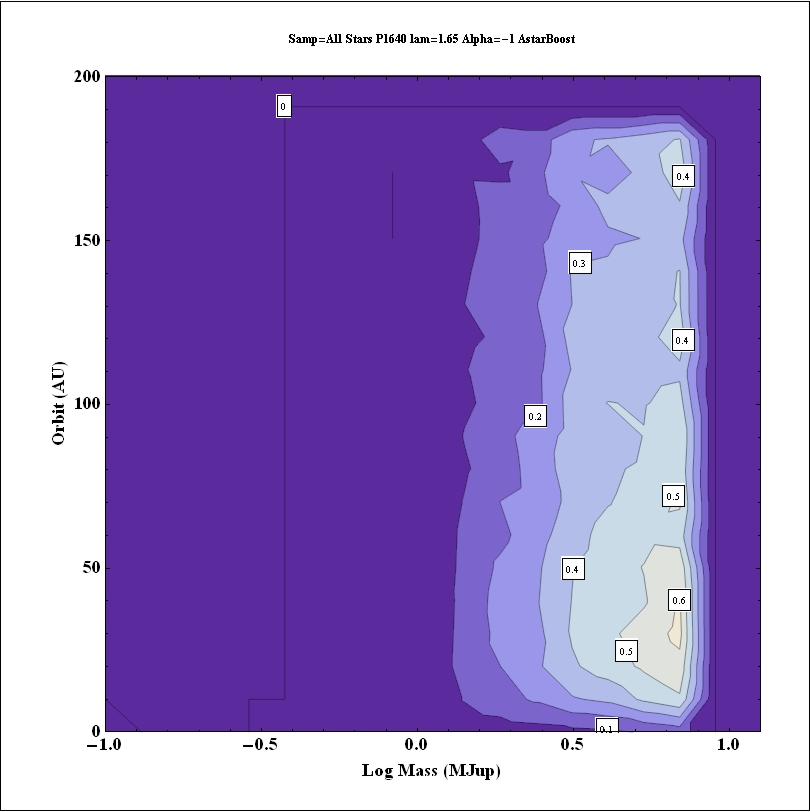} 
\end{center}
\caption{The plot shows the fractional detectability of planets orbiting nearby young stars ($\alpha=-1$)  for the P1640 coronagraph operating at 1.65 $\mu$m. The vertical axis represents orbital semi-major axis (AU) and the horizontal axis Log(planet mass) in (M$_{Jup}$). A comparable instrument operating on an 8 m telescope (GPI/SPHERE) would find planets at 5/8$\times$ smaller separations. The contours (displayed in white boxes) represent the probability of detection of a planet, averaged over all stars, in a specific Mass-SMA bin, i.e. the number of planets detected in that bin divided by the number of planets generated in the simulation. }\label{groundHP1640}
\end{figure}

\clearpage
\begin{figure}
\begin{center}
\includegraphics[height=.6\textheight]{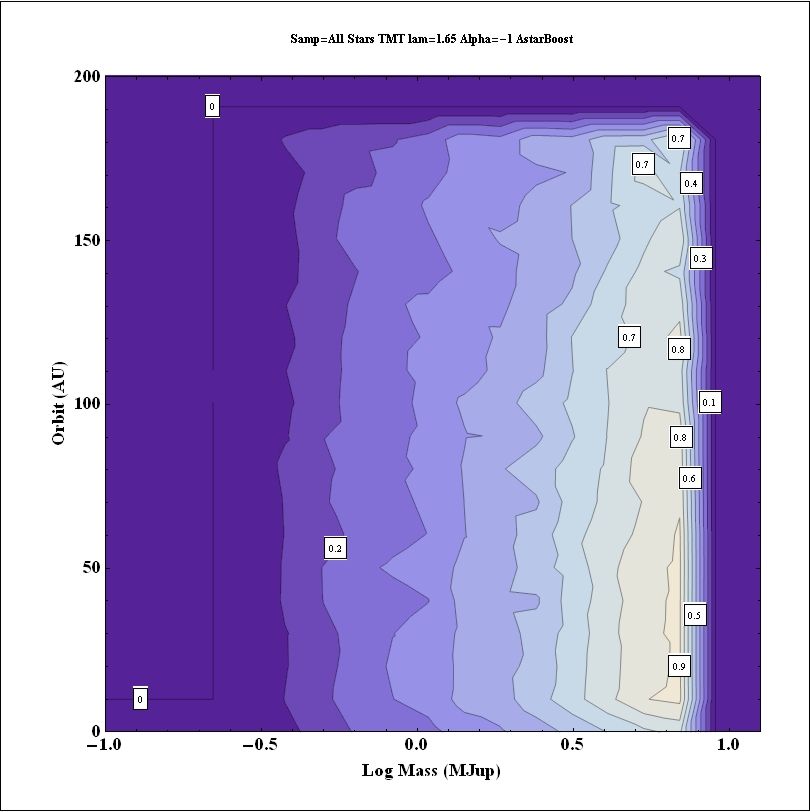}\\
\end{center}
\caption{The plot shows the fractional detectability of planets orbiting nearby young stars for a coronagraph operating at 1.65 $\mu$m on a 30 m telescope (TMT). The vertical axis represents orbital semi-major axis (AU) and the horizontal axis Log(planet mass) in (M$_{Jup}$). The contours represent the probability of detection of a planet, averaged over all stars, in a specific Mass-SMA bin, i.e. the number of planets detected in that bin divided by the number of planets generated in the simulation.}\label{groundHTMT}
\end{figure}

\clearpage
\begin{figure}
\begin{center}
\includegraphics[height=.6\textheight]{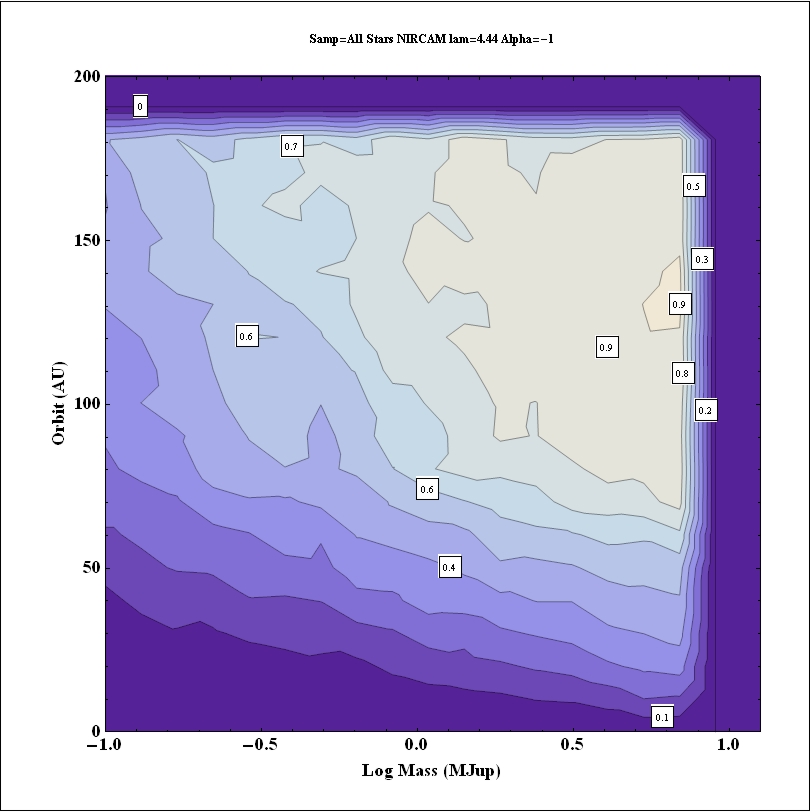}
\end{center}
\caption{The plot shows the fractional detectability of planets orbiting nearby young stars ($\alpha=-1$) for the NIRCam coronagraph at 4.4 $\mu$m. The vertical axis represents orbital semi-major axis (AU) and the horizontal axis Log(planet mass) in (M$_{Jup}$). The contours represent the probability of detection of a planet, averaged over all stars, in a specific Mass-SMA bin, i.e. the number of planets detected in that bin divided by the number of planets generated in the simulation.}\label{nircam}
\end{figure}

\clearpage
\begin{figure} 
\begin{center}
\includegraphics[height=.6\textheight]{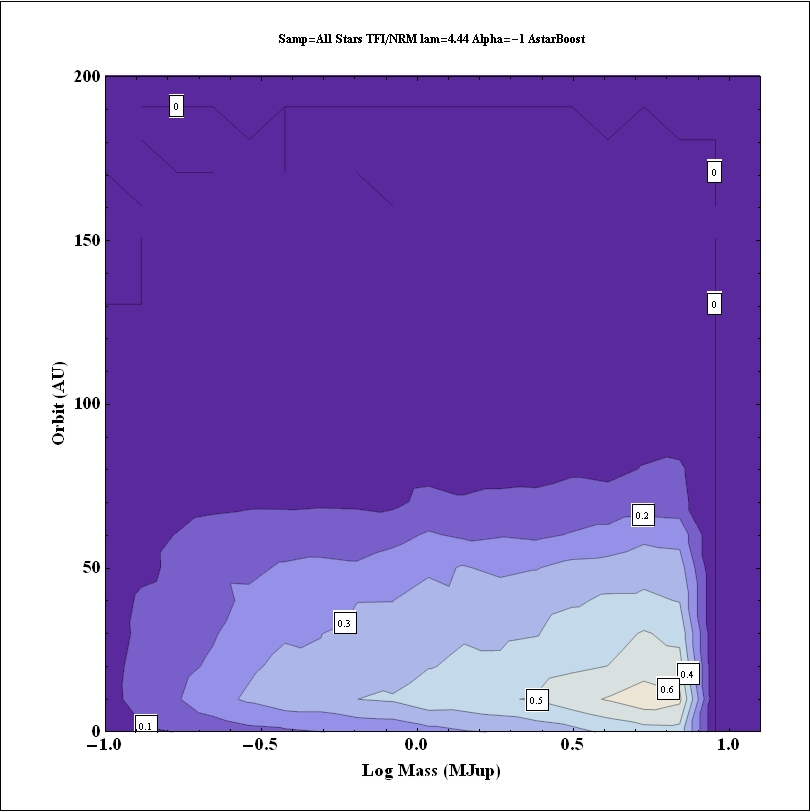} 
\end{center}
\caption{The plot shows the fractional detectability of planets orbiting nearby young stars ($\alpha=-1$)  for the TFI/NRM imager at 4.4 $\mu$m. The vertical axis represents orbital semi-major axis (AU) and the horizontal axis Log(planet mass) in (M$_{Jup}$). The contours represent the probability of detection of a planet, averaged over all stars, in a specific Mass-SMA bin, i.e. the number of planets detected in that bin divided by the number of planets generated in the simulation. The restriction to small separations is due to the restricted field of view of the NRM imager.}\label{NRMdetect}
\end{figure}

\clearpage
\begin{figure} 
\begin{center}
\includegraphics[height=.6\textheight]{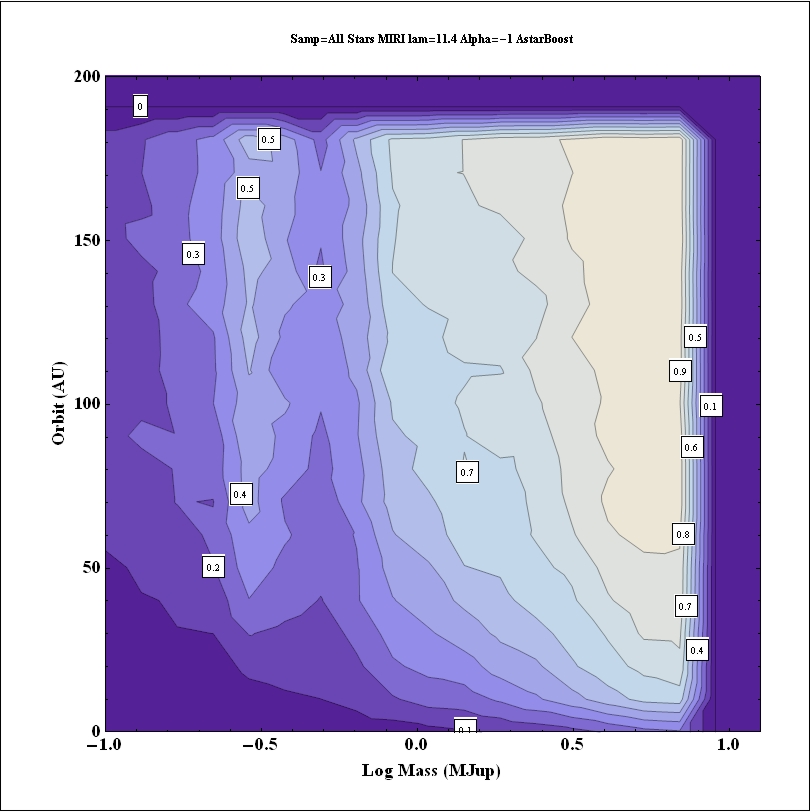}
\end{center}
\caption{ The plot shows the fractional detectability of planets orbiting nearby young stars ($\alpha=-1$) for the MIRI/FQPM interferometer at 11.4 $\mu$m. The vertical axis represents orbital semi-major axis (AU) and the horizontal axis Log(planet mass) in (M$_{Jup}$). The contours represent the probability of detection of a planet, averaged over all stars, in a specific Mass-SMA bin, i.e. the number of planets detected in that bin divided by the number of planets generated in the simulation. }\label{MIRIdetect}
\end{figure}

\clearpage
\begin{figure} 
\begin{center}
\includegraphics[height=.6\textheight]{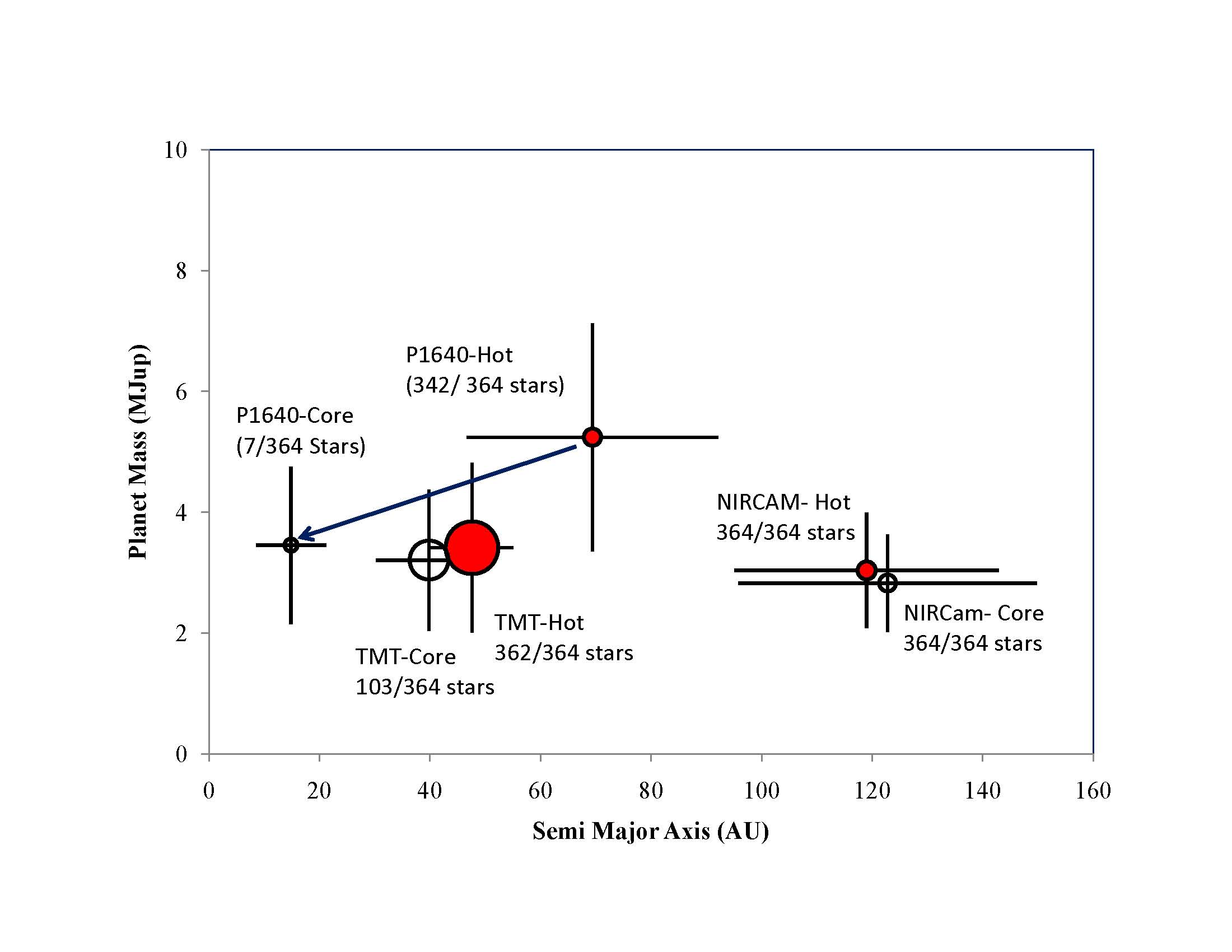}
\end{center}
\caption{The plot shows the average values of planet mass and semi-major axis for planets detected with P1640, TMT and NIRcam (Table~\ref{Fortneydetectionsummary}) for the ``hot start'' \citep{baraffe} and ``core accretion" \citep{fortney, marley} evolutionary scenarios. Stellar ages are less than 100 Myr and planet masses restricted to 1-10 M$_{Jup}$. The most dramatic change comes for P1640 class instruments for which detectability drops to very small numbers. TMT and NIRCam have sensitivity to detect planets in this age-mass range with some independence of the evolutionary model; lower mass planets would start to become hard to detect however. The size of the symbol is proportional the fraction of stars around which at least 1 planet was detected. The vertical and horizontal bars denote the 1 $\sigma$ dispersion in these quantities.}\label{Fortney}
\end{figure}

\clearpage
\begin{figure} 
\begin{center}$
\begin{array}{c}
\includegraphics[height=.4\textheight]{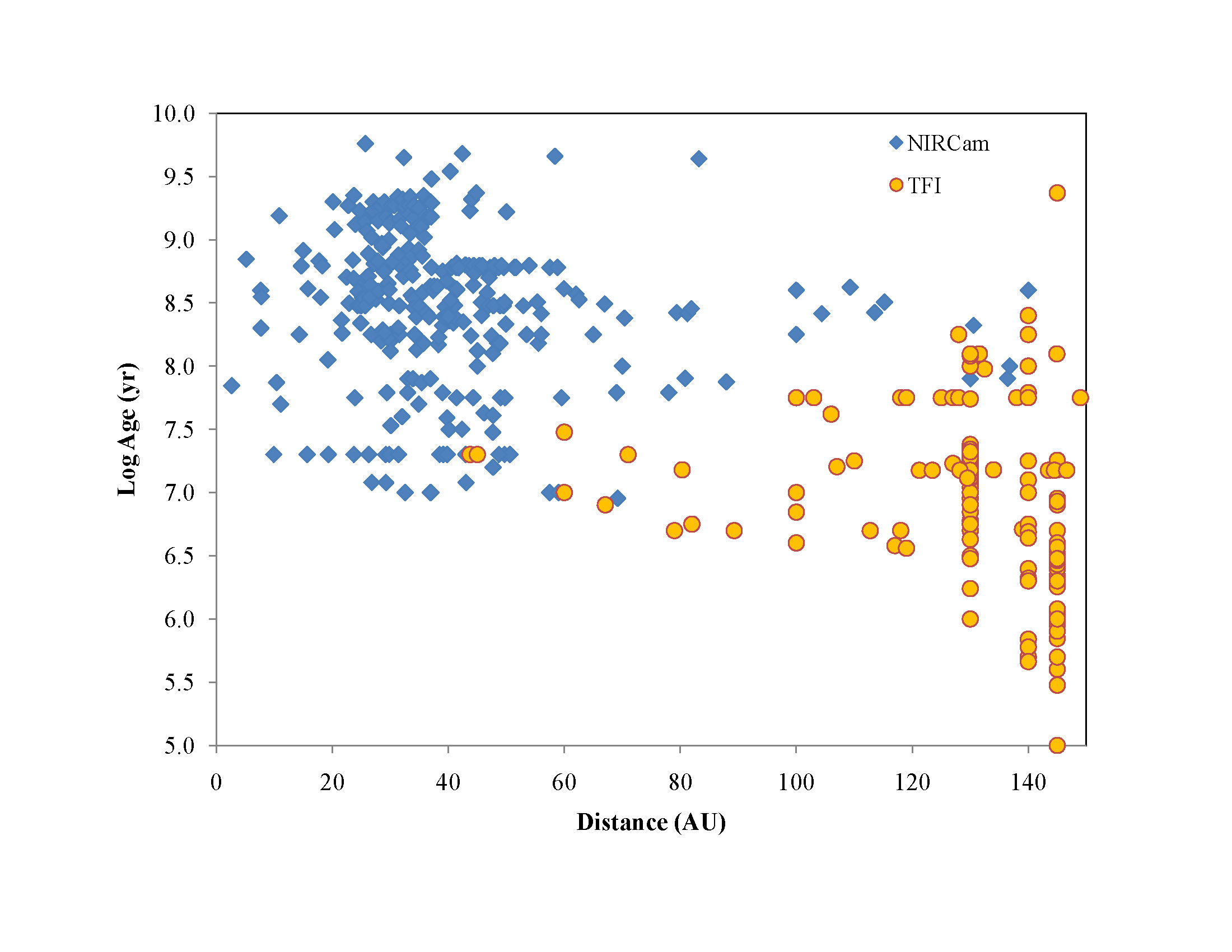}\\
\includegraphics[height=.4\textheight]{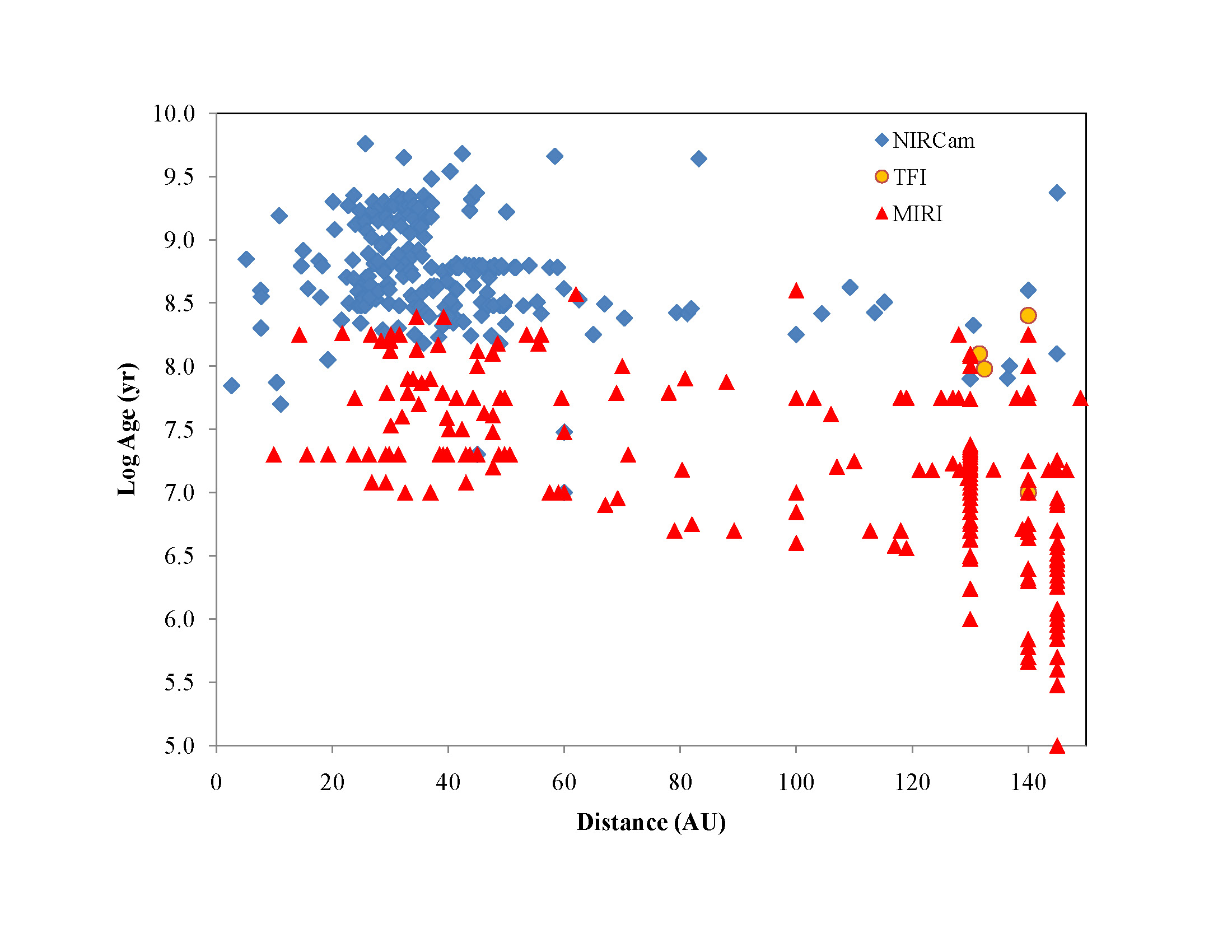}\\
\end{array}$
\end{center}
\caption{top) The instrument achieving the highest detectability score for the young stars sample ($\alpha=-1$) is shown on a star by star basis in the distance-age plane for JWST's NIRCam and TFI/NRM at 4.4 $\mu$m. bottom) comparable plot but for with the addition of MIRI at 11.4 $\mu$m.}\label{StarProps}
\end{figure}

\clearpage
\begin{figure} 
\begin{center}
\includegraphics[height=.4\textheight]{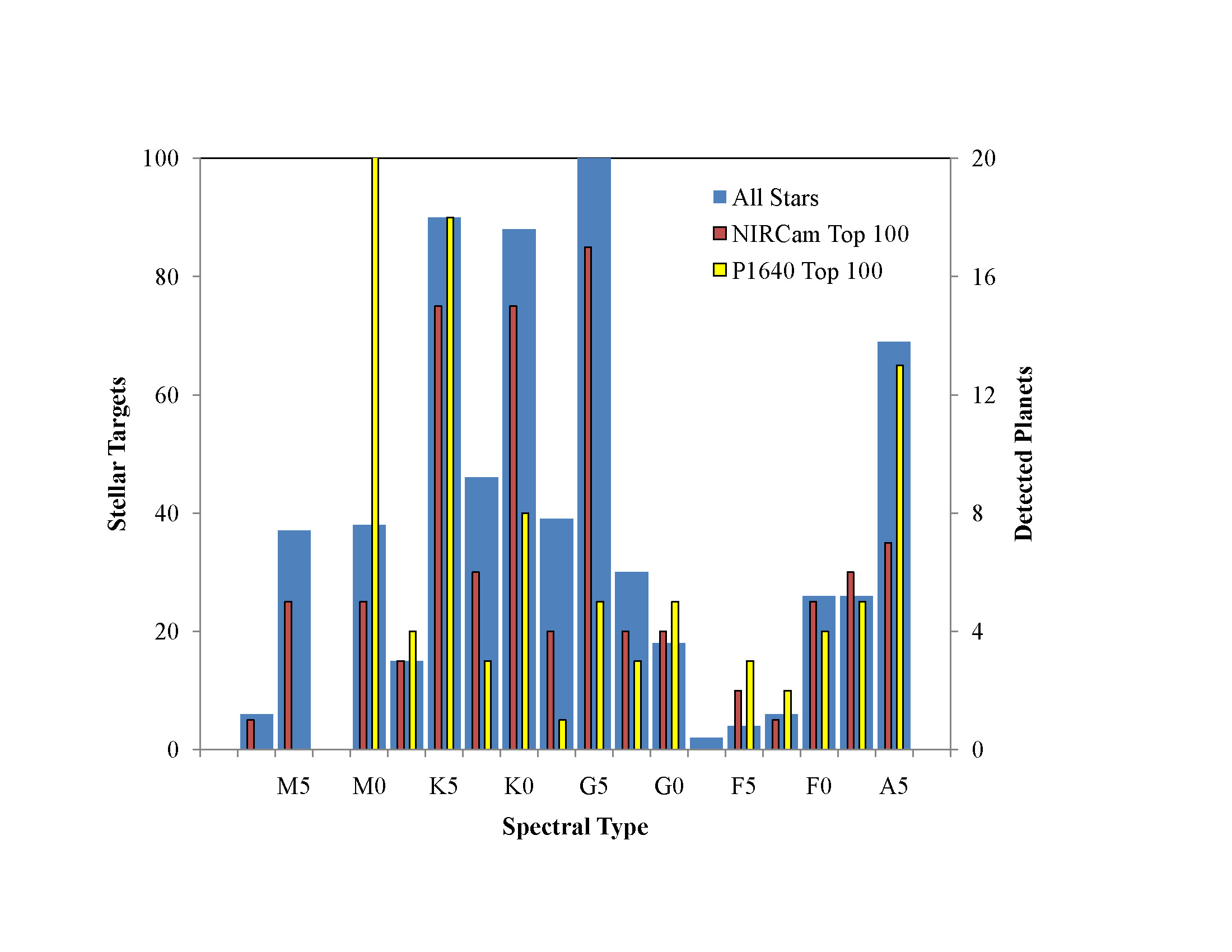}
\end{center}
\caption{top) The blue histogram bins  (left hand scale)  denote the distribution of spectral types for the 641 stars in the young star sample. The red and yellow bins (right hand scale) give the distribution of the top 100 ranked stars detected by NIRCam and P1640, respectively, for the $\alpha=-1$ planet distribution. }\label{SpHisto}
\end{figure}

\clearpage
\begin{figure} 
\begin{center}$
\begin{array}{c}
\includegraphics[height=.4\textheight]{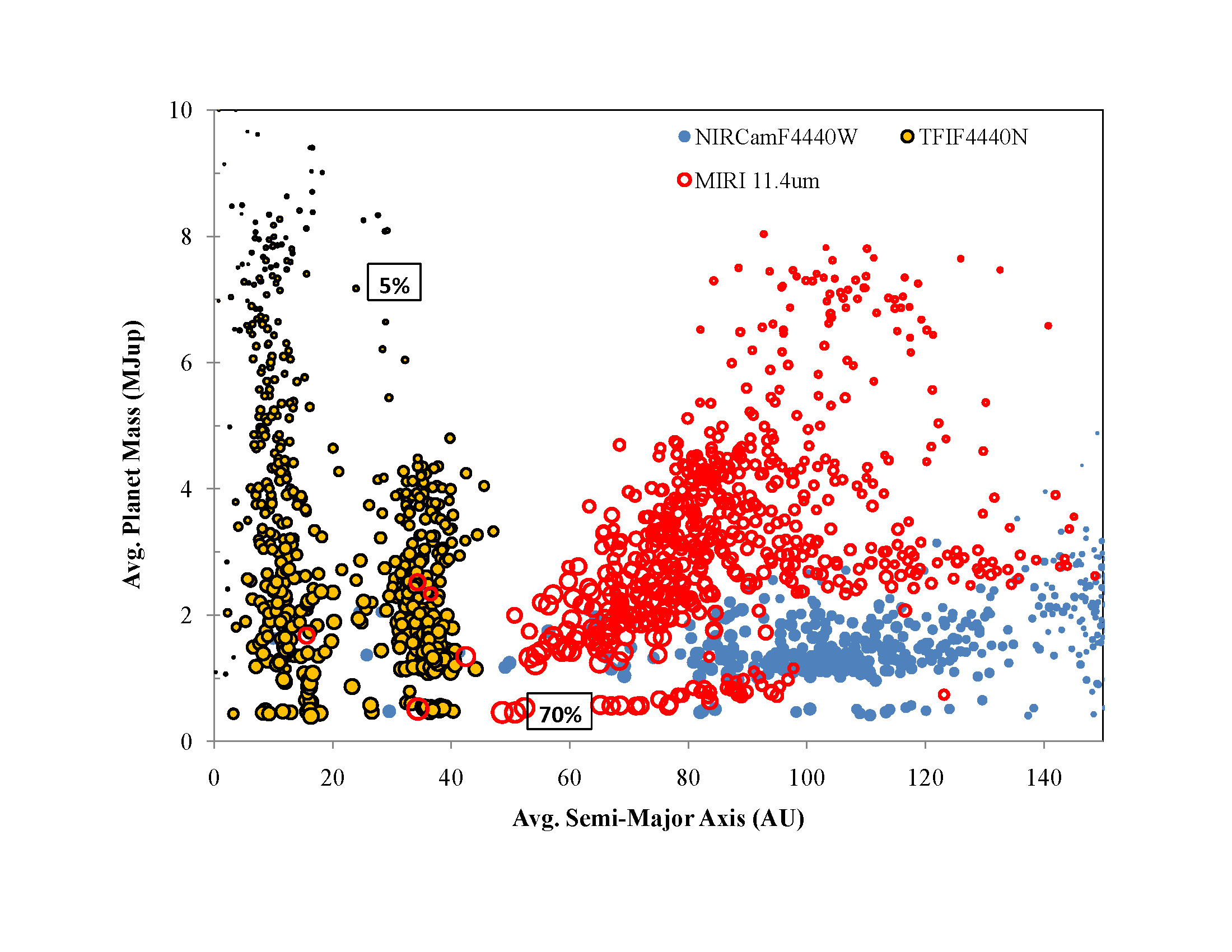}\\
\includegraphics[height=.4\textheight]{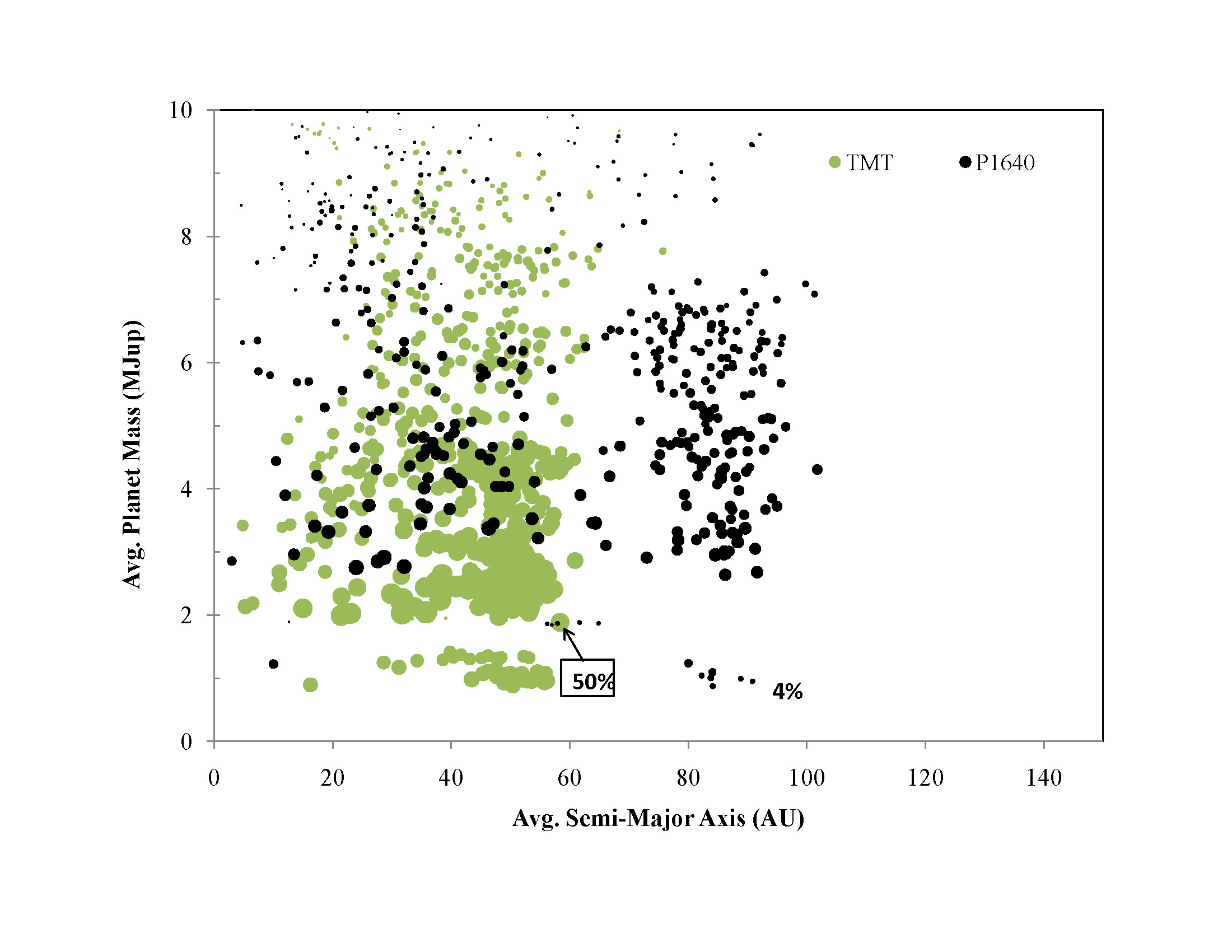} \\
\end{array}$
\end{center}
\caption{top) The average Mass and Semi Major Axis (SMA) detected for each  star in the young star sample ($\alpha=-1$)is shown with the size of the point proportional the fractional detectability of planets around that star (the percentages in boxes denote the achieved success rate corresponding to  the adjacent symbol size). Results for three JWST instruments are shown: NIRCam (light blue circles) and TFI/NRM (yellow circles with black border) at 4.4 $\mu$m, MIRI at 11.4 $\mu$m (red open circles). bottom) Comparable plot for P1640 (black circles) and TMT (green circles) at 1.65 $\mu$m. For GPI and SPHERE on 8 m telescopes, the black dots would shift inward by roughly a factor of 5/8 in orbital radius. The detections below $\sim$ 1 M$_{Jup}$ correspond to planets orbiting young M stars for which the contrast ratio is particularly favorable.}\label{PlanetProp}
\end{figure}

\clearpage
\begin{figure}
\begin{center}
\includegraphics[height=.6\textheight]{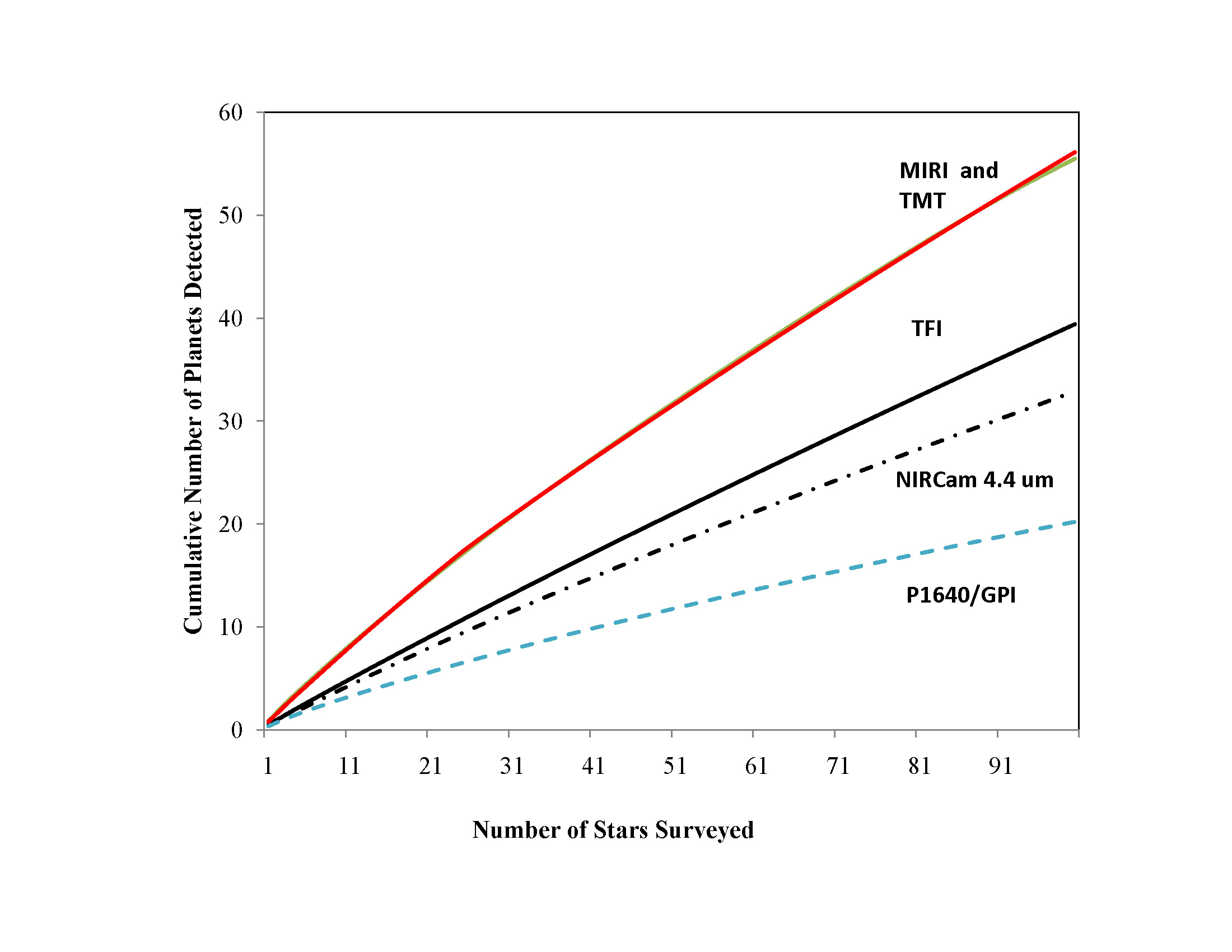} 
\end{center}
\caption{ The cumulative number of planets detected as a function of the number of stars surveyed for the $\alpha=-1$ young star sample for  different instruments. The stars are rank ordered according to the score in the Monte Carlo simulation (Appendix 2).}\label{Cumulative}
\end{figure}

\clearpage
\begin{figure}
\begin{center}
\includegraphics[height=.6\textheight]{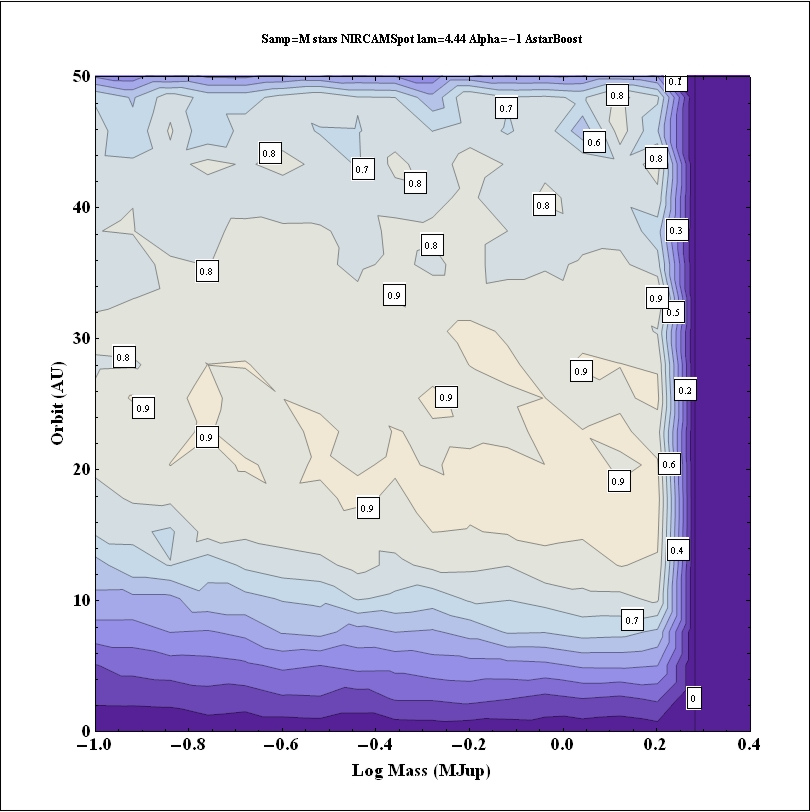} 
\end{center}
\caption{ The ability for NIRCam to detect planets orbiting nearby M stars at 4.4 $\mu$m. The vertical axis represents orbital semi-major axis (AU) and the horizontal axis Log(planet mass) in (M$_{Jup}$). The contours represent the probability of detection of a planet, averaged over all stars, in a specific Mass-SMA bin, i.e. the number of planets detected in that bin divided by the number of planets generated in the simulation. NIRCam can detect planets uniformly across the mass range 0.1-2 M$_{Jup}$ assumed in the simulation but their detectability drops off at distances beyond 50  AU which  falls outside  the 10\arcsec\ field of view for stars beyond 5-10 pc. }\label{NIRCAMdetectmstar}
\end{figure}

\clearpage
\begin{figure}
\begin{center}
\includegraphics[height=.6\textheight]{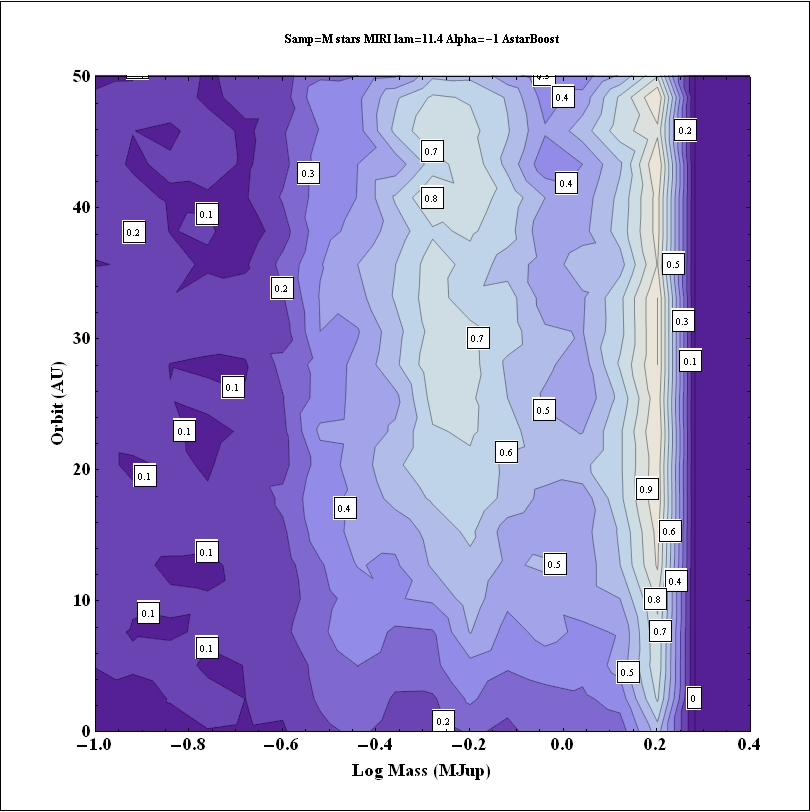}
\end{center}
\caption{The ability for MIRI to detect planets orbiting nearby M stars at 11.4 $\mu$m. The vertical axis represents orbital semi-major axis (AU) and the horizontal axis Log(planet mass) in (M$_{Jup}$). The contours represent the probability of detection of a planet, averaged over all stars, in a specific Mass-SMA bin, i.e. the number of planets detected in that bin divided by the number of planets generated in the simulation. The  vertical locus at  of contours represent  M  stars younger than 100 Myr  for which $<1$ M$_{Jup}$ planets can be detected while the second locus ($log(M)\sim +0.2$ ) represents older stars for which only more massive planets can be detected. The detectability of planets drops off at distances beyond 50 AU due to the 13\arcsec\ field of view for stars beyond 5-10 pc.}\label{MIRIdetectmstar}
\end{figure}

\clearpage
\section{Appendix I. Broad Band Colors for Planets}

The following tables list absolute magnitudes (10 pc) for planets as observed through standard ground-based filters (J, H, K, L, M), four JWST/NIRCAM broad-band filters, and three JWST/MIRI narrow band filters. For the NIRCam filters we have used measured transmissions curves while for
the MIRI filters we have taken a square (R=20) passband centered on 10.65, 11.4, and 15.5 $\mu$m as a reasonable approximation for these narrow band filters. The evolutionary tracks represent an extension of CONDO3 models \cite{baraffe} for planetary masses down to 0.1 M$_{Jup}$ and for ages up to 10 Gyr. Tracks are dropped as the effective temperature for a given mass-age combination drops below 100 K at which point the model atmosphere calculations become unreliable. The synthetic spectra utilize an updated grid (the GAIA grid) which include minor corrections to the older "AMES" spectra. The changes to the colors are modest and are too small to impact the cooling tracks in any noticeable way. There are ten tables for ages ranging from 1 Myr to 10 Gyr. These are available electronically.

\clearpage
\begin{table}
\caption{Calculated Planet Magnitudes (10 pc) For 1 Myr }\footnotesize
\tiny
\begin{tabular}{lrrr|rrrrr|rrrr|rrr}
M$_{pl}$&T$_{eff}$&L$_{pl}$&log(g)&J&H&K&L&M&F150W&F277W&F356W&F444W&MIRI-1&MIRI-2&MIRI-3\\
M$\odot$&(K)&L$_\odot$&cgs&(mag)&(mag)&(mag)&(mag)&(mag)&(mag)&(mag)&(mag)&(mag)&(mag)&(mag)&(mag)\\ \hline
0.0001&247&-7.201&2.15&28.26&26.07&35.45&18.83&17.08&23.1&21.58&18.76&15.86&14.66&13.04&12.59\\
0.0002&384&-6.184&2.2&20.55&19.87&23.05&15.34&14.14&20.67&19.72&16.97&14.25&13.21&11.73&11.32\\
0.0003&450&-5.9&2.37&19.11&18.57&20.41&14.57&13.49&18.56&18.55&15.76&13.27&12.45&11.06&10.78\\
0.0004&517&-5.68&2.52&17.87&17.49&18.33&13.93&12.97&17.74&18.32&15.38&13.01&12.33&10.9&10.67\\
0.0005&628&-5.37&2.64&16.55&16.28&16.43&13.15&12.45&16.73&17.29&14.51&12.42&11.41&10.33&10.17\\
0.001&942&-4.715&3&14.52&14.11&13.69&11.52&11.45&14.77&14.67&12.49&11.2&9.84&9.47&9.45\\
0.002&1285&-4.137&3.26&12.87&12.38&11.95&10.27&10.64&13.13&12.58&10.87&10.28&9.07&8.96&8.98\\
0.003&1553&-3.746&3.37&11.81&11.36&10.89&9.57&10.09&12.1&11.59&9.95&9.67&8.68&8.62&8.62\\
0.004&1747&-3.482&3.44&11.12&10.71&10.21&9.12&9.66&11.41&10.93&9.41&9.23&8.39&8.36&8.35\\
0.005&1901&-3.273&3.47&10.6&10.22&9.73&8.77&9.28&10.84&10.36&9&8.85&8.13&8.1&8.1\\
0.006&2004&-3.129&3.5&10.23&9.87&9.4&8.52&8.97&10.43&9.93&8.72&8.56&7.92&7.9&7.9\\
0.007&2098&-2.998&3.52&9.91&9.55&9.11&8.29&8.69&10.06&9.52&8.45&8.29&7.72&7.7&7.7\\
0.008&2159&-2.893&3.52&9.67&9.3&8.87&8.08&8.45&9.78&9.22&8.23&8.06&7.53&7.52&7.51\\
0.009&2207&-2.811&3.53&9.48&9.11&8.68&7.92&8.26&9.56&8.99&8.06&7.89&7.39&7.37&7.37\\
0.010&2251&-2.735&3.53&9.3&8.93&8.52&7.78&8.09&9.36&8.77&7.9&7.72&7.25&7.23&7.22\\
0.012&2321&-2.615&3.54&9.02&8.65&8.25&7.54&7.81&9.05&8.43&7.65&7.46&7.04&7.02&7.01\\
0.015&2400&-2.455&3.54&8.64&8.27&7.89&7.21&7.45&8.64&8&7.31&7.12&6.72&6.71&6.69\\
0.020&2484&-2.28&3.55&8.22&7.86&7.49&6.85&7.05&8.19&7.54&6.92&6.73&6.37&6.36&6.34\\
0.030&2598&-2.174&3.69&8&7.62&7.27&6.68&6.85&7.91&7.29&6.74&6.55&6.23&6.22&6.2\\
0.040&2746&-1.939&3.68&7.5&7.1&6.76&6.22&6.31&7.33&6.73&6.26&6.08&5.8&5.79&5.77\\
0.050&2768&-1.637&3.49&6.74&6.35&6.02&5.48&5.57&6.58&5.98&5.52&5.34&5.07&5.05&5.03\\
0.060&2824&-1.604&3.57&6.68&6.28&5.96&5.44&5.52&6.49&5.92&5.48&5.3&5.04&5.03&5.01\\
0.070&2853&-1.478&3.53&6.37&5.96&5.65&5.15&5.23&6.17&5.61&5.18&5.01&4.76&4.74&4.72\\
0.072&2858&-1.455&3.52&6.31&5.9&5.6&5.1&5.18&6.11&5.56&5.13&4.95&4.71&4.69&4.67\\
0.075&2833&-1.389&3.46&6.14&5.73&5.42&4.91&5&5.95&5.38&4.95&4.77&4.51&4.5&4.47\\
0.080&2869&-1.373&3.49&6.11&5.7&5.4&4.9&4.98&5.91&5.36&4.93&4.76&4.51&4.5&4.48\\
0.090&2867&-1.289&3.46&5.89&5.48&5.18&4.69&4.77&5.7&5.15&4.72&4.55&4.3&4.29&4.26\\
0.100&2856&-1.187&3.39&5.63&5.22&4.92&4.43&4.51&5.44&4.89&4.46&4.28&4.04&4.02&4\\
\hline
\end{tabular}
\tablecomments{$^1$ Assumed central wavelengths for the MIRI/FQPM narrow band filters
are 10.65, 11.4, and 15.5 $\mu$m with a simple square passband with resolution=20.}
\label{baraffe}
\end{table}

\begin{table}
\caption{Calculated Planet Magnitudes (10 pc) For 5 Myr}\footnotesize
\tiny
\begin{tabular}{lrrr|rrrrr|rrrr|rrr}
M$_{pl}$&T$_{eff}$&L$_{pl}$&log(g)&J&H&K&L&M&F150W&F277W&F356W&F444W&MIRI-1&MIRI-2&MIRI-3\\
M$\odot$&(K)&L$_\odot$&cgs&(mag)&(mag)&(mag)&(mag)&(mag)&(mag)&(mag)&(mag)&(mag)&(mag)&(mag)&(mag)\\ \hline
0.0001&217&-7.514&2.24&31.39&28.37&39.89&20.08&18.13&23.72&22.08&19.24&16.31&15.07&13.43&12.97\\
0.0002&308&-6.769&2.41&22.75&22.04&27.36&16.77&15.38&22.18&20.92&18.13&15.31&14.18&12.63&12.2\\
0.0003&360&-6.46&2.55&21.47&20.8&24.72&16.05&14.78&21.39&20.34&17.58&14.83&13.77&12.27&11.85\\
0.0004&400&-6.284&2.68&20.51&19.9&22.81&15.58&14.39&19.66&19.1&16.5&13.91&12.94&11.65&11.34\\
0.0005&455&-6.079&2.79&19.37&18.91&20.96&15.03&13.92&18.91&19.01&16.26&13.74&13.03&11.62&11.34\\
0.001&644&-5.522&3.14&16.84&16.57&16.97&13.56&12.85&16.98&17.6&14.92&12.79&11.92&10.85&10.67\\
0.002&901&-4.92&3.42&14.94&14.66&14.3&12.07&11.83&15.31&15.42&13.18&11.68&10.46&9.94&9.88\\
0.003&1098&-4.552&3.58&13.89&13.56&13.21&11.21&11.31&14.23&13.96&12.06&11.03&9.74&9.52&9.51\\
0.004&1263&-4.283&3.67&13.16&12.79&12.46&10.66&10.96&13.48&13.05&11.33&10.62&9.42&9.29&9.28\\
0.005&1413&-4.061&3.74&12.58&12.17&11.85&10.23&10.67&12.9&12.39&10.71&10.27&9.18&9.1&9.09\\
0.006&1543&-3.886&3.8&12.11&11.72&11.36&9.92&10.44&12.43&12.01&10.36&10.01&9.02&8.96&8.93\\
0.007&1668&-3.725&3.84&11.69&11.3&10.9&9.64&10.2&12.02&11.61&10.01&9.76&8.86&8.82&8.79\\
0.008&1786&-3.579&3.87&11.33&10.95&10.52&9.4&9.97&11.63&11.24&9.71&9.53&8.71&8.68&8.66\\
0.009&1885&-3.46&3.9&11.04&10.68&10.24&9.22&9.76&11.32&10.94&9.49&9.34&8.58&8.56&8.54\\
0.01&1965&-3.363&3.92&10.8&10.44&10&9.06&9.57&11.04&10.64&9.3&9.15&8.46&8.44&8.43\\
0.012&2102&-3.194&3.95&10.39&10.03&9.61&8.77&9.21&10.55&10.11&8.96&8.81&8.23&8.22&8.21\\
0.015&2264&-2.981&3.96&9.92&9.53&9.14&8.39&8.72&9.96&9.45&8.53&8.37&7.9&7.88&7.87\\
0.020&2452&-2.661&3.9&9.18&8.79&8.42&7.77&8&9.13&8.55&7.86&7.68&7.31&7.29&7.28\\
0.030&2616&-2.262&3.8&8.23&7.85&7.5&6.91&7.07&8.13&7.52&6.97&6.79&6.48&6.46&6.44\\
0.040&2754&-2.068&3.81&7.82&7.42&7.08&6.54&6.63&7.65&7.06&6.59&6.41&6.13&6.12&6.1\\
0.050&2806&-1.846&3.72&7.28&6.88&6.55&6.03&6.11&7.1&6.52&6.07&5.9&5.63&5.62&5.59\\
0.060&2875&-1.902&3.9&7.45&7.03&6.72&6.22&6.3&7.23&6.68&6.26&6.09&5.84&5.83&5.81\\
0.070&2916&-1.829&3.92&7.28&6.85&6.55&6.07&6.14&7.04&6.51&6.1&5.94&5.7&5.69&5.67\\
0.072&2921&-1.814&3.92&7.24&6.82&6.51&6.03&6.11&7&6.48&6.06&5.9&5.67&5.65&5.63\\
0.075&2912&-1.716&3.83&6.99&6.57&6.27&5.78&5.86&6.76&6.23&5.81&5.65&5.41&5.4&5.38\\
0.080&2946&-1.76&3.92&7.11&6.68&6.39&5.92&6&6.87&6.35&5.94&5.79&5.56&5.55&5.53\\
0.090&2974&-1.696&3.93&6.96&6.53&6.24&5.78&5.85&6.7&6.2&5.79&5.64&5.42&5.41&5.39\\
0.100&2983&-1.582&3.87&6.67&6.24&5.95&5.5&5.58&6.42&5.92&5.52&5.37&5.15&5.14&5.12\\
\end{tabular}
\tablecomments{$^1$ Assumed central wavelengths for the MIRI/FQPM narrow band filters
are 10.65, 11.4, and 15.5 $\mu$m with a simple square passband with resolution=20.}
\end{table}

\clearpage
\begin{table}
\caption{Calculated Planet Magnitudes (10 pc) For 10 Myr}\footnotesize
\tiny
\begin{tabular}{lrrr|rrrrr|rrrr|rrr}
M$_{pl}$&T$_{eff}$&L$_{pl}$&log(g)&J&H&K&L&M&F150W&F277W&F356W&F444W&MIRI-1&MIRI-2&MIRI-3\\
M$\odot$&(K)&L$_\odot$&cgs&(mag)&(mag)&(mag)&(mag)&(mag)&(mag)&(mag)&(mag)&(mag)&(mag)&(mag)&(mag)\\ \hline
0.0001&196&-7.739&2.29&33.53&29.94&42.93&20.92&18.84&24.12&22.39&19.55&16.59&15.32&13.66&13.19\\
0.0002&274&-7.048&2.49&25.63&24.2&31.59&17.95&16.38&22.82&21.41&18.61&15.74&14.58&12.99&12.54\\
0.0003&326&-6.713&2.62&22.36&21.75&26.77&16.68&15.3&22.01&20.8&18.05&15.26&14.17&12.64&12.21\\
0.0004&366&-6.499&2.74&21.36&20.8&24.78&16.16&14.87&21.44&20.41&17.68&14.95&13.92&12.44&12.01\\
0.0005&394&-6.38&2.85&20.7&20.16&23.4&15.83&14.61&21.07&20.16&17.45&14.76&13.76&12.32&11.91\\
0.001&540&-5.87&3.18&18.12&17.79&19.05&14.48&13.5&18.02&18.67&15.91&13.51&13&11.61&11.34\\
0.002&746&-5.298&3.47&16.02&15.79&15.86&13.02&12.47&16.29&16.77&14.29&12.36&11.35&10.53&10.39\\
0.003&920&-4.914&3.63&14.88&14.65&14.33&12.09&11.84&15.26&15.35&13.17&11.67&10.48&9.98&9.92\\
0.004&1064&-4.645&3.74&14.12&13.85&13.51&11.46&11.45&14.5&14.33&12.37&11.21&9.97&9.68&9.65\\
0.005&1193&-4.429&3.82&13.52&13.22&12.92&10.99&11.18&13.88&13.52&11.75&10.87&9.66&9.48&9.47\\
0.006&1316&-4.244&3.88&13.03&12.68&12.41&10.62&10.95&13.35&12.93&11.25&10.59&9.44&9.32&9.3\\
0.007&1426&-4.091&3.94&12.64&12.27&11.99&10.33&10.77&12.94&12.51&10.85&10.36&9.3&9.21&9.17\\
0.008&1536&-3.945&3.98&12.25&11.87&11.56&10.07&10.58&12.57&12.17&10.53&10.15&9.16&9.09&9.05\\
0.009&1635&-3.82&4.01&11.92&11.54&11.19&9.84&10.4&12.26&11.88&10.26&9.97&9.04&8.99&8.95\\
0.010&1731&-3.704&4.04&11.63&11.26&10.87&9.65&10.23&11.94&11.57&10&9.78&8.92&8.89&8.85\\
0.012&1915&-3.488&4.08&11.1&10.75&10.33&9.32&9.87&11.38&11.02&9.59&9.44&8.71&8.68&8.66\\
0.015&2188&-3.125&4.05&10.25&9.87&9.47&8.68&9.07&10.33&9.88&8.85&8.69&8.17&8.16&8.15\\
0.020&2442&-2.709&3.94&9.3&8.91&8.54&7.88&8.11&9.25&8.68&7.98&7.8&7.42&7.41&7.39\\
0.030&2622&-2.435&3.97&8.68&8.28&7.93&7.35&7.5&8.56&7.97&7.41&7.24&6.92&6.91&6.89\\
0.040&2734&-2.411&4.14&8.67&8.25&7.92&7.37&7.5&8.5&7.93&7.43&7.26&6.98&6.97&6.95\\
0.050&2821&-2.233&4.12&8.26&7.84&7.52&7&7.1&8.06&7.5&7.05&6.88&6.62&6.61&6.59\\
0.060&2871&-2.188&4.18&8.16&7.73&7.43&6.93&7.03&7.94&7.41&6.97&6.81&6.56&6.55&6.53\\
0.070&2919&-2.099&4.19&7.95&7.52&7.22&6.74&6.83&7.71&7.19&6.77&6.62&6.38&6.37&6.35\\
0.072&2927&-2.083&4.19&7.91&7.48&7.18&6.7&6.8&7.67&7.15&6.74&6.58&6.35&6.33&6.31\\
0.075&2942&-2.022&4.16&7.76&7.33&7.04&6.56&6.65&7.52&7.01&6.59&6.44&6.21&6.2&6.18\\
0.080&2958&-2.016&4.19&7.75&7.32&7.03&6.56&6.65&7.51&7&6.59&6.44&6.21&6.2&6.18\\
0.090&2993&-1.943&4.19&7.58&7.14&6.86&6.41&6.49&7.32&6.83&6.42&6.28&6.06&6.05&6.03\\
0.100&3024&-1.854&4.16&7.37&6.93&6.65&6.21&6.29&7.1&6.61&6.22&6.08&5.87&5.86&5.84\\
\end{tabular}
\tablecomments{$^1$ Assumed central wavelengths for the MIRI/FQPM narrow band filters
are 10.65, 11.4, and 15.5 $\mu$m with a simple square passband with resolution=20.}
\end{table}

\clearpage
\begin{table}
\caption{Calculated Planet Magnitudes (10 pc) For 50 Myr}\footnotesize
\tiny
\begin{tabular}{lrrr|rrrrr|rrrr|rrr}
M$_{pl}$&T$_{eff}$&L$_{pl}$&log(g)&J&H&K&L&M&F150W&F277W&F356W&F444W&MIRI-1&MIRI-2&MIRI-3\\
M$\odot$&(K)&L$_\odot$&cgs&(mag)&(mag)&(mag)&(mag)&(mag)&(mag)&(mag)&(mag)&(mag)&(mag)&(mag)&(mag)\\ \hline
0.0001&139&-8.453&2.41&39.4&34.24&51.32&23.18&20.73&25.17&23.2&20.33&17.29&15.96&14.24&13.76\\
0.0002&196&-7.794&2.65&33.33&29.9&43.33&21.08&18.97&24.22&22.42&19.65&16.7&15.47&13.8&13.32\\
0.0003&232&-7.465&2.79&29.54&27.25&38.39&19.78&17.86&23.62&21.92&19.23&16.32&15.17&13.54&13.06\\
0.0004&262&-7.232&2.89&26.54&25.17&34.3&18.72&16.95&23.14&21.56&18.9&16.04&14.94&13.34&12.87\\
0.0005&285&-7.069&2.97&24.23&23.59&31.08&17.89&16.24&22.81&21.32&18.68&15.85&14.78&13.21&12.75\\
0.001&375&-6.587&3.27&21.1&20.8&25.14&16.44&15.02&21.51&20.45&17.88&15.16&14.25&12.77&12.35\\
0.002&491&-6.1&3.55&18.92&18.55&20.7&15.15&13.94&18.68&19.17&16.51&13.97&13.64&12.2&11.83\\
0.003&585&-5.789&3.72&17.63&17.39&18.67&14.31&13.34&17.68&18.3&15.75&13.36&12.86&11.62&11.34\\
0.004&676&-5.531&3.84&16.72&16.51&17.21&13.64&12.85&16.91&17.45&15.01&12.82&12.09&11.09&10.88\\
0.005&756&-5.336&3.93&16.09&15.91&16.2&13.16&12.51&16.37&16.77&14.44&12.44&11.55&10.74&10.58\\
0.006&840&-5.152&4.01&15.51&15.36&15.34&12.72&12.21&15.88&16.12&13.91&12.11&11.11&10.45&10.33\\
0.007&928&-4.978&4.08&14.98&14.85&14.64&12.31&11.94&15.4&15.48&13.39&11.8&10.71&10.19&10.11\\
0.008&1010&-4.832&4.14&14.55&14.41&14.14&11.97&11.74&14.99&14.94&12.96&11.55&10.42&10.02&9.96\\
0.009&1085&-4.706&4.19&14.19&14.04&13.76&11.69&11.59&14.63&14.48&12.59&11.36&10.21&9.9&9.86\\
0.010&1171&-4.569&4.23&13.82&13.64&13.38&11.38&11.43&14.26&13.99&12.21&11.17&10.02&9.8&9.77\\
0.012&1520&-4.052&4.24&12.49&12.17&11.95&10.34&10.81&12.83&12.45&10.82&10.39&9.41&9.33&9.28\\
0.015&1903&-3.568&4.25&11.29&10.94&10.56&9.5&10.06&11.57&11.24&9.77&9.62&8.88&8.86&8.82\\
0.020&1909&-3.642&4.45&11.47&11.13&10.79&9.69&10.26&11.74&11.45&9.97&9.82&9.08&9.06&9.02\\
0.030&2246&-3.311&4.58&10.7&10.33&9.98&9.19&9.63&10.77&10.44&9.36&9.23&8.72&8.71&8.69\\
0.040&2480&-3.08&4.64&10.21&9.81&9.48&8.82&9.12&10.14&9.72&8.93&8.78&8.41&8.4&8.38\\
0.050&2651&-2.884&4.66&9.79&9.38&9.06&8.48&8.7&9.64&9.18&8.56&8.4&8.1&8.09&8.07\\
0.060&2759&-2.755&4.68&9.51&9.09&8.79&8.24&8.42&9.34&8.86&8.32&8.16&7.89&7.88&7.86\\
0.070&2837&-2.647&4.69&9.27&8.84&8.55&8.04&8.19&9.08&8.6&8.1&7.95&7.69&7.68&7.67\\
0.072&2852&-2.623&4.68&9.22&8.79&8.5&7.99&8.14&9.01&8.53&8.04&7.89&7.64&7.63&7.61\\
0.075&2872&-2.591&4.68&9.15&8.71&8.42&7.92&8.07&8.93&8.45&7.98&7.83&7.58&7.57&7.56\\
0.080&2901&-2.551&4.69&9.06&8.62&8.34&7.85&7.99&8.82&8.34&7.89&7.74&7.5&7.49&7.47\\
0.090&2948&-2.472&4.69&8.88&8.44&8.16&7.68&7.82&8.63&8.15&7.72&7.57&7.35&7.34&7.32\\
0.100&2989&-2.397&4.68&8.7&8.26&7.99&7.53&7.65&8.45&7.98&7.56&7.42&7.2&7.19&7.18\\
\end{tabular}
\tablecomments{$^1$ Assumed central wavelengths for the MIRI/FQPM narrow band filters
are 10.65, 11.4, and 15.5 $\mu$m with a simple square passband with resolution=20.}
\end{table}

\clearpage
\begin{table}
\caption{Calculated Planet Magnitudes (10 pc) For 100 Myr}\footnotesize
\tiny
\begin{tabular}{lrrr|rrrrr|rrrr|rrr}
M$_{pl}$&T$_{eff}$&L$_{pl}$&log(g)&J&H&K&L&M&F150W&F277W&F356W&F444W&MIRI-1&MIRI-2&MIRI-3\\
M$\odot$&(K)&L$_\odot$&cgs&(mag)&(mag)&(mag)&(mag)&(mag)&(mag)&(mag)&(mag)&(mag)&(mag)&(mag)&(mag)\\ \hline
0.0001&115&-8.816&2.45&41.81&35.99&54.78&24.1&21.49&25.57&23.5&20.62&17.55&16.2&14.45&13.96\\
0.0002&160&-8.188&2.69&36.66&32.33&48.49&22.42&20.08&24.8&22.81&20.07&17.07&15.82&14.11&13.61\\
0.0003&190&-7.858&2.84&33.38&30.03&44.4&21.32&19.15&24.29&22.37&19.7&16.75&15.57&13.89&13.39\\
0.0004&218&-7.599&2.94&30.57&28.07&40.63&20.34&18.3&23.82&21.99&19.37&16.46&15.33&13.68&13.18\\
0.0005&240&-7.418&3.02&28.42&26.59&37.66&19.57&17.64&23.5&21.75&19.16&16.27&15.18&13.56&13.06\\
0.001&309&-6.957&3.3&22.43&22.38&29.11&17.41&15.69&22.47&21.03&18.51&15.71&14.76&13.19&12.73\\
0.002&425&-6.383&3.58&20.05&19.76&23.13&15.94&14.55&19.62&19.3&16.89&14.28&13.67&12.31&11.98\\
0.003&493&-6.112&3.75&18.88&18.57&20.88&15.21&13.93&18.71&19.12&16.58&14.01&13.71&12.3&11.92\\
0.004&563&-5.88&3.87&17.95&17.71&19.35&14.59&13.5&17.96&18.49&16.02&13.56&13.15&11.87&11.55\\
0.005&630&-5.686&3.96&17.23&17.02&18.15&14.06&13.14&17.34&17.89&15.48&13.15&12.58&11.46&11.2\\
0.006&688&-5.534&4.05&16.71&16.51&17.26&13.67&12.83&16.91&17.36&15.02&12.82&12.1&11.15&10.94\\
0.007&760&-5.365&4.12&16.16&16.01&16.38&13.26&12.55&16.45&16.76&14.52&12.5&11.64&10.84&10.68\\
0.008&816&-5.246&4.18&15.76&15.65&15.79&12.97&12.35&16.13&16.34&14.17&12.28&11.35&10.66&10.52\\
0.009&886&-5.103&4.23&15.32&15.23&15.16&12.63&12.13&15.74&15.85&13.77&12.04&11.04&10.45&10.34\\
0.010&953&-4.978&4.28&14.94&14.86&14.69&12.34&11.96&15.4&15.42&13.41&11.83&10.78&10.29&10.2\\
0.012&1335&-4.332&4.3&13.2&12.97&12.76&10.9&11.17&13.58&13.19&11.55&10.82&9.74&9.6&9.56\\
0.015&1399&-4.281&4.42&13.05&12.82&12.65&10.83&11.15&13.44&13.05&11.44&10.8&9.75&9.64&9.58\\
0.020&1561&-4.11&4.57&12.6&12.34&12.17&10.53&10.99&12.95&12.6&11.01&10.57&9.63&9.56&9.49\\
0.030&1979&-3.668&4.72&11.52&11.2&10.9&9.82&10.38&11.77&11.5&10.07&9.93&9.24&9.22&9.18\\
0.040&2270&-3.386&4.8&10.89&10.52&10.19&9.39&9.84&10.95&10.65&9.56&9.44&8.94&8.93&8.9\\
0.050&2493&-3.167&4.84&10.43&10.02&9.71&9.04&9.37&10.35&9.98&9.17&9.03&8.65&8.64&8.62\\
0.060&2648&-3.008&4.86&10.1&9.68&9.37&8.78&9.03&9.95&9.54&8.87&8.73&8.42&8.41&8.39\\
0.070&2762&-2.879&4.87&9.82&9.39&9.1&8.55&8.75&9.65&9.21&8.64&8.49&8.21&8.2&8.19\\
0.072&2782&-2.856&4.88&9.77&9.34&9.05&8.51&8.7&9.6&9.16&8.6&8.45&8.18&8.17&8.15\\
0.075&2809&-2.821&4.88&9.69&9.26&8.97&8.44&8.63&9.51&9.07&8.52&8.37&8.11&8.1&8.08\\
0.080&2846&-2.776&4.88&9.6&9.16&8.87&8.36&8.53&9.39&8.95&8.43&8.28&8.03&8.02&8\\
0.090&2910&-2.689&4.88&9.4&8.96&8.68&8.19&8.35&9.16&8.7&8.24&8.09&7.86&7.85&7.83\\
0.100&2960&-2.617&4.89&9.24&8.8&8.52&8.05&8.19&8.98&8.53&8.09&7.94&7.72&7.71&7.7\\
\end{tabular}
\tablecomments{$^1$ Assumed central wavelengths for the MIRI/FQPM narrow band filters
are 10.65, 11.4, and 15.5 $\mu$m with a simple square passband with resolution=20.}
\end{table}

\clearpage
\begin{table}
\caption{Calculated Planet Magnitudes (10 pc) For 120 Myr}\footnotesize
\tiny
\begin{tabular}{lrrr|rrrrr|rrrr|rrr}
M$_{pl}$&T$_{eff}$&L$_{pl}$&log(g)&J&H&K&L&M&F150W&F277W&F356W&F444W&MIRI-1&MIRI-2&MIRI-3\\
M$\odot$&(K)&L$_\odot$&cgs&(mag)&(mag)&(mag)&(mag)&(mag)&(mag)&(mag)&(mag)&(mag)&(mag)&(mag)&(mag)\\ \hline
0.0001&109&-8.919&2.46&42.43&36.44&55.66&24.33&21.69&25.67&23.58&20.69&17.62&16.25&14.5&14.01\\
0.0002&152&-8.294&2.71&37.47&32.91&49.74&22.74&20.35&24.92&22.89&20.15&17.14&15.89&14.17&13.67\\
0.0003&179&-7.971&2.85&34.38&30.75&45.96&21.72&19.48&24.45&22.47&19.81&16.84&15.66&13.96&13.46\\
0.0004&205&-7.719&2.95&31.76&28.92&42.51&20.81&18.7&24.03&22.12&19.51&16.59&15.45&13.78&13.28\\
0.0005&227&-7.526&3.03&29.59&27.43&39.5&20.03&18.03&23.68&21.87&19.28&16.38&15.29&13.64&13.14\\
0.001&295&-7.043&3.31&23.06&22.92&30.33&17.71&15.92&22.68&21.16&18.64&15.83&14.87&13.28&12.81\\
0.002&403&-6.481&3.59&20.41&20.14&23.91&16.18&14.75&19.93&19.34&17.01&14.38&13.68&12.34&12.02\\
0.003&475&-6.183&3.75&19.17&18.88&21.55&15.42&14.09&18.97&19.16&16.68&14.1&13.72&12.33&11.95\\
0.004&537&-5.97&3.88&18.29&18.03&19.95&14.84&13.66&18.25&18.72&16.25&13.75&13.4&12.07&11.72\\
0.005&599&-5.781&3.97&17.53&17.32&18.69&14.3&13.31&17.6&18.17&15.76&13.35&12.87&11.67&11.38\\
0.006&665&-5.602&4.05&16.94&16.75&17.67&13.85&12.97&17.11&17.57&15.23&12.98&12.32&11.31&11.07\\
0.007&716&-5.477&4.13&16.51&16.35&16.95&13.53&12.73&16.74&17.11&14.84&12.7&11.93&11.04&10.85\\
0.008&783&-5.326&4.19&16.02&15.9&16.18&13.17&12.48&16.34&16.59&14.4&12.43&11.55&10.8&10.64\\
0.009&835&-5.219&4.24&15.67&15.58&15.68&12.92&12.31&16.05&16.21&14.09&12.23&11.29&10.63&10.49\\
0.010&900&-5.088&4.29&15.26&15.18&15.09&12.6&12.11&15.69&15.77&13.71&12&11&10.43&10.32\\
0.012&1273&-4.432&4.31&13.45&13.25&13.03&11.11&11.29&13.86&13.51&11.83&10.99&9.87&9.71&9.68\\
0.015&1298&-4.436&4.45&13.43&13.27&13.07&11.15&11.34&13.86&13.49&11.85&11.02&9.93&9.77&9.74\\
0.020&1484&-4.218&4.59&12.87&12.65&12.51&10.75&11.13&13.24&12.87&11.29&10.74&9.76&9.67&9.6\\
0.030&1905&-3.764&4.74&11.75&11.44&11.17&9.98&10.56&12.02&11.75&10.26&10.1&9.36&9.34&9.29\\
0.040&2204&-3.472&4.83&11.08&10.72&10.4&9.54&10.03&11.18&10.91&9.73&9.61&9.07&9.06&9.03\\
0.050&2441&-3.244&4.88&10.6&10.2&9.89&9.19&9.55&10.54&10.2&9.32&9.18&8.78&8.77&8.75\\
0.060&2606&-3.083&4.91&10.26&9.84&9.54&8.93&9.2&10.13&9.75&9.04&8.9&8.57&8.56&8.54\\
0.070&2733&-2.944&4.92&9.97&9.54&9.25&8.69&8.91&9.79&9.37&8.77&8.63&8.34&8.33&8.31\\
0.072&2753&-2.922&4.92&9.92&9.49&9.2&8.65&8.86&9.76&9.33&8.74&8.6&8.32&8.31&8.29\\
0.075&2783&-2.886&4.92&9.84&9.41&9.12&8.58&8.78&9.67&9.24&8.67&8.52&8.25&8.24&8.22\\
0.080&2824&-2.838&4.93&9.74&9.3&9.02&8.5&8.68&9.55&9.12&8.57&8.43&8.17&8.16&8.14\\
0.090&2894&-2.753&4.94&9.56&9.11&8.83&8.34&8.5&9.31&8.87&8.39&8.24&8.01&8&7.98\\
0.100&2949&-2.673&4.94&9.38&8.93&8.66&8.18&8.33&9.12&8.68&8.23&8.08&7.86&7.85&7.83\\
\end{tabular}
\tablecomments{$^1$ Assumed central wavelengths for the MIRI/FQPM narrow band filters
are 10.65, 11.4, and 15.5 $\mu$m with a simple square passband with resolution=20.}
\end{table}

\clearpage
\begin{table}
\caption{Calculated Planet Magnitudes (10 pc) For 500 Myr}\footnotesize
\tiny
\begin{tabular}{lrrr|rrrrr|rrrr|rrr}
M$_{pl}$&T$_{eff}$&L$_{pl}$&log(g)&J&H&K&L&M&F150W&F277W&F356W&F444W&MIRI-1&MIRI-2&MIRI-3\\
M$\odot$&(K)&L$_\odot$&cgs&(mag)&(mag)&(mag)&(mag)&(mag)&(mag)&(mag)&(mag)&(mag)&(mag)&(mag)&(mag)\\ \hline
0.0004&128&-8.602&3.01&38.69&33.88&53.5&23.59&21.03&25.18&22.84&20.3&17.28&16.1&14.33&13.79\\
0.0005&141&-8.415&3.1&37.42&33.07&51.62&23.09&20.59&24.97&22.66&20.14&17.14&16&14.23&13.69\\
0.001&203&-7.753&3.37&31.61&29.15&43.23&20.93&18.68&24.04&21.98&19.52&16.61&15.59&13.87&13.35\\
0.002&272&-7.218&3.64&25.07&24.62&34.02&18.66&16.58&22.99&21.3&18.9&16.04&15.21&13.55&13.05\\
0.003&322&-6.913&3.81&22.05&22.27&29.03&17.52&15.53&22.27&20.82&18.48&15.65&14.92&13.33&12.85\\
0.004&370&-6.67&3.93&21.01&21.06&26.45&16.85&15.05&21.62&20.34&18.06&15.28&14.58&13.08&12.62\\
0.005&409&-6.496&4.03&20.2&20.11&24.54&16.32&14.67&19.97&19.32&17.13&14.46&13.85&12.53&12.16\\
0.006&449&-6.34&4.11&19.64&19.51&23.1&15.92&14.36&19.48&19.17&16.97&14.31&13.87&12.53&12.15\\
0.007&488&-6.2&4.18&19.1&18.91&21.68&15.52&14.06&19.04&19.04&16.83&14.19&13.88&12.54&12.14\\
0.008&525&-6.08&4.25&18.65&18.46&20.74&15.19&13.83&18.68&18.78&16.6&14&13.67&12.38&12\\
0.009&564&-5.963&4.31&18.21&18.04&19.95&14.88&13.63&18.3&18.44&16.3&13.76&13.37&12.15&11.8\\
0.010&599&-5.864&4.36&17.8&17.66&19.22&14.59&13.43&17.93&18.13&16&13.52&13.08&11.92&11.61\\
0.012&759&-5.447&4.43&16.38&16.29&16.82&13.51&12.68&16.7&16.84&14.76&12.67&11.89&11.11&10.92\\
0.015&791&-5.404&4.56&16.2&16.16&16.58&13.41&12.6&16.56&16.62&14.61&12.58&11.78&11.06&10.88\\
0.020&936&-5.133&4.7&15.33&15.34&15.37&12.78&12.21&15.82&15.72&13.85&12.14&11.23&10.71&10.58\\
0.030&1264&-4.636&4.91&13.9&13.87&13.69&11.7&11.68&14.41&14.07&12.48&11.44&10.43&10.22&10.18\\
0.040&1583&-4.255&5.04&12.92&12.76&12.68&10.95&11.33&13.32&12.96&11.46&10.95&10.06&9.98&9.9\\
0.050&1875&-3.955&5.13&12.21&11.95&11.79&10.44&10.99&12.49&12.22&10.74&10.53&9.79&9.76&9.69\\
0.060&2116&-3.729&5.19&11.69&11.36&11.13&10.1&10.63&11.87&11.65&10.32&10.21&9.6&9.58&9.54\\
0.070&2329&-3.534&5.23&11.27&10.9&10.63&9.81&10.26&11.28&11.04&9.96&9.85&9.37&9.36&9.33\\
0.072&2369&-3.498&5.24&11.19&10.81&10.54&9.75&10.19&11.18&10.93&9.9&9.78&9.33&9.31&9.29\\
0.075&2426&-3.445&5.24&11.08&10.69&10.42&9.67&10.08&11.04&10.78&9.82&9.7&9.27&9.26&9.23\\
0.080&2518&-3.356&5.25&10.89&10.49&10.22&9.53&9.89&10.81&10.52&9.66&9.54&9.16&9.14&9.12\\
0.090&2680&-3.189&5.24&10.55&10.12&9.85&9.25&9.53&10.37&10.04&9.35&9.22&8.9&8.89&8.88\\
0.100&2804&-3.047&5.22&10.25&9.8&9.54&9&9.22&10.05&9.68&9.09&8.95&8.68&8.67&8.65\\
\end{tabular}
\tablecomments{$^1$ Assumed central wavelengths for the MIRI/FQPM narrow band filters
are 10.65, 11.4, and 15.5 $\mu$m with a simple square passband with resolution=20.}
\end{table}

\clearpage
\begin{table}
\caption{Calculated Planet Magnitudes (10 pc) For 1 Gyr}\footnotesize
\tiny
\begin{tabular}{lrrr|rrrrr|rrrr|rrr}M$_{pl}$&T$_{eff}$&L$_{pl}$&log(g)&J&H&K&L&M&F150W&F277W&F356W&F444W&MIRI-1&MIRI-2&MIRI-3\\
M$\odot$&(K)&L$_\odot$&cgs&(mag)&(mag)&(mag)&(mag)&(mag)&(mag)&(mag)&(mag)&(mag)&(mag)&(mag)&(mag)\\ \hline
0.0005&111&-8.851&3.12&40.19&35.07&55.87&24.15&21.49&25.42&22.94&20.45&17.41&16.26&14.44&13.89\\
0.001&160&-8.185&3.39&35.58&32.06&49.17&22.4&19.95&24.67&22.36&19.93&16.97&15.93&14.14&13.6\\
0.002&226&-7.56&3.66&29.28&27.8&40.31&20.18&17.94&23.65&21.73&19.35&16.42&15.58&13.86&13.32\\
0.003&270&-7.244&3.83&25.29&24.99&34.51&18.84&16.66&22.98&21.3&18.96&16.05&15.33&13.66&13.14\\
0.004&304&-7.031&3.95&22.49&22.91&30.25&17.91&15.73&22.52&20.98&18.69&15.8&15.14&13.52&13.01\\
0.005&342&-6.831&4.05&21.71&21.98&28.28&17.38&15.36&22.03&20.61&18.38&15.53&14.91&13.35&12.87\\
0.006&377&-6.664&4.13&20.96&21.07&26.39&16.87&15.01&21.62&20.27&18.12&15.29&14.68&13.17&12.72\\
0.007&403&-6.556&4.21&20.41&20.43&25.04&16.54&14.76&20.14&19.32&17.27&14.54&13.98&12.63&12.29\\
0.008&438&-6.417&4.27&19.93&19.89&23.77&16.18&14.5&19.75&19.2&17.14&14.44&13.98&12.64&12.28\\
0.009&464&-6.325&4.33&19.59&19.49&22.85&15.93&14.31&19.49&19.11&17.06&14.37&13.98&12.65&12.27\\
0.010&491&-6.235&4.38&19.23&19.07&21.9&15.65&14.11&19.2&18.99&16.95&14.27&13.94&12.64&12.24\\
0.012&578&-5.955&4.47&18.15&18.03&19.86&14.88&13.6&18.28&18.3&16.28&13.73&13.33&12.14&11.81\\
0.015&628&-5.835&4.59&17.7&17.62&19.05&14.56&13.38&17.89&17.88&15.92&13.47&12.97&11.92&11.63\\
0.020&766&-5.514&4.74&16.56&16.55&17.19&13.72&12.8&16.91&16.83&14.94&12.81&12.08&11.34&11.14\\
0.030&1009&-5.071&4.95&15.1&15.16&15.14&12.67&12.15&15.64&15.4&13.67&12.07&11.17&10.74&10.62\\
0.040&1271&-4.696&5.1&14.04&14.04&13.9&11.88&11.8&14.57&14.2&12.66&11.58&10.62&10.41&10.36\\
0.050&1543&-4.374&5.21&13.21&13.12&13.04&11.25&11.53&13.63&13.24&11.8&11.18&10.29&10.19&10.12\\
0.060&1801&-4.106&5.29&12.56&12.36&12.27&10.77&11.26&12.87&12.56&11.12&10.82&10.05&10.01&9.94\\
0.070&2082&-3.829&5.33&11.93&11.62&11.43&10.32&10.85&12.14&11.91&10.56&10.43&9.8&9.78&9.73\\
0.072&2140&-3.772&5.34&11.8&11.48&11.27&10.23&10.75&11.97&11.75&10.45&10.33&9.74&9.72&9.67\\
0.075&2234&-3.679&5.33&11.6&11.25&11.02&10.08&10.58&11.66&11.44&10.24&10.14&9.6&9.59&9.55\\
0.080&2383&-3.527&5.32&11.26&10.89&10.63&9.84&10.27&11.25&11.01&9.98&9.87&9.42&9.41&9.38\\
0.090&2627&-3.268&5.28&10.72&10.3&10.03&9.4&9.72&10.57&10.27&9.51&9.38&9.05&9.04&9.02\\
0.100&2784&-3.083&5.25&10.33&9.89&9.62&9.07&9.31&10.13&9.77&9.15&9.02&8.74&8.73&8.71\\
\end{tabular}
\tablecomments{$^1$ Assumed central wavelengths for the MIRI/FQPM narrow band filters
are 10.65, 11.4, and 15.5 $\mu$m with a simple square passband with resolution=20.}
\end{table}

\clearpage
\begin{table}
\caption{Calculated Planet Magnitudes (10 pc) For 5 Gyr}\footnotesize
\tiny
\begin{tabular}{lrrr|rrrrr|rrrr|rrr}M$_{pl}$&T$_{eff}$&L$_{pl}$&log(g)&J&H&K&L&M&F150W&F277W&F356W&F444W&MIRI-1&MIRI-2&MIRI-3\\
M$\odot$&(K)&L$_\odot$&cgs&(mag)&(mag)&(mag)&(mag)&(mag)&(mag)&(mag)&(mag)&(mag)&(mag)&(mag)&(mag)\\ \hline
0.002&129&-8.57&3.7&38.16&34.52&53.62&23.33&20.8&25&22.61&20.24&17.19&16.35&14.46&13.86\\
0.003&162&-8.166&3.87&34.96&32.46&49.21&22.22&19.81&24.49&22.35&20.01&16.94&16.26&14.41&13.79\\
0.004&193&-7.867&3.99&32.14&30.52&45.15&21.28&18.92&24.04&22.09&19.77&16.71&16.12&14.31&13.7\\
0.005&220&-7.644&4.09&29.83&28.84&41.71&20.48&18.16&23.47&21.71&19.43&16.42&15.97&14.18&13.57\\
0.006&244&-7.469&4.18&27.8&27.34&38.67&19.78&17.49&23.06&21.42&19.19&16.22&15.85&14.08&13.48\\
0.007&265&-7.328&4.25&26.02&26&35.96&19.16&16.88&22.75&21.19&19.02&16.06&15.75&14&13.41\\
0.008&284&-7.217&4.32&24.5&24.85&33.65&18.64&16.36&22.5&20.99&18.86&15.91&15.63&13.91&13.33\\
0.009&301&-7.124&4.38&23.21&23.85&31.64&18.18&15.91&22.33&20.83&18.76&15.82&15.54&13.84&13.29\\
0.010&322&-7.015&4.43&22.74&23.3&30.49&17.91&15.71&22.17&20.68&18.66&15.71&15.42&13.76&13.24\\
0.012&361&-6.823&4.52&21.77&22.17&28.16&17.41&15.32&21.97&20.41&18.5&15.54&15.16&13.57&13.12\\
0.015&399&-6.671&4.63&20.85&21.14&26.01&16.96&14.97&21.83&20.15&18.35&15.39&14.9&13.43&13\\
0.020&473&-6.401&4.79&19.89&19.92&23.31&16.18&14.45&19.87&19.07&17.31&14.56&14.18&12.94&12.53\\
0.030&610&-6.008&5.01&18.34&18.31&20.04&15.04&13.71&18.48&18.01&16.34&13.82&13.34&12.36&12.02\\
0.040&760&-5.67&5.18&17.04&17.11&17.99&14.17&13.13&17.38&16.97&15.32&13.17&12.5&11.8&11.57\\
0.050&931&-5.353&5.31&15.94&16.05&16.38&13.38&12.63&16.41&15.97&14.38&12.6&11.82&11.34&11.17\\
0.060&1120&-5.058&5.42&15.01&15.13&15.15&12.73&12.27&15.57&15.11&13.6&12.18&11.35&11.02&10.9\\
0.070&1524&-4.504&5.47&13.52&13.5&13.44&11.6&11.77&13.97&13.56&12.19&11.47&10.6&10.49&10.42\\
0.072&1712&-4.278&5.45&12.97&12.85&12.8&11.16&11.54&13.33&12.97&11.58&11.15&10.36&10.29&10.21\\
0.075&2006&-3.942&5.41&12.2&11.92&11.78&10.54&11.06&12.42&12.17&10.78&10.63&9.97&9.94&9.89\\
0.080&2320&-3.603&5.35&11.44&11.07&10.82&9.97&10.43&11.45&11.22&10.12&10.01&9.52&9.51&9.48\\
0.090&2622&-3.275&5.29&10.73&10.31&10.04&9.42&9.73&10.58&10.28&9.52&9.39&9.05&9.04&9.02\\
0.100&2785&-3.083&5.25&10.33&9.89&9.62&9.07&9.31&10.12&9.77&9.15&9.02&8.74&8.73&8.71\\
\end{tabular}
\tablecomments{$^1$ Assumed central wavelengths for the MIRI/FQPM narrow band filters
are 10.65, 11.4, and 15.5 $\mu$m with a simple square passband with resolution=20.}
\end{table}

\clearpage
\begin{table}
\caption{Calculated Planet Magnitudes (10 pc) For 10 Gyr}\footnotesize
\tiny
\begin{tabular}{lrrr|rrrrr|rrrr|rrr}
M$_{pl}$&T$_{eff}$&L$_{pl}$&log(g)&J&H&K&L&M&F150W&F277W&F356W&F444W&MIRI-1&MIRI-2&MIRI-3\\
M$\odot$&(K)&L$_\odot$&cgs&(mag)&(mag)&(mag)&(mag)&(mag)&(mag)&(mag)&(mag)&(mag)&(mag)&(mag)&(mag)\\ \hline
0.003&125&-8.629&3.88&38.29&35.04&54.28&23.36&20.89&24.98&22.7&20.34&17.22&16.56&14.64&13.99\\
0.004&149&-8.325&4.01&36.01&33.56&51.11&22.59&20.19&24.59&22.49&20.16&17.03&16.49&14.6&13.94\\
0.005&172&-8.087&4.11&34.06&32.18&48.26&21.92&19.57&24&22.13&19.83&16.76&16.39&14.51&13.84\\
0.006&193&-7.888&4.2&32.23&30.85&45.53&21.29&18.97&23.56&21.84&19.58&16.55&16.3&14.43&13.76\\
0.007&213&-7.724&4.27&30.55&29.62&43.01&20.71&18.41&23.18&21.57&19.35&16.36&16.18&14.33&13.67\\
0.008&232&-7.584&4.34&29&28.46&40.66&20.18&17.89&22.91&21.37&19.19&16.22&16.09&14.26&13.61\\
0.009&249&-7.469&4.39&27.62&27.41&38.55&19.7&17.42&22.69&21.2&19.07&16.11&16&14.2&13.56\\
0.01&265&-7.368&4.45&26.34&26.44&36.59&19.26&16.98&22.53&21.06&18.97&16.01&15.93&14.14&13.52\\
0.012&293&-7.204&4.54&24.04&24.64&33.06&18.48&16.2&22.4&20.86&18.85&15.89&15.74&14.02&13.43\\
0.015&330&-7.016&4.65&22.7&23.33&30.46&17.93&15.76&22.41&20.65&18.75&15.78&15.46&13.88&13.33\\
0.020&389&-6.759&4.8&21.28&21.73&27.02&17.21&15.19&22.06&20.24&18.48&15.51&15.07&13.65&13.14\\
0.030&504&-6.358&5.03&19.72&19.68&22.59&16.04&14.37&19.83&18.96&17.32&14.6&14.2&13.04&12.62\\
0.040&634&-6.004&5.2&18.33&18.33&19.99&15.03&13.75&18.49&17.9&16.28&13.86&13.33&12.44&12.1\\
0.050&776&-5.695&5.34&17.1&17.18&18.07&14.23&13.21&17.43&16.94&15.37&13.26&12.59&11.93&11.69\\
0.060&941&-5.393&5.45&16.04&16.18&16.55&13.5&12.73&16.53&16.01&14.48&12.72&11.96&11.5&11.33\\
0.070&1289&-4.832&5.5&14.37&14.43&14.36&12.27&12.08&14.89&14.43&13&11.88&11.01&10.8&10.73\\
0.072&1556&-4.472&5.48&13.44&13.41&13.36&11.55&11.75&13.87&13.46&12.1&11.42&10.57&10.46&10.39\\
0.075&1997&-3.954&5.41&12.23&11.95&11.81&10.56&11.08&12.44&12.19&10.8&10.64&9.97&9.95&9.89\\
0.080&2322&-3.602&5.35&11.43&11.06&10.82&9.97&10.43&11.44&11.21&10.12&10.01&9.52&9.51&9.48\\
0.090&2624&-3.274&5.29&10.73&10.31&10.04&9.41&9.73&10.58&10.28&9.51&9.39&9.05&9.04&9.02\\
0.100&2786&-3.082&5.25&10.32&9.88&9.62&9.07&9.3&10.12&9.76&9.15&9.02&8.74&8.73&8.71\\
\end{tabular}
\tablecomments{$^1$ Assumed central wavelengths for the MIRI/FQPM narrow band filters
are 10.65, 11.4, and 15.5 $\mu$m with a simple square passband with resolution=20.}
\end{table}

\clearpage
\section{Appendix II. Selected Samples of Stars}

The following tables present information on the stars with highest fraction of
Monte Carlo runs resulting in a detected planet. Up to 25 stars are presented for each instrument configuration. The columns include: (1)-(5) characterize the spectral type, distance, age (Myr) and H magnitude (typically from 2MASS) of the star; (6) and (7) give the average mass of the detected planets; and (8) gives the fraction of runs on which a planet was detected. Columns (8)-(16) repeat this information for a different instrument.

\begin{sidewaystable}
\footnotesize
\tiny
\caption{Young Stars With Highest Probability of Planet Detection}
\begin{tabular}{lrrrr|rrr||lrrrr|rrr}
(1)&(2)&(3)&(4)&(5)&(6)&(7)&(8)&(9)&(10)&(11)&(12)&(13)&(14)(15)&(16)\\
 &Spec&Dist&Log&H& Mass & SMA&Score&Star&Spec&Dist&Log(Age)&H&Mass&SMA&Score\\
Star &Type&(pc)&Age(yr)&(mag)&(M$_{Jup}$)&(AU)&(\%)&Star &Type&(pc)&(yr) &(mag)&(M$_{Jup}$)&(AU)&(\%)\\ \hline
\multicolumn{8}{c}{\bf P1640/GPI}&\multicolumn{8}{c}{ \bf NIRCam 4.4 $\mu$m}\\ 
HD 31295& A3V&37.0&7.0&4.4&2.55&29&43&HD 172555& A5 IV/V&29.2&7.1&4.3&1.61&84&40\\
HD 46190& A0V&79.0&6.7&6.4&2.33&62&42&HD172555& A7&29.2&7.3&4.3&1.96&85&40\\
HD 110411& A3 Va&36.9&7.0&4.7&2.69&29&42&HD35850& F7&26.8&7.1&5.1&1.39&82&40\\
HD 172555& A5 IV/V&29.2&7.1&4.3&2.54&23&40&HD164249& F5&23.7&7.3&6.0&1.03&69&40\\
TWA11A& A0&60.0&7.0&5.8&2.91&45&40&HD 39060& A5V&19.3&7.3&3.5&1.74&56&40\\
BetaPic& A3V&19.3&7.3&3.5&2.74&16&40&V383 LAC& K0VIV&33.0&7.8&6.6&1.06&87&39\\
HD 109573& A0V&67.1&6.9&5.8&2.86&54&40&CD-64d1208& K7&29.2&7.3&6.3&1.03&82&39\\
HD 146624& A0 (V)&43.1&7.1&4.7&2.67&32&40&HD37572& K0V&23.9&7.8&5.9&1.19&69&38\\
HD 39060& A5V&19.3&7.3&3.5&2.59&16&38&BetaPic& A3V&19.3&7.3&3.5&1.76&64&38\\
HD 188228& A0V&32.5&7.0&4.0&2.45&26&38&HD207129& G2&15.6&7.3&4.3&1.24&50&38\\
HD 183324& A0V&59.0&7.0&5.5&3.30&43&37&51Eri& F0V&29.8&7.3&4.8&1.67&87&38\\
HD 30422& A3IV&57.5&7.0&5.7&2.82&43&37&HD139813& G5&21.7&8.3&5.6&1.15&67&38\\
PreibZinn99-67& K5&145.0&6.0&8.7&2.44&74&36&HD17925& K1V&10.4&7.9&4.2&1.16&34&38\\
HD 38206& A0V&69.2&7.0&5.8&2.95&48&34&HIP23309& M0/1&26.3&7.3&6.4&0.46&82&38\\
TYC7349-2191-1& K2-IV&130.0&6.0&8.4&2.35&77&32&HD217343& G3V&32.0&7.6&6.0&1.27&94&37\\
CITau& K0?&140.0&5.8&7.7&2.32&81&32&GJ3305& M0.5&29.8&7.3&9.4&0.48&82&37\\
PreibZinn99-84& K3&145.0&6.0&9.1&2.46&74&31&HD135363& G5(V)&29.4&7.8&6.3&1.14&84&37\\
HR9& F2IV&39.1&7.3&5.3&3.51&27&31&RE J0723+20& K3(V)&30.0&8.1&7.0&1.09&87&37\\
PreibZinn99-19& K6&145.0&5.9&8.3&2.60&76&31&GJ803& M1&9.9&7.3&4.8&0.47&30&37\\
PreibZinn99-68& K2&145.0&6.0&8.0&2.57&83&31&HD 188228& A0V&32.5&7.0&4.0&2.09&89&36\\
HD172555& A7&29.2&7.3&4.3&3.28&24&31&HD155555C& M4.5&31.4&7.3&7.7&0.51&85&36\\
BP-Tau& K7&140.0&5.8&8.2&2.28&78&30&HD76218& G9-V&26.2&8.7&5.9&1.25&81&36\\
RX-J1844.3-3541& K5&130.0&6.2&8.6&2.59&77&30&HD25457& F7V&19.2&8.1&4.3&1.33&66&36\\
HD207129& G2&15.6&7.3&4.3&2.93&14&30&HD105& G0V&40.1&7.5&6.2&1.30&96&35\\
HR136& A0&45.0&7.3&5.2&3.62&35&30&HD25300& K0&30.0&8.2&6.9&0.97&88&35\\

\end{tabular}
\tablecomments{Columns (1)-(5) and (9)- (13) describe the name, spectral type, distance, age, and H magnitude of the star. Columns (6)-(8) and (14)-(16) characterize the average mass, semi-major axis, and fraction of planets of detected for that star. ``PreibZinn'' sources refer to objects cited in  Preibisch \& Zinnecker(1999).}

\label{BestStars1}
\end{sidewaystable}

\clearpage
\begin{sidewaystable}
\footnotesize
\tiny
\caption{Young Stars With Highest Probability of Planet Detection}
\begin{tabular}{lrrrr|rrr||lrrrr|rrr}
(1)&(2)&(3)&(4)&(5)&(6)&(7)&(8)&(9)&(10)&(11)&(12)&(13)&(14)(15)&(16)\\
 &Spec&Dist&Log(Age)&H& Mass & SMA&Score&Star&Spec&Dist&Log(Age)&H&Mass&SMA&Score\\
Star &Type&(pc)&(yr) &(mag)&(M$_{Jup}$)&(AU)&(\%)&Star &Type&(pc)&(yr) &(mag)&(M$_{Jup}$)&(AU)&(\%)\\ \hline
\multicolumn{8}{c}{ \bf JWST TFI/NRM 4.4 $\mu$m}&\multicolumn{8}{c}{ \bf JWST/MIRI 11.4 $\mu$m}\\ 
TWA9B& M1&60.0&7.0&9.4&0.40&16&49&GJ803& M1&9.9&7.3&4.8&0.51&34&85\\
PreibZinn99-84& K3&145.0&6.0&9.1&1.29&37&45&GJ3305& M0.5&29.8&7.3&9.4&0.46&49&82\\
TWA17& K5&60.0&7.0&9.2&1.08&17&45&HD164249& F5&23.7&7.3&6.0&1.23&54&82\\
GI-Tau& K7&140.0&6.3&9.4&1.34&36&45&HD155555C& M4.5&31.4&7.3&7.7&0.46&51&76\\
PreibZinn99-11& K7:&145.0&5.9&8.5&1.12&40&45&HD207129& G2&15.6&7.3&4.3&1.37&54&75\\
TYC7349-2191-1& K2-IV&130.0&6.0&8.4&1.16&35&45&HIP23309& M0/1&26.3&7.3&6.4&0.53&52&72\\
GSC8862-0019& K4V&60.0&7.5&9.0&0.92&16&44&CD-64d1208& K7&29.2&7.3&6.3&1.32&53&71\\
TWA18& M0.5&60.0&7.0&9.1&0.45&18&44&AOMen& K6&38.5&7.3&7.0&1.43&59&71\\
HD285751/v1200-Tau& K2(V)&82.0&6.8&8.9&0.87&23&43&HIP1993& K7&45.0&7.3&7.9&1.42&55&69\\
TWA12& M2&60.0&7.0&8.3&0.43&16&43&HD37572& K0V&23.9&7.8&5.9&1.54&58&69\\
PreibZinn99-33& M1&145.0&5.9&9.1&0.52&35&43&2RE-J0255+474& K5Ve&33.0&7.9&7.3&1.58&57&68\\
RX-J1844.3-3541& K5&130.0&6.2&8.6&1.15&32&43&PPM366328& K0&45.0&7.3&7.7&1.24&65&68\\
PreibZinn99-69& K5&145.0&6.3&8.8&1.18&37&42&HD17925& K1V&10.4&7.9&4.2&1.34&42&67\\
PreibZinn99-27& M3&145.0&5.7&9.0&0.50&38&42&QT-AND& G0?&39.0&7.8&7.5&1.60&61&66\\
PreibZinn99-67& K5&145.0&6.0&8.7&1.18&35&42&HD 172555& A5 IV/V&29.2&7.1&4.3&2.16&56&66\\
TWA9A& K5&60.0&7.0&8.0&0.87&16&42&HD987& G6&43.7&7.3&7.1&1.48&65&65\\
V830-Tau& K7&140.0&6.3&8.6&1.11&37&42&RE-J0723+20& K3(V)&30.0&8.1&7.0&1.55&67&64\\
1RXS-J043243.2-152003& G4V&117.0&6.6&8.7&1.14&32&42&HD35850& F7&26.8&7.1&5.1&1.66&57&64\\
PreibZinn99-70& M1&145.0&5.7&9.1&0.52&39&42&HD 39060& A5V&19.3&7.3&3.5&2.27&61&64\\
V836Tau& K7&140.0&6.3&9.1&1.38&35&42&HD200798& A5&45.0&7.3&6.1&2.32&57&64\\
TWA10& M2.5&60.0&7.0&8.5&0.47&16&42&TWA9A& K5&60.0&7.0&8.0&1.45&60&64\\
PreibZinn99-59& M0&145.0&6.3&9.2&0.55&35&41&TWA17& K5&60.0&7.0&9.2&1.42&60&63\\
TWA13& M1&60.0&7.0&7.7&0.48&16&41&TWA6& K7&60.0&7.0&8.2&1.41&69&63\\
HIP1993& K7&45.0&7.3&7.9&0.98&12&41&BetaPic& A3V&19.3&7.3&3.5&1.99&70&62\\
RECX10& M0&100.0&6.9&8.9&0.57&26&41&TWA19B& K7&60.0&7.0&8.5&1.59&64&61\\
\end{tabular}
\tablecomments{Columns (1)-(5) and (9)- (13) describe the name, spectral type, distance, age, and H magnitude of the star. Columns (6)-(8) and (14)-(16) characterize the average mass, semi-major axis, and fraction of planets of detected for that star. ``PreibZinn'' sources refer to objects cited in  Preibisch \& Zinnecker(1999).}
\label{BestStars2}
\end{sidewaystable}

\clearpage
\begin{sidewaystable}
\footnotesize
\tiny
\caption{M Stars With Highest Probability of Planet Detection}
\begin{tabular}{lrrrr|rrr||lrrrr|rrr}
(1)&(2)&(3)&(4)&(5)&(6)&(7)&(8)&(9)&(10)&(11)&(12)&(13)&(14)(15)&(16)\\
 &Spec&Dist&Log(Age)&H& Mass & SMA&Score&Star&Spec&Dist&Log(Age)&H&Mass&SMA&Score\\
Star &Type&(pc)&(yr) &(mag)&(M$_{Jup}$)&(AU)&(\%)&Star &Type&(pc)&(yr) &(mag)&(M$_{Jup}$)&(AU)&(\%)\\ \hline
\multicolumn{8}{c}{ \bf JWST NIRCam 4.4 $\mu$m}&\multicolumn{8}{c}{ \bf JWST/TFI 4.4 $\mu$m}\\ 
GJ 268& M4.5&6.4&8.8&6.2&0.44&20&0.47&GJ 4040& M3&13.6&9.1&7.3&0.43&4&0.28\\
GJ 875.1& M3&14.2&7.8&7.1&0.44&43&0.46&GJ 2151& M3.0&14.2&9.7&8.2&0.50&5&0.28\\
GJ 494& M0.5&11.4&7.3&5.8&0.43&35&0.45&GJ 3148A& M3 V &14.3&8.3&7.3&0.49&5&0.27\\
GJ 595& M3&8.2&9.7&7.4&0.44&26&0.45&GJ 875.1& M3&14.2&7.8&7.1&0.46&5&0.27\\
GJ 735& M3&11.6&7.4&5.7&0.51&33&0.45&GJ 813& M2&13.7&9.7&7.9&0.45&5&0.27\\
GJ 781.1A& M3.0&14.6&8.2&7.7&0.48&48&0.44&GJ 206& M3.5&12.8&7.9&6.9&0.46&4&0.27\\
GJ 793& M2.5&8.0&8.9&6.1&0.46&25&0.44&GJ 781.1A& M3.0&14.6&8.2&7.7&0.41&5&0.27\\
GJ 896A& M3.5&6.3&8.2&5.6&0.47&20&0.44&GJ 1284& M3&14.8&7.7&6.6&0.44&5&0.26\\
GJ 277B& M3.5&11.5&7.7&7.0&0.47&36&0.44&GJ 360& M2&11.7&8.5&6.3&0.47&4&0.26\\
GJ 465& M2&8.9&9.7&7.3&0.51&31&0.43&GJ 799& M4.5&10.2&7.3&5.2&0.37&4&0.26\\
GJ 402& M4&5.6&9.7&6.7&0.45&18&0.43&GJ 3846& M2.5&14.3&9.7&7.8&0.49&5&0.26\\
GJ 54.1& M4.5&3.7&9.1&6.8&0.47&12&0.43&GJ 82& M4&12.0&8.0&7.2&0.38&4&0.25\\
GJ 9492& M1.5&9.9&9.1&6.7&0.44&31&0.43&GJ 9724& M1.5&14.3&9.7&7.9&0.49&5&0.25\\
GJ 285& M4.5&5.9&8.1&6.0&0.41&19&0.43&GJ 3459& M3&10.9&9.7&7.3&0.44&4&0.25\\
GJ 597& M3&13.1&9.7&7.7&0.47&43&0.43&GJ 597& M3&13.1&9.7&7.7&0.52&5&0.25\\
GJ 9520& M1.5&11.4&7.7&6.0&0.45&35&0.43&GJ 606& M1&13.9&8.9&6.6&0.63&5&0.24\\
GJ 273& M3.5&3.8&9.8&5.2&0.49&12&0.43&HIP 12261& M3 V&14.9&9.7&7.8&0.49&5&0.24\\
GJ 431& M3.5&10.5&8.0&6.8&0.50&31&0.42&GJ 9520& M1.5&11.4&7.7&6.0&0.46&4&0.23\\
GJ 3135& M2.5+ &9.5&9.7&7.9&0.48&28&0.42&GJ 816& M3&13.8&9.2&7.0&0.61&4&0.23\\
GJ 4247& M4.5 V&9.0&8.3&7.0&0.47&29&0.42&GJ 9773& M3&13.9&9.7&7.7&0.54&5&0.22\\
GJ 103& M0 VP&11.5&6.4&5.1&0.43&37&0.42&GJ 179& M3.5&12.1&9.7&7.2&0.58&4&0.22\\
GJ 203& M3.5&8.7&9.7&7.8&0.45&27&0.42&GJ 145& M2.5 V&10.9&9.2&7.2&0.46&4&0.22\\
GJ 1105& M3.5&8.2&9.7&7.1&0.46&26&0.42&GJ 469& M3.5&13.6&9.7&7.2&0.66&5&0.22\\
GJ 362& M3&11.5&8.5&6.7&0.46&37&0.42&GJ 553.1& M3.5&11.1&9.7&7.3&0.51&4&0.22\\
GJ 745A& M1.5&8.6&9.7&6.7&0.38&29&0.42&GJ 821& M1&12.2&9.7&7.1&0.56&4&0.21\\
\end{tabular}
\tablecomments{Columns (1)-(5) and (9)- (13) describe the name, spectral type, distance, age, and H magnitude of the star.
Columns (6)-(8) and (14)-(16) characterize the average mass, semi-major axis, and fraction of planets of detected for that star.}
\label{BestStars3}
\end{sidewaystable}

\clearpage
\begin{sidewaystable}
\footnotesize
\tiny
\caption{M Stars With Highest Probability of Planet Detection}
\begin{tabular}{lrrrr|rrr||lrrrr|rrr}
(1)&(2)&(3)&(4)&(5)&(6)&(7)&(8)&(9)&(10)&(11)&(12)&(13)&(14)(15)&(16)\\
 &Spec&Dist&Log(Age)&H& Mass & SMA&Score&Star&Spec&Dist&Log(Age)&H&Mass&SMA&Score\\
Star &Type&(pc)&(yr) &(mag)&(M$_{Jup}$)&(AU)&(\%)&Star &Type&(pc)&(yr) &(mag)&(M$_{Jup}$)&(AU)&(\%)\\ \hline
\multicolumn{8}{c}{ \bf JWST MIRI 11.4 $\mu$m}&\multicolumn{8}{c}{ \bf TMT 1.65 $\mu$m}\\
GJ 206& M3.5&12.8&7.9&6.9&0.46&39&93&GJ 103& M0 VP&11.5&6.4&5.1&0.54&15&63\\
GJ 494& M0.5&11.4&7.3&5.8&0.42&36&93&GJ 799& M4.5&10.2&7.3&5.2&0.80&16&40\\
GJ 1284& M3&14.8&7.7&6.6&0.44&45&92&GJ 803& M1&9.9&7.3&4.8&0.85&14&39\\
GJ 82& M4&12.0&8.0&7.2&0.46&38&92&GJ 494& M0.5&11.4&7.3&5.8&0.86&15&37\\
GJ 9520& M1.5&11.4&7.7&6.0&0.47&37&91&GJ 735& M3&11.6&7.4&5.7&1.02&17&33\\
GJ 103& M0 VP&11.5&6.4&5.1&0.45&39&91&GJ 643& M3.5&6.5&7.6&7.1&0.92&10&29\\
GJ 277B& M3.5&11.5&7.7&7.0&0.44&36&91&GJ 1284& M3&14.8&7.7&6.6&1.23&18&26\\
GJ 875.1& M3&14.2&7.8&7.1&0.46&40&91&GJ 644A& M2 V&5.7&7.7&4.8&1.01&8&26\\
GJ 277A& M2.5&11.4&7.7&6.2&0.44&34&91&GJ 277B& M3.5&11.5&7.7&7.0&1.07&16&25\\
GJ 781.1A& M3.0&14.6&8.2&7.7&0.44&42&90&GJ 277A& M2.5&11.4&7.7&6.2&1.13&15&25\\
GJ 735& M3&11.6&7.4&5.7&0.44&37&88&GJ 9520& M1.5&11.4&7.7&6.0&1.10&16&24\\
GJ 431& M3.5&10.5&8.0&6.8&0.44&36&88&GJ 875.1& M3&14.2&7.8&7.1&1.35&19&21\\
GJ 643& M3.5&6.5&7.6&7.1&0.47&26&86&GJ 206& M3.5&12.8&7.9&6.9&1.40&15&16\\
GJ 803& M1&9.9&7.3&4.8&0.55&36&85&GJ 82& M4&12.0&8.0&7.2&1.45&15&14\\
GJ 799& M4.5&10.2&7.3&5.2&0.48&33&84&GJ 431& M3.5&10.5&8.0&6.8&1.48&12&14\\
GJ 285& M4.5&5.9&8.1&6.0&0.44&23&81&GJ 285& M4.5&5.9&8.1&6.0&1.35&11&13\\
GJ 644A& M2 V&5.7&7.7&4.8&0.47&24&80&GJ 896A& M3.5&6.3&8.2&5.6&1.39&10&12\\
GJ 896A& M3.5&6.3&8.2&5.6&0.46&26&80&GJ 4247& M4.5 V&9.0&8.3&7.0&1.75&14&7\\
GJ 3148A& M3 V &14.3&8.3&7.3&0.51&48&75&GJ 781.1A& M3.0&14.6&8.2&7.7&1.80&17&6\\
GJ 4247& M4.5 V&9.0&8.3&7.0&0.49&38&70&GJ 3148A& M3 V&14.3&8.3&7.3&1.81&13&4\\
GJ 208& M0.5 V &11.4&8.4&5.4&0.51&56&58&GJ 208& M0.5 V&11.4&8.4&5.4&1.87&14&3\\
GJ 487& M3&10.2&8.5&6.3&0.60&47&51&GJ 725A& K5&3.6&9.7&4.7&0.40&0&2\\
GJ 699& M4&1.8&9.9&4.8&0.49&12&46&GJ 725B& K5&3.5&9.7&5.2&0.43&0&1\\
GJ 447& M4&3.3&9.5&6.0&0.52&21&45&GJ 729& M3.5&3.0&8.9&5.7&0.34&0&1\\
GJ 362& M3&11.5&8.5&6.7&0.84&41&42&GJ 411& M2V&2.6&9.5&3.7&1.40&0&0\\
\end{tabular}
\tablecomments{Columns (1)-(5) and (9)- (13) describe the name, spectral type, distance, age, and H magnitude of the star.
Columns (6)-(8) and (14)-(16) characterize the average mass, semi-major axis, and fraction of planets of detected for that star.}
\label{BestStars4}
\end{sidewaystable}

\end{document}